\documentstyle[preprint,aps,eqsecnum]{revtex}
\tightenlines
\begin{document}
\draft

\title{The Rest-Frame Instant Form of Relativistic Perfect Fluids with
Equation of State $\rho = \rho (n,s)$ and of Non-Dissipative Elastic Materials.}

\author{Luca Lusanna}

\address{
Sezione INFN di Firenze\\
L.go E.Fermi 2 (Arcetri)\\
50125 Firenze, Italy\\
E-mail LUSANNA@FI.INFN.IT}

\author{and}

\author{Dobromila Nowak-Szczepaniak}

\address
{Institute of Theoretical Physics\\
 University of Wroc\l aw\\
 pl. M.Borna 9\\
 50-204 Wroc\l aw, Poland\\
 E-mail dobno@ift.uni.wroc.pl}

\maketitle
\begin{abstract}

For perfect fluids with equation of state $\rho =\rho (n,s)$, Brown
\cite{brown} gave an action principle depending only on their Lagrange
coordinates $\alpha^i(x)$ without Clebsch potentials. After a
reformulation on arbitrary spacelike hypersurfaces in Minkowski
spacetime, the Wigner-covariant rest-frame instant form of these
perfect fluids is given. Their Hamiltonian invariant mass can be given
in closed form for the dust and the photon gas. The action for the
coupling to tetrad gravity is given. Dixon's multipoles for the
perfect fluids are studied on the rest-frame Wigner hyperplane. It is
also shown that the same formalism can be applied to non-dissipative
relativistic elastic materials described in terms of Lagrangian
coordinates.

\vskip 1truecm
\noindent \today
\vskip 1truecm

\end{abstract}

\newpage

\vfill\eject

\section{Introduction}

\newcommand{\beq}{\begin{equation}}
\newcommand{\eeq}{\end{equation}}
\newcommand{\bea}{\begin{eqnarray}}
\newcommand{\eea}{\end{eqnarray}}

Stability of stellar models for rotating stars, gravity-fluid models,
neutron stars, accretion discs around compact objects, collapse of
stars, merging of compact objects are only some of the many topics in
astrophysics and cosmology in which relativistic hydrodynamics is the
basic underlying theory. This theory is also needed in heavy-ions
collisions.

As shown in Ref.\cite{brown} there are many ways to describe
relativistic perfect fluids by means of action functionals both in
special and general relativity. Usually, besides the thermodynamical
variables $n$ (particle number density), $\rho$ (energy density), $p$
(pressure), $T$ (temperature), $s$ (entropy per particle), which are
spacetime scalar fields whose values represent measurements made in
the rest frame of the fluid (Eulerian observers), one characterizes
the fluid motion by its unit timelike 4-velocity vector field
$U^{\mu}$ (see Appendix A for a review of the relations among the
local thermodynamical variables and Appendix B for a review of
covariant relativistic thermodynamics following Ref.\cite{israel}).
However, these variables are constrained due to [we use a general
relativistic notation: ``$;\mu$" denotes a covariant
derivative]\hfill\break i) particle number conservation, $(n
U^{\mu})_{;\mu}=0$;\hfill\break ii) absence of entropy exchange
between neighbouring flow lines, $(n s U^{\mu})_{;\mu}=0$;\hfill\break
iii) the requirement that the fluid flow lines should be fixed on the
boundary.\hfill\break Therefore one needs Lagrange multipliers to
incorporate i) anf ii) into the action and this leads to use Clebsch
(or velocity-potential) representations of the 4-velocity and action
functionals depending on many redundant variables, generating first
and second class constraints at the Hamiltonian level (see Appendix
A).

Following Ref.\cite{lin} the previous constraint iii) may be enforced by
replacing the unit 4-velocity $U^{\mu}$ with a set of spacetime scalar fields
${\tilde \alpha}^i(z)$, $i=1,2,3,$ interpreted as ``Lagrangian (or comoving)
coordinates for the fluid" labelling the fluid flow lines (physically
determined by the average particle motions) passing through the points inside
the boundary (on the boundary they are fixed: either the ${\tilde \alpha}
^i(z^o,\vec z)$'s have a compact boundary $V_{\alpha}(z^o)$ or they have
assigned boundary conditions at spatial infinity).
This requires the choice of an arbitrary
spacelike hypersurface on which the $\alpha^i$'s are the 3-coordinates.
A similar point of view is contained in the concept of ``material space" of
Refs.\cite{tul,tul1}, describing the collection of all the idealized points of
the material; besides to non-dissipative isentropic fluids the scheme can be
applied to isotropic elastic media and anisotropic (crystalline) materials,
namely to an arbitrary non-dissipative relativistic continuum
\cite{magli}. See Ref.\cite{adler} for the study of the transformation from
Eulerian to Lagrangian coordinates (in the non-relativistic framework of the
Euler-Newton equations).

Notice that the use of Lagrangian (comoving) coordinates in place of
Eulerian quantities allows the use of standard Poisson brackets in the
Hamiltonian description, avoiding the formulation with Lie-Poisson
brackets of Ref.\cite{bao}, which could be recovered by a so-called
Lagrangian to Eulerian map.

Let $M^4$ be a curved globally hyperbolic spacetime [with signature $\epsilon
(+---)$, $\epsilon =\pm$]
whose points have locally
coordinates $z^{\mu}$. Let ${}^4g_{\mu\nu}(z)$ be its 4-metric with determinant
${}^4g= |det\, {}^4g_{\mu\nu}|$. Given a perfect fluid with Lagrangian
coordinates ${\bf {\tilde \alpha}}(z)=\{ {\tilde \alpha}^i(z)\}$, unit
4-velocity vector field $U^{\mu}(z)$ and particle number density $n(z)$, let
us introduce the number flux vector

\beq
n(z) U^{\mu}(z) = {{J^{\mu}({\tilde \alpha}^i(z))}\over
{\sqrt{{}^4g(z)}}},
\label{I1}
\eeq

\noindent and the densitized fluid number flux vector
or material current [$\epsilon^{0123}=1/
\sqrt{{}^4g}$; $\partial_{\alpha}(\sqrt{{}^4g}\epsilon^{\mu\nu\rho\sigma})=0$;
$\eta_{123}({\tilde \alpha}^i)$ describes the orientation of the
volume in thematerial space]

\bea
J^{\mu}({\tilde \alpha}^i(z)) &=&-\sqrt{{}^4g}
\epsilon^{\mu\nu\rho\sigma}
\eta_{123}({\bf {\tilde
\alpha}^i}(z)) \partial_{\nu}{\tilde \alpha}^1(z)\partial_{\rho}{\tilde \alpha}
^2(z)\partial_{\sigma}{\tilde \alpha}^3(z),\nonumber \\
\Rightarrow && n(z)={{|J({\tilde \alpha}^i(z))|}\over {\sqrt{{}^4g}}}=\eta_{123}({\tilde \alpha}^i
(z)){{ \sqrt{\epsilon {}^4g_{\mu\nu}(z)J^{\mu}({\tilde
\alpha}^i(z))J^{\nu}({\tilde \alpha}^i(z))} }\over {\sqrt{{}^4g(z)}}},\nonumber
\\
 \Rightarrow && \partial_{\mu}J^{\mu}({\tilde \alpha}^i(z)) =\sqrt{{}^4g}\,
 [n(z)U^{\mu}(z)]_{;\mu}=0,\nonumber \\
\Rightarrow && J^{\mu}({\tilde \alpha}^i(z))\partial_{\mu}{\tilde \alpha}^i(z)=[\sqrt{{}^4g} n
U^{\mu}](z)\partial_{\mu}{\tilde \alpha}^i(z)=0.
\label{I2}
\eea

\noindent This shows that the fluid flow lines, whose tangent vector field is
the fluid 4-velocity timelike vector field $U^{\mu}$, are identified
by ${\bf {\tilde
\alpha}^i}=const.$ and that the particle number conservation is automatic.
Moreover, if the entropy for particle is a function only of the fluid
Lagrangian coordinates, $s=s({\bf {\tilde \alpha}^i})$, the assumed
form of $J^{\mu}$ also implies automatically the absence of entropy
exchange between neighbouring flow lines, $(nsU^{\mu})_{,\mu}=0$.
Since $U^{\mu}
\partial_{\mu} s({\tilde \alpha}^i )=0$, the perfect fluid is locally adiabatic;
instead for an isentropic fluid we have $\partial_{\mu} s=0$, namely $s=const.$.

Even if in general the timelike
vector field $U^{\mu}(z)$ is not surface forming (namely
has a non-vanishing vorticity, see for instance Ref.\cite{brk}), in each point
$z$ we can consider the spacelike hypersurface orthogonal to the fluid flow line
in that point (namely we split the tangent space $T_zM^4$ at $z$ in the
$U^{\mu}(z)$ direction and in the orthogonal complement) and consider
${1\over {3!}} [U^{\mu}\epsilon_{\mu\nu\rho\sigma}dz^{\nu}\wedge dz^{\rho}
\wedge dz^{\sigma}](z)$ as the infinitesimal 3-volume on it at $z$. Then the
3-form

\beq
\eta [z]=
[\eta_{123}({\bf {\tilde \alpha}}) d{\tilde \alpha}^1\wedge d{\tilde \alpha}^2
\wedge d{\tilde \alpha}^3](z)=
{1\over {3!}}n(z) [U^{\mu}\epsilon_{\mu\nu\rho\sigma}dz^{\nu}\wedge dz^{\rho}
\wedge dz^{\sigma}](z),
\label{I3}
\eeq

\noindent may be interpreted as the number of particles in this 3-volume. If V
is a volume around $z$ on the spacelike hypersurface, then $\int_V \eta$ is the
number of particle in V and $\int_V s\eta$ is the total entropy contained in
the flow lines included in the volume V. Note that locally $\eta_{123}$ can be
set to unity by an appropriate choice of coordinates.

In Ref.\cite{brown} it is shown that the action functional

\beq
S[{}^4g_{\mu\nu}, {\bf {\tilde \alpha}}]
= -\int d^4z \sqrt{{}^4g(z)} \, \rho ({{|J ({\tilde \alpha}^i(z))|}\over
{\sqrt{{}^4g(z)}}},\,  s({\bf {\tilde \alpha}^i}(z)) ),
\label{I4}
\eeq

\noindent has a variation with respect to the 4-metric, which gives rise to
the correct stress tensor $T^{\mu\nu}=(\rho +p)U^{\mu}U
^{\nu}-\epsilon \, p\, {}^4g^{\mu\nu}$ with $p=n{{\partial \rho}\over
{\partial n}}{|}_s-\rho$ for a perfect fluid [see Appendix A].

The Euler-Lagrange equations associated to the variation of the Lagrangian
coordinates are\cite{brown} [$V_{\mu}=\mu U_{\mu}$ is the Taub current, see
Appendix A]

\bea
{1\over {\sqrt{{}^4g}}} {{\delta S}\over {\delta {\tilde
\alpha}^i}}&=&{1\over 2}\epsilon^{\mu\nu\rho\sigma} V_{\mu ;\nu}\,
\eta_{ijk}\, \partial_{\rho}{\tilde
\alpha}^j\, \partial_{\sigma}{\tilde \alpha}^k - n\,  T\, {{\partial s}\over
{\partial {\tilde \alpha}^i}}=0,\nonumber \\
&&{}\nonumber \\
{1\over {\sqrt{{}^4g}}} {{\delta S}\over {\delta {\tilde \alpha}^i}}\partial
_{\mu}{\tilde \alpha}^i&=&{1\over 2} \epsilon^{\alpha\beta\gamma\delta}V_{\alpha
;\beta}\, U^{\nu}\, \epsilon_{\nu\mu\gamma\delta}-T\, \partial_{\mu}s=2V_{[\mu ;\nu]}
U^{\nu} -T\, \partial_{\mu} s=0.
\label{I5}
\eea

As shown in Appendix A, these equations together with the entropy
exchange constraint imply the Euler equations implied from the
conservation of the stress-energy-momentum tensor.

Therefore, with this description the conservation laws are
automatically satisfied and the Euler-Lagrange equations are
equivalent to the Euler equations. In Minkowski spacetime the
conserved particle number is ${\cal N}=\int_{V_{\alpha}(z^o)} d^3z\,
n(z)\, U^o(z)=\int_{V_{\alpha}(z^o)} d^3z\, J^o({\tilde
\alpha}^i(z))$, while the conserved entropy per particle is
$\int_{V_{\alpha}(z^o)} d^3z\, s(z)\, n(z)\, U^o(z)\,
=\int_{V_{\alpha}(z^o)} d^3z\, s(z)\, J^o({\tilde \alpha}^i)$. Moreover, the conservation
laws $T^{\mu\nu}{}_{,\nu}=0$ will generate the conserved 4-momentum
and angular momentum of the fluid.

However, in Ref.\cite{brown} there are only some comments on the
Hamiltonian description implied by this particular action.

This description of perfect fluids fits naturally with parametrized Minkowski
theories\cite{lus} for arbitrary isolated relativistic systems [see
Ref.\cite{india} for a review] on arbitrary spacelike hypersurfaces, leaves of
the foliation of Minkowski spacetime $M^4$ associated with one of its 3+1
splittings.

Therefore, the aim of this paper is to find the Wigner covariant
rest-frame instant form of the dynamics of a perfect fluid, on the
special Wigner hyperplanes orthogonal to the total 4-momentum of the
fluid. In this way we will get the description of the global rest
frame of the fluid as a whole; instead, the 4-velocity vector field
$U^{\mu}$ defines the local rest frame in each point of the fluid by
means of the projector ${}^4g^{\mu\nu}-\epsilon U^{\mu}U^{\nu}$. This
approach will also produce automatically the coupling of the fluid to
ADM metric and tetrad gravity with the extra property of allowing a
well defined deparametrization of the theory leading to the rest-frame
instant form in Minkowski spacetime with Cartesian coordinates when we
put equal to zero the Newton constant G \cite{india}. In this paper we
will consider the perfect fluid only in Minkowski spacetime, except
for some comments on its coupling to gravity.

The starting point is the foliation of Minkowski spacetime $M^4$, which
is defined by an embedding $R\times \Sigma \rightarrow M^4$, $(\tau ,\vec
\sigma ) \mapsto z^{\mu}(\tau ,\vec \sigma )\in \Sigma_{\tau}$ and with
$\Sigma$ an abstract 3-surface diffeomorphic to $R^3$, with
$\Sigma_{\tau}$ its copy embedded in $M^4$ labelled by the value
$\tau$ (the scalar mathematical ``time" parameter $\tau$ labels the
leaves of the foliation, $\vec \sigma$ are curvilinear coordinates on
$\Sigma_{\tau}$ and $\sigma^{\check A}=(\sigma^{\tau}=\tau
,\sigma^{\check r} )$ are $\Sigma_{\tau}$-adapted holonomic
coordinates for $M^4$). See Appendix C for the notations on spacelike
hypersurfaces.

In this way one gets a parametrized field theory with a covariant 3+1
splitting of Minkowski spacetime and already in a form suited to the
transition to general relativity in its ADM canonical formulation (see
also Ref.\cite{ku}, where a theoretical study of this problem is done
in curved spacetimes). The price is that one has to add as new
independent configuration variables  the embedding coordinates
$z^{\mu}(\tau ,\vec \sigma )$ of the points of the spacelike
hypersurface $\Sigma_{\tau}$ [the only ones carrying Lorentz indices]
and then to define the fields on $\Sigma_{\tau}$ so that they know
the hypersurface $\Sigma_{\tau}$ of $\tau$-simultaneity [for a
Klein-Gordon field $\phi (x)$, this new field is $\tilde \phi (\tau
,\vec \sigma )=\phi (z(\tau ,\vec \sigma ))$: it contains the
non-local information about the embedding]. Then one rewrites the
Lagrangian of the given isolated system in the form required by the
coupling to an external gravitational field, makes the previous 3+1
splitting of Minkowski spacetime and interpretes all the fields of the
system as the new fields on $\Sigma_{\tau}$ (they are Lorentz scalars,
having only surface indices). Instead of considering the 4-metric as
describing a gravitational field (and therefore as an independent
field as it is done in metric gravity, where one adds the Hilbert
action to the action for the matter fields), here one replaces the
4-metric with the the induced metric $g_{\check A\check B}[z]
=z^{\mu}_{\check A}\eta_{\mu\nu}z^{\nu}_{\check B}$ on
$\Sigma_{\tau}$ [a functional of $z^{\mu}$; $z^{\mu}_{\check
A}=\partial z^{\mu}/\partial \sigma^{\check A}$ are flat inverse
tetrad fields on Minkowski spacetime with the $z^{\mu}_{\check r}$'s
tangent to $\Sigma_{\tau}$] and considers the embedding coordinates
$z^{\mu}(\tau ,\vec \sigma )$ as independent fields [this is not
possible in metric gravity, because in curved spacetimes
$z^{\mu}_{\check A}\not= \partial z^{\mu}/\partial \sigma^{\check A} $
are not tetrad fields so that holonomic coordinates $z^{\mu}(\tau
,\vec
\sigma )$ do not exist]. From this Lagrangian,
besides a Lorentz-scalar form of the constraints of the given system,
we get four extra primary first class constraints

\beq
{\cal H}
_{\mu}(\tau ,\vec \sigma )=\rho_{\mu}(\tau ,\vec \sigma )-l_{\mu}(\tau
,\vec \sigma )T_{sys}^{\tau\tau}(\tau ,\vec \sigma )-z_{\check r
\mu}(\tau ,\vec \sigma )T_{sys}^{\check r\tau}(\tau ,\vec \sigma )
\approx 0,
\label{I6}
\eeq

\noindent
[ here $T_{sys}^{\tau\tau}(\tau ,\vec \sigma )={\cal M}(\tau ,\vec \sigma )$,
$T_{sys}^{\check r\tau}(\tau ,\vec \sigma )={\cal M}^{\check r}(\tau ,\vec
\sigma )$, are the components of
the energy-momentum tensor in the holonomic coordinate system,
corresponding to the energy- and momentum-density of the
isolated system; one has $\lbrace {\cal H}_{(\mu )}(\tau ,\vec \sigma ),
{\cal H}_{(\nu )}(\tau ,{\vec \sigma}^{'}) \rbrace =0$]
implying the independence of the description from the choice of the 3+1
splitting, i.e. from the choice of the foliation with spacelike hypersufaces.
As shown in Appendix C
the evolution vector is given by $z^{\mu}_{\tau}=N_{[z](flat)}l^{\mu}+
N^{\check r}_{[z](flat)}z^{\mu}_{\check r}$,
where $l^{\mu}(\tau ,\vec \sigma )$ is
the normal to $\Sigma_{\tau}$ in $z^{\mu}(\tau ,\vec \sigma )$ and
$N_{[z](flat)}(\tau ,\vec \sigma )$, $N_{[z](flat)}^{\check r
}(\tau ,\vec \sigma )$
are the flat lapse and shift functions defined through the metric like in
general relativity: however, now they are not
independent variables but functionals of $z^{\mu}(\tau ,\vec \sigma )$.

The Dirac Hamiltonian contains the piece $\int d^3\sigma
\lambda^{\mu}(\tau ,\vec \sigma ){\cal H}_{\mu} (\tau ,\vec \sigma )$
with $\lambda^{\mu}(\tau ,\vec \sigma )$ Dirac multipliers. It is
possible to rewrite the integrand in the form [$\gamma^{\check r\check
s}=-\epsilon \, {}^3g^{\check r\check s}$ is the inverse of the
spatial metric $g_{\check r\check s}={}^4g_{\check r\check
s}=-\epsilon
\, {}^3g_{\check r\check s}$, with ${}^3g_{\check r\check s}$ of
positive signature $(+++)$]

\bea
\lambda_{\mu}(\tau ,\vec
\sigma ){\cal H}^{\mu}(\tau ,\vec \sigma )&=&[(\lambda_{\mu}l^{\mu})
(l_{\nu}{\cal H}^{\nu})-(\lambda_{\mu}z^{\mu}_{r})(\gamma
^{\check r\check s} z_{\check s\nu}{\cal H}^{\nu})](\tau ,\vec \sigma )\,
{\buildrel {def} \over =}\nonumber \\
 &{\buildrel {def} \over =}\,&
N_{(flat)}(\tau ,\vec \sigma ) (l_{\mu}{\cal H}^{\mu})(\tau ,\vec
\sigma )-N_{(flat)\check r}(\tau ,\vec \sigma ) (\gamma^{\check r\check s}
z_{\check s \nu}{\cal H}^{\nu})(\tau ,\vec \sigma ),
\label{I7}
\eea

\noindent with the (non-holonomic form of
the) constraints $(l_{\mu}{\cal H}^{\mu })(\tau ,\vec \sigma )\approx
0$, $(\gamma^{\check r\check s} z_{\check s \mu} {\cal H}^{\mu}) (\tau
,\vec \sigma )\approx 0$, satisfying the universal Dirac algebra of
the ADM constraints. In this way we have defined new  flat lapse and
shift functions

\bea
N_{(flat)}(\tau ,\vec \sigma )&=& \lambda_{\mu}(\tau ,\vec \sigma )
l^{\mu}(\tau ,\vec \sigma ),\nonumber \\
 N_{(flat)\check r}(\tau ,\vec
\sigma )&=& \lambda_{\mu}(\tau ,\vec \sigma ) z^{\mu}_{\check r}(\tau
,\vec \sigma ).
\label{I8}
\eea

\noindent
which have the same content of the arbitrary Dirac multipliers
$\lambda_{\mu}(\tau ,\vec \sigma )$, namely they multiply primary
first class constraints satisfying the Dirac algebra. In Minkowski
spacetime they are quite distinct from the previous lapse and shift
functions $N_{[z](flat)}$, $N_{[z](flat)\check r}$, defined starting
from the metric. Since the Hamilton equations imply
$z^{\mu}_{\tau}(\tau ,\vec \sigma ) = \lambda^{\mu}(\tau ,\vec \sigma
)$, it is only through the equations of motion that the two types of
functions are identified. Instead in general relativity the lapse and
shift functions defined starting from the 4-metric are the
coefficients (in the canonical part $H_c$ of the Hamiltonian) of
secondary first class constraints satisfying the Dirac algebra
independently from the equations of motion.

For the relativistic perfect fluid with equation of state $\rho =\rho
(n,s)$ in Minkowski spacetime, we have only to replace the external
4-metric ${}^4g_{\mu\nu}$ with $g_{\check A\check B}(\tau ,\vec \sigma
)={}^4g_{\check A\check B}(\tau ,\vec \sigma )$ and the scalar fields
for the Lagrangian coordinates with $\alpha^i(\tau ,\vec \sigma
)={\tilde \alpha}^i(z (\tau ,\vec \sigma )) $; now either the
$\alpha^i(\tau ,\vec \sigma )$'s have a compact boundary
$V_{\alpha}(\tau ) \subset
\Sigma_{\tau}$ or have boundary conditions at spatial infinity. For
each value of $\tau$, one could invert $\alpha^i=\alpha^i(\tau ,\vec
\sigma )$ to $\vec \sigma =\vec \sigma (\tau ,\alpha^i)$ and use the
$\alpha^i$'s as a special coordinate system on $\Sigma_{\tau}$ inside
the support $V_{\alpha}(\tau )\subset \Sigma_{\tau}$: $z^{\mu}(\tau
,\vec \sigma (\tau ,\alpha^i))={\check z}^{\mu}(\tau ,\alpha^i)$.

By going to $\Sigma_{\tau}$-adapted coordinates such that
$\eta_{123}({\bf \alpha})=1$ we get [$\gamma =| det\, g_{\check
r\check s}|$; $\,\, \sqrt{g}=\sqrt{{}^4g}=\sqrt{|det\, g_{\check
A\check B}|}=N \sqrt{\gamma}$]

\bea
J^{\check A}(\alpha^i(\tau ,\vec \sigma ))&=&[ N \sqrt{\gamma} n
U^{\check A}](\tau ,\vec \sigma ),\nonumber \\ &&{}\nonumber \\
J^{\tau}(\alpha^i(\tau ,\vec \sigma ))&=&[-\epsilon^{\check r\check
u\check v}\partial
_{\check r}\alpha^1 \partial _{\check u}\alpha^2\partial _{\check v}\alpha
^3](\tau ,\vec \sigma )=- det\, ( \partial_{\check r}\alpha^i )(\tau ,\vec \sigma ),\nonumber \\
J^{\check r}(\alpha^i(\tau ,\vec \sigma ))&=&[\sum_{i=1; i,j,k\,
cyclic}^3
\partial
_{\tau}\alpha^i \epsilon^{\check r\check u\check v}\partial _{\check u}\alpha^j
\partial _{\check v}\alpha^k](\tau ,\vec \sigma )=\nonumber \\
&&={1\over 2} \epsilon^{\check r\check u\check v}\epsilon_{ijk}[\partial_{\tau}
\alpha^i\partial _{\check u}\alpha^j
\partial _{\check v}\alpha^k](\tau ,\vec \sigma ),\nonumber \\
&&{}\nonumber \\
\Rightarrow && n(\tau ,\vec \sigma )={{|J|}\over {N\sqrt{\gamma}}}
(\tau ,\vec \sigma )={{\sqrt{\epsilon g_{\check A\check B}J^{\check A}J^{\check
B}}}\over {N\sqrt{\gamma}}}
(\tau ,\vec \sigma ),
\label{I9}
\eea

\noindent with ${\cal N}=\int_{V_{\alpha}(\tau )} d^3\sigma J^{\tau}(\alpha^i(\tau
,\vec \sigma ))$
giving the conserved particle number and $\int_{V
_{\alpha}(\tau )} d^3\sigma (s\, J^{\tau})(\tau ,\vec \sigma )$ giving the
conserved entropy per particle.

The action becomes

\bea
S&=& \int d\tau d^3\sigma L(z^{\mu}(\tau ,\vec \sigma ),\alpha^i(\tau ,\vec
\sigma )) =\nonumber \\
&=& -\int d\tau d^3\sigma (N\sqrt{\gamma})(\tau ,\vec \sigma ) \rho (
{{|J(\alpha^i(\tau ,\vec \sigma ))|}
\over  {(N\sqrt{\gamma})(\tau \vec \sigma )}},\,  s({\bf \alpha}^i
(\tau ,\vec \sigma )) )=\nonumber \\
 &=&-\int d\tau d^3\sigma
(N\sqrt{\gamma})(\tau ,\vec \sigma ) \nonumber \\
 &&\rho ({1\over {\sqrt{\gamma (\tau ,\vec \sigma )}}}\sqrt{\Big[
(J^{\tau})^2-{}^3g_{\check u\check v}{{J^{\check u}+N^{\check u}
J^{\tau}}\over N}{{J^{\check v}+N^{\check v}J^{\tau}}\over N}\Big]
(\tau ,\vec \sigma ;\alpha^i (\tau ,\vec \sigma ))},\,  s({\bf
\alpha}^i(\tau ,\vec \sigma )) ),\nonumber \\
 &&
\label{I10}
\eea

\noindent with $N=N_{[z](flat)}$, $N^{\check r}=N^{\check r}_{[z](flat)}$.

This is the form of the action whose Hamiltonian formulation will be studied
in this paper.

We shall begin in Section II with the simple case of dust, whose equation of
state is $\rho =\mu n$.

In Section III we will define the ``external" and ``internal" centers
of mass of the dust.

In Section IV we will study Dixon's multipoles of a perfect fluid on
the Wigner hyperplane in Minkowski spacetime using the dust as an
example.

Then in Section V we will consider some equations of state for
isentropic fluids and we will make some comments on non-isentropic
fluids.

In Section VI we will define the coupling to ADM metric and tetrad
gravity.

In Section VII we will describe with the same technology isentropic
elastic media.

In the Conclusions, after some general remarks, we will delineate the
treatment of perfect fluids in tetrad gravity (this will be the
subject of a future paper).

In Appendix A there is a review of some of the results of Ref.\cite{brown}
for relativistic perfect fluids.

In Appendix  B there is a review of covariant relativistic
thermodynamics of equilibrium and non-equilibrium.

In Appendix C there is some notation on spacelike hypersurfaces.

In Appendix D there is the definition of other types of Dixon's
multipoles.

\vfill\eject

\section{Dust.}

Let us consider first the simplest case of an isentropic perfect
fluid, a dust with $p=0$, $s=const.$, and equation of state $\rho =
\mu n$. In this case the chemical potential $\mu$ is the rest
mass-energy of a fluid particle: $\mu = m$ (see Appendix A).

Eq.(\ref{I10}) implies that the Lagrangian density is [we shall use
the notation $g_{\check A\check B}={}^4g_{\check A\check B}$ with
signature $\epsilon (+---)$, $\epsilon =\pm 1$; by using the notation
with lapse and shift functions given in Appendix C we get:
$g_{\tau\tau}=\epsilon (N^2-{}^3g_{\check r\check s}N^{\check
r}N^{\check s})$, $g_{\tau \check r}=-\epsilon \, {}^3g_{\check
r\check s}N^{\check s}$, $g_{\check r\check s}=-\epsilon
\, {}^3g_{\check r\check s}$ with ${}^3g_{\check r\check s}$
of positive signature (+++), $g^{\tau\tau}={{\epsilon}\over {N^2}}$,
$g^{\tau \check r}=-\epsilon {{N^{\check r}}\over {N^2}}$, $g^{\check
r\check s}=-\epsilon ({}^3g^{\check r\check s}-{{N^{\check r}N^{\check
s}}\over {N^2}})$; the inverse of the spatial 4-metric ${}^4g_{\check
r\check s}$ is denoted $\gamma^{\check r\check s}={}^4\gamma^{\check
r\check s}=-\epsilon \, {}^3g^{\check r\check s}$, where
${}^3g^{\check r\check s}$ is the inverse of the 3-metric
${}^3g_{\check r\check s}$ and we use
$\sqrt{\gamma}=\sqrt{{}^3g_{\check r\check s}}$]

\bea
L({\alpha}^{i},z^{\mu})&=& -\sqrt{g} \rho =-\mu \sqrt{g} n=
-\mu \sqrt{\epsilon g_{\check A\check B} J^{\check A}
J^{\check B}}=\nonumber \\
 &=&-{\mu} N \sqrt{(J^{{\tau}})^{2}-{}^3g_{\check
r\check s}Y^{\check r}Y^{\check s}}=-\mu N X,\nonumber \\
&&{}\nonumber \\
 Y^{\check r}&=&\frac{1}{N}(J^{\check r}+N^{\check
r}J^{{\tau}}),\nonumber \\
 X&=&\sqrt{(J^{{\tau}})^{2}-{}^3g_{\check
r\check s}Y^{\check r}Y^{\check s}}= {{\sqrt{g}}\over N} n =
\sqrt{\gamma} n,
\label{II1}
\eea

\noindent with $J^{\tau}$, $J^{\check r}$ given in Eqs.(\ref{I9}).

The momentum conjugate to ${\alpha}^{i}$ is

\bea
{\Pi}_i&=&\frac{\partial L}{\partial \partial_{\tau}\alpha^i}=\mu
\frac{Y^{\check t}\, {}^3g_{\check t\check r}\, \epsilon^{\check r\check u\check v}\,
\partial_{\check u} \alpha^j \partial_{\check v}\alpha^k}{X}
{|}_{i, j, k\quad cyclic}=\nonumber \\
 &=&\mu {{Y^{\check t}}\over
{2X}}\,\, {}^3g_{\check t\check r}\, \epsilon^{\check r\check u\check
v}\epsilon_{ijk}\, \partial_{\check u}\alpha^j\partial_{\check v}
\alpha^k=\mu \, {{Y^{\check r}}\over X}\, T_{\check ri},\nonumber \\,
 &&{}\nonumber \\
 T_{\check ti}\, &{\buildrel {def} \over =}\,&{1\over 2}g_{\check t\check
r}\, \epsilon^{\check r\check u\check v}\epsilon_{ijk}\, \partial
_{\check u}\alpha^j \partial _{\check v}\alpha^k=g_{\check t\check
r}\,\, ( ad\, J_{i\check r}),
\label{II2}
\eea

\noindent where $ad\, J_{i\check r}= (det\, J) J^{-1}_{i\check r}$ is the
adjoint matrix of the Jacobian $J=(J_{i\check r}=\partial_{\check r}\alpha^i)$
of the transformation from the Lagrangian coordinates $\alpha^i(\tau ,\vec
\sigma )$ to the Eulerian ones $\vec \sigma$ on $\Sigma_{\tau}$.

The momentum conjugate to $z^{{\mu}}$ is

\beq
{\rho}_{\mu}(\tau ,\vec \sigma )=-\frac{\partial L}{\partial
{z_{{\tau}}^{{\mu}}}}(\tau ,\vec \sigma )=\Big[ \mu \, l_{\mu}\,
\frac{(J^{\tau})^{2}}{X}+\mu \, z_{r\mu}\, J^{\tau}\,
\frac{Y^{r}}{X}\Big] (\tau ,\vec \sigma ).
\label{II3}
\eeq

The  following Poisson brackets are assumed

\begin{eqnarray}
&&\lbrace z^{\mu}(\tau ,\vec \sigma ),\rho_{\nu}(\tau ,{\vec \sigma}^{'}\rbrace
=-\eta^{\mu}_{\nu}\delta^3(\vec \sigma -{\vec \sigma}^{'}),\nonumber \\
&&\lbrace \alpha^i(\tau ,\vec \sigma ),\Pi_j(\tau ,{\vec
\sigma}^{'})\rbrace =
\delta^i_j\delta^3(\vec \sigma -{\vec \sigma}^{'}).
\label{II4}
\end{eqnarray}

We can express $Y^{\check r}/X$ in terms of $\Pi_i$ with the help of the
inverse $(T^{-1})^{\check ri}$ of the matrix $T_{\check ti}$

\beq
{{Y^{\check r}}\over X}=\frac{1}{\mu} (T^{-1})^{\check ri}\Pi_{i},
\label{II5}
\eeq

\noindent where

\beq
(T^{-1})^{\check ri}={{{}^3g^{\check r\check s}\, \partial_{\check
s}\alpha
^i}\over {det\, (\partial_{\check u}\alpha^k)}}.
\label{II6}
\eeq

From the definition of $X$  we find

\beq
X=\frac{\mu J^{\tau}}{\sqrt{\mu^2+ {}^3g_{\check u\check
v}(T^{-1})^{\check ui} (T^{-1})^{\check vj}{\Pi}_{i}{\Pi}_{j} }}.
\label{II7}
\eeq

\noindent
Consequently, we can get the expression of the velocities of the Lagrangian
coordinates in terms of the momenta

\beq
\partial_{\tau}\alpha^i=-\frac{J^{\check r}\partial_{\check r}\alpha^i}
{J^{\tau}}=\frac{(N^{\check r}J^{\tau}-NY^{\check r})\,
\partial_{\check r}
\alpha^i}{J^{\tau}},
\label{II8}
\eeq

\noindent namely

\beq
\partial_{\tau}\alpha^i=\partial_{\check r}\alpha^i\, \Big[ N^{\check r}- N
(T^{-1})^{\check ri} \Pi_i\,
\sqrt{\mu^2+{}^3g_{\check u\check v}(T^{-1})^{\check ui}(T^{-1})^{\check vj}
\Pi_i \Pi_j } \Big] .
\label{II9}
\eeq

Now $\rho_{\mu}$ can be expressed as a function of the $z$'s,
$\alpha$'s and $\Pi$'s:

\beq
{\rho}_{\mu}=l_{{\mu}}\, J^{\tau}\,
\sqrt{\mu^2+{}^3g_{\check u\check v}(T^{-1})^{\check ui}\Pi_i(T^{-1})
^{\check vj}\Pi_j}+z_{\check r{\mu}}\, J^{\tau}\,
(T^{-1})^{\check ri}\Pi_i.
\label{II10}
\eeq

Since the Lagrangian is homogenous in the velocities, the Hamiltonian
is only

\beq
H_{D}=\int d^{3}{\sigma}{\lambda}^{\mu}(\tau ,\vec \sigma ) {\cal
H}_{\mu}(\tau ,\vec \sigma ),
\label{II11}
\eeq

\noindent where the ${\cal H}_{\mu}$ are the primary constraints

\bea
{\cal H}_{\mu}&=&{\rho}_{\mu}-l_{\mu}\, {\cal M} +z_{\check r{\mu}}\,
{\cal M}^{\check R} \approx 0,\nonumber \\
 &&{}\nonumber \\
 {\cal M} &=& T^{\tau\tau} =  J^{\tau} \,
\sqrt{\mu^2+{}^3g_{\check u\check v}(T^{-1})^{\check ui}\Pi_i(T^{-1})^{\check
vj}\Pi_j},\nonumber \\
 {\cal M}^{\check r} &=& T^{\tau \check r} = J^{\tau}\, (T^{-1})^{\check
ri}\Pi_i.
\label{II12}
\eea

\noindent  satisfying

\begin{equation}
\lbrace {\cal H}_{\mu}(\tau ,\vec \sigma ),{\cal H}_{\nu}(\tau ,{\vec \sigma}
^{'})\rbrace =0.
\label{II13}
\end{equation}

\noindent One finds that $\lbrace {\cal H}_{\mu}(\tau ,\vec \sigma ),H_D
\rbrace =0$. Therefore, there are only  the four first class constraints
 ${\cal H}_{\mu}(\tau ,\vec \sigma )\approx 0$. They
describe the arbitrariness of the
foliation: physical results do not depend on its choice.

The conserved Poincar\'e generators are (the suffix ``s" denotes the
hypersurface $\Sigma_{\tau}$)

\begin{eqnarray}
&&p^{\mu}_s=\int d^3\sigma \rho^{\mu}(\tau ,\vec \sigma ),\nonumber \\
&&J_s^{\mu\nu}=\int d^3\sigma [z^{\mu}(\tau ,\vec \sigma )\rho^{\nu}(\tau ,
\vec \sigma )-z^{\nu}(\tau ,\vec \sigma )\rho^{\mu}(\tau ,\vec \sigma )],
\label{II14}
\end{eqnarray}

\noindent and one has

\begin{equation}
\lbrace z^{\mu}(\tau ,\vec \sigma ),p^{\nu}_s\rbrace =-\eta^{\mu\nu},
\label{II15}
\end{equation}

\begin{eqnarray}
\int d^3\sigma {\cal H}_{\mu}(\tau ,\vec \sigma )&=& p_s^{\mu}-\int d^3\sigma
\Big[l_{\mu}\, J^{\tau}\,
\sqrt{\mu^2+{}^3g_{\check u\check v}(T^{-1})^{\check ui}\Pi_i(T^{-1})^{\check
vj}\Pi_j}\Big] (\tau ,\vec \sigma )+\nonumber \\ &+&\int d^3\sigma
\Big[ z_{\check r{\mu}}\, J^{\tau}\, (T^{-1})^{\check ri}\Pi_i\Big] (\tau
,\vec \sigma )\approx 0.
\label{II16}
\end{eqnarray}

Let us now restrict ourselves to spacelike hyperplanes $\Sigma_{\tau}$
by imposing the gauge-fixings

\begin{eqnarray}
\zeta^{\mu}(\tau ,\vec \sigma )&=&z^{\mu}(\tau ,\vec \sigma )-x_s^{\mu}(\tau )-
b^{\mu}_{\check r}(\tau )\sigma^{\check r}\approx 0,\nonumber \\
&&{}\nonumber \\
&&\lbrace \zeta^{\mu}(\tau ,\vec \sigma ),{\cal H}_{\nu}(\tau ,{\vec \sigma}
^{'})\rbrace =-\eta^{\mu}_{\nu}\delta^3(\vec \sigma -{\vec \sigma}^{'}),
\label{II17}
\end{eqnarray}

\noindent where $x^{\mu}_s(\tau )$ is an arbitrary point of $\Sigma_{\tau}$,
chosen as origin of the coordinates $\sigma^{\check r}$, and
$b^{\mu}_{\check r}(\tau )$, ${\check r}=1,2,3$, are three orthonormal
vectors such that the constant (future pointing) normal to the
hyperplane is

\begin{equation}
l^{\mu}(\tau ,\vec \sigma )\approx l^{\mu}=b^{\mu}_{\tau}=\epsilon^{\mu}
{}_{\alpha\beta\gamma}b^{\alpha}_{\check 1}(\tau )b^{\beta}_{\check 2}(\tau )
b^{\gamma}_{\check 3}(\tau ).
\label{II18}
\end{equation}

Therefore, we get

\begin{eqnarray}
&&z^{\mu}_{\check r}(\tau ,\vec \sigma )\approx b^{\mu}_{\check r}(\tau ),
\nonumber \\
&&z^{\mu}_{\tau}(\tau ,\vec \sigma )\approx {\dot x}^{\mu}_s(\tau
)+{\dot b}^{\mu}_{\check r}(\tau )\sigma^{\check r},\nonumber \\
&&g_{{\check r}{\check s}}(\tau ,\vec \sigma )\approx
-\epsilon \delta_{{\check r}{\check s}},\quad\quad \gamma^{{\check r}{\check
s}}(\tau ,\vec \sigma )
\approx -\epsilon \delta^{{\check r}{\check s}},\quad\quad \gamma (\tau ,\vec \sigma )
\approx 1.
\label{II19}
\end{eqnarray}

By introducing the Dirac brackets for the resulting second class
constraints

\begin{equation}
\lbrace A,B\rbrace {}^{*}=\lbrace A,B\rbrace -\int d^3\sigma [\lbrace A,\zeta
^{\mu}(\tau ,\vec \sigma )\rbrace \lbrace {\cal H}_{\mu}(\tau ,\vec \sigma ),
B\rbrace -\lbrace A,{\cal H}_{\mu}(\tau ,\vec \sigma )\rbrace \lbrace \zeta
^{\mu}(\tau ,\vec \sigma ),B\rbrace ],
\label{II20}
\end{equation}

\noindent we find that,
by using Eq.(\ref{II15}) and (\ref{II16}) [with $x^{\mu}_s (\tau
)=z^{\mu}(\tau ,\vec \sigma )-b^{\mu}_{\check r}(\tau )\sigma^{\check
r}-\zeta^{\mu}(\tau ,\vec \sigma )$ and with the assumption $\lbrace
b^{\mu}_{\check r}(\tau ), p^{\nu}_s\rbrace =0$], we get

\begin{equation}
\lbrace x_s^{\mu}(\tau ),p^{\nu}_s(\tau )\rbrace {}^{*}=-\eta^{\mu\nu}.
\label{II21}
\end{equation}

The ten degrees of freedom describing the hyperplane are $x^{\mu}_s(\tau )$
with conjugate momentum $p^{\mu}_s$ and six variables $\phi_{\lambda}(\tau )$,
$\lambda =1,..,6$, which parametrize the orthonormal tetrad $b^{\mu}_{\check A}
(\tau )$, with their conjugate momenta $T_{\lambda}(\tau )$.

The preservation of the gauge-fixings $\zeta^{\mu}(\tau ,\vec \sigma )
\approx 0$ in time implies

\begin{equation}
{d\over {d\tau}}\zeta^{\mu}(\tau ,\vec \sigma )=\lbrace \zeta^{\mu}(\tau ,\vec
\sigma ),H_D\rbrace =-\lambda^{\mu}(\tau ,\vec \sigma )-{\dot x}^{\mu}_s(\tau
)-{\dot b}^{\mu}_{\check r}(\tau )\sigma^{\check r}\approx 0,
\label{II22}
\end{equation}

\noindent so that one has [by using ${\dot b}^{\mu}_{\tau}=0$ and ${\dot b}
^{\mu}_{\check r}(\tau )b^{\nu}_{\check r}(\tau )=-b^{\mu}_{\check r}(\tau )
{\dot b}^{\nu}_{\check r}(\tau )$]

\begin{eqnarray}
\lambda^{\mu}(\tau ,\vec \sigma )&\approx&{\tilde \lambda}^{\mu}(\tau )+
{\tilde \lambda}^{\mu}{}_{\nu}(\tau )b^{\nu}_{\check r}(\tau )
\sigma^{\check r},\nonumber \\
&&{\tilde \lambda}^{\mu}(\tau )=-{\dot x}^{\mu}_s(\tau ),\nonumber \\
&&{\tilde \lambda}^{\mu\nu}(\tau )=-{\tilde \lambda}^{\nu\mu}(\tau )={1\over 2}
[{\dot b}^{\mu}_{\check r}(\tau )b^{\nu}_{\check r}(\tau )-b^{\mu}_{\check r}
(\tau ){\dot b}^{\nu}_{\check r}(\tau )].
\label{II23}
\end{eqnarray}

Thus, the Dirac Hamiltonian becomes

\begin{equation}
H_D={\tilde \lambda}^{\mu}(\tau ){\tilde {\cal H}}_{\mu}(\tau )-{1\over 2}
{\tilde \lambda}^{\mu\nu}(\tau ){\tilde {\cal H}}_{\mu\nu}(\tau ),
\label{II24}
\end{equation}

\noindent and this shows that the gauge fixings $\zeta^{\mu}(\tau ,\vec
\sigma )\approx 0$ do not transform completely the constraints ${\cal H}
_{\mu}(\tau ,\vec \sigma )\approx 0$ in their second class partners; still the
following ten first class constraints are left

\begin{eqnarray}
{\tilde {\cal H}}^{\mu}(\tau )&=&\int d^3\sigma {\cal H}^{\mu}(\tau ,\vec
\sigma )=\, p^{\mu}_s-\nonumber \\
&-&\int d^{3}{\sigma} \Big(  J^{{\tau}} \left[ l^{\mu}
\sqrt{\mu^2+ \delta_{\check u\check v}
(T^{-1})^{\check ui}{\Pi}_{i}(T^{-1})^{\check vj}{\Pi}_{j}
}+b^{{\mu}}_{\check r}(T^{-1})^{\check rl} {\Pi}_{l}\right] \Big)
(\tau ,\vec \sigma ) \approx 0,\nonumber \\
 &&{}\nonumber \\
{\tilde {\cal H}}^{\mu\nu}(\tau )&=&b^{\mu}_{\check r}(\tau )\int d^3\sigma
\sigma^{\check r}\, {\cal H}^{\nu}(\tau ,\vec \sigma )-b^{\nu}_{\check r}(\tau )
\int d^3\sigma \sigma^{\check r}\, {\cal H}^{\mu}(\tau ,\vec \sigma )=
\nonumber \\
&=&S_s^{\mu\nu}(\tau )-\nonumber \\
 &-&[b^{\mu}_{\check r}(\tau
)b^{\nu}_{\tau}-b^{\nu}_{\check r}(\tau )b^{\mu}_{\tau}]
\int d^{3}{\sigma}  {\sigma}^{\check r} \Big( J^{{\tau}}
\sqrt{\mu^2+ \delta_{\check u\check v}
(T^{-1})^{\check ui}{\Pi}_{i}(T^{-1})^{\check vj}{\Pi}_{j}}\Big)
(\tau ,\vec \sigma )+\nonumber \\
&+&[b^{\mu}_{\check r}(\tau )b^{\nu}_{\check s}(\tau )-b^{\nu}_{\check r}(\tau )
b^{\mu}_{\check s}(\tau )]
\int d^{3}{\sigma}\sigma^{\check r} \Big( J^{{\tau}}(T^{-1})
^{\check sl}{\Pi}_{l}\Big) (\tau ,\vec \sigma )\approx 0.
\label{II25}
\end{eqnarray}

Here $S^{\mu\nu}_s$ is the spin part of the Lorentz generators

\begin{eqnarray}
J^{\mu\nu}_s&=&x^{\mu}_sp^{\nu}_s-x^{\nu}_sp^{\mu}_s+S^{\mu\nu}_s,\nonumber \\
&&S^{\mu\nu}_s=b^{\mu}_{\check r}(\tau )\int d^3\sigma \sigma^{\check r}
\rho^{\nu}(\tau ,\vec \sigma )-b^{\nu}_{\check r}(\tau )\int d^3\sigma
\sigma^{\check r}\rho^{\mu}(\tau ,\vec \sigma ).
\label{II26}
\end{eqnarray}

As shown in Ref.\cite{lus} instead of finding
$\phi_{\lambda}(\tau ), T_{\lambda}(\tau )$, one can use the redundant
variables $b^{\mu}_{\check A}(\tau ), S_s^{\mu\nu}(\tau )$, with the
following Dirac brackets assuring the validity of the orthonormality
condition $\eta^{\mu\nu}-b^{\mu}_{\check A}\eta^{{\check A}{\check b}}b^{\nu}
_{\check B}=0$ [$C^{\mu\nu\alpha\beta}_{\gamma\delta}=\eta^{\nu}_{\gamma}
\eta^{\alpha}_{\delta}\eta^{\mu\beta}+\eta^{\mu}_{\gamma}\eta^{\beta}_{\delta}
\eta^{\nu\alpha}-\eta^{\nu}_{\gamma}\eta^{\beta}_{\delta}\eta^{\mu\alpha}-
\eta^{\mu}_{\gamma}\eta^{\alpha}_{\delta}\eta^{\nu\beta}$
are the structure constants of the Lorentz group]

\begin{eqnarray}
&&\lbrace S_s^{\mu\nu},b^{\rho}_{\check A}\rbrace {}^{*}=\eta^{\rho\nu}
b^{\mu}_{\check A}-\eta^{\rho\mu}b^{\nu}_{\check A}\nonumber \\
&&\lbrace S^{\mu\nu}_s,S_s^{\alpha\beta}\rbrace {}^{*}=C^{\mu\nu\alpha\beta}
_{\gamma\delta}S_s^{\gamma\delta},
\label{II27}
\end{eqnarray}

\noindent so that, while ${\tilde {\cal H}}^{\mu}(\tau )\approx 0$ has zero
Dirac bracket with itself and with ${\tilde {\cal H}}^{\mu\nu}(\tau )
\approx 0$, these last six constraints have the Dirac brackets

\begin{equation}
\lbrace {\tilde {\cal H}}^{\mu\nu}(\tau ),{\tilde {\cal H}}^{\alpha\beta}
(\tau )\rbrace {}^{*}=C^{\mu\nu\alpha\beta}_{\gamma\delta}{\tilde {\cal H}}
^{\gamma\delta}(\tau )\approx 0.
\label{II28}
\end{equation}

We have now only the  variables: $x^{\mu}_{s}$, $p^{\mu}_{s}$,
$b^{\mu}_{\check A}$, $S^{\mu\nu}_s$, $\alpha^i$, $\Pi_i$ with the
following Dirac brackets:
\bea
\{ x^{\mu}_{s}(\tau ),p_{s}^{\nu}(\tau ) \}^{*} &=& -\eta^{\mu \nu}, \nonumber \\
\{ S^{\mu\nu}_{s}(\tau ),b_{\check A}^{\rho}(\tau ) \}^{*} &=& \eta^{\rho \nu}
b_{\check A}^{\mu}(\tau )-\eta^{\rho \mu}b_{\check A}^{\nu}(\tau ),
\nonumber  \\
\{ S^{\mu\nu}_{s}(\tau ),S^{\alpha\beta}_{s}(\tau ) \}^{*} &=& C^{{\mu}{\nu}\alpha
\beta}_{\gamma \delta}S^{{\gamma}{\delta}}_{s}(\tau ),\nonumber \\
 \{\alpha ^{i}(\tau ,\vec \sigma ),
\Pi_{j}(\tau ,{\vec \sigma}^{'}) \}^{*}&=&\delta_{j}^{i}
\delta^3(\vec \sigma -{\vec \sigma}^{'}).
\label{II29}
\eea

After the restriction to spacelike hyperplanes
we have $z^{\mu}_{\check r}(\tau
,\vec \sigma )\approx b^{\mu}_{\check r}(\tau )$, so that $z^{\mu}
_{\tau}(\tau ,\vec \sigma )\approx N_{[z](flat)}(\tau ,\vec \sigma )l
^{\mu}(\tau ,\vec \sigma )+N^{\check r}_{[z](flat)}(\tau ,\vec \sigma )$
$b^{(\mu )}_{\check r}(\tau ,\vec
\sigma )\approx {\dot x}^{\mu}_s(\tau )+{\dot b}^{\mu}_{\check r}(\tau )
\sigma^{\check r}=-{\tilde \lambda}^{\mu}(\tau )-{\tilde
\lambda}^{\mu\nu}(\tau )b_{\check r \nu}(\tau )\sigma^{\check r}$.
As  said in the Introduction only now we get the coincidence of the
two definitions of flat lapse and shift functions (this point was
missed in the older treatments of parametrized Minkowski theories):

\bea
N_{[z](flat)}(\tau ,\vec \sigma )&\approx& N_{(flat)}(\tau ,\vec
\sigma )=-{\tilde \lambda}
_{\mu}(\tau )l^{\mu}-l^{\mu}{\tilde \lambda}_{\mu\nu}(\tau )b
^{\nu}_{\check s}(\tau ) \sigma^{\check s}=N(\tau ,\vec \sigma ),\nonumber \\
N_{[z](flat)\check r}(\tau ,\vec \sigma )&\approx& N_{(flat )\check
r}(\tau ,\vec \sigma )=-{\tilde \lambda}
_{\mu}(\tau )b^{\mu}_{\check r}(\tau )-b^{\mu}_{\check r}(\tau ){\tilde
\lambda}_{\mu\nu}(\tau ) b^{\nu}_{\check s}(\tau ) \sigma^{\check s}=
N_{\check r}(\tau ,\vec \sigma ).\nonumber \\
 &&{}
\label{II30}
\eea

Let us now restrict ourselves to configurations with $\epsilon p_s^2 >
0$ and let us use the Wigner boost
$L^{\mu}{}_{\nu}(\stackrel{\circ}{p_s}, p_s)$ to boost to rest the
variables $b^{\mu}_{\check A}$, $S_s^{\mu\nu}$ of the following
non-Darboux basis

${}$

$x^{\mu}_s, p^{\mu}_s, b^{\mu}_{\check A}, S_s^{\mu\nu}, \alpha^i,
\Pi_i$

${}$

\noindent of the Dirac brackets $\lbrace .,.\rbrace {}^{*}$.
The following new non-Darboux basis is obtained [${\tilde x}^{\mu}_s$
is no more a fourvector; we choose the sign $\eta =sign\, p^o_s$
positive]

\begin{eqnarray}
&&{\tilde x}^{\mu}_s=x_s^{\mu}+{1\over 2}\,
\epsilon^A_{\nu}(u(p_s))\eta_{AB} { {\partial
\epsilon^B_{\rho}(u(p_s))}\over {\partial p_{s\mu}} }\,
S^{\nu\rho}_s=\nonumber \\
 &&=x_s^{\mu}-{ 1\over {\sqrt{\epsilon
p_s^2}(p_s^o+\eta \sqrt{p_s^2})} }\, [p_{s\nu} S^{\nu\mu}_s+
\sqrt{\epsilon p_s^2} (S_s^{o\mu}-S_s^{o\nu}{ {p_{s\nu}p_s^{\mu}}\over {\epsilon p_s^2}
})]=\nonumber \\
 &&=x_s^{\mu}-{1\over {\sqrt{\epsilon p_s^2}}
}[\eta^{\mu}_A({\bar S}_s^{\bar oA}
-{ {{\bar S}_s^{Ar}p_s^r}\over {p_s^o+\sqrt{\epsilon p_s^2}} })+{ {p_s^{\mu}+2
\sqrt{\epsilon p_s^2}\eta^{\mu o}}\over {\sqrt{\epsilon p_s^2}(p_s^o+
\sqrt{\epsilon p_s^2})} }{\bar S}_s^{\bar or}\, p_s^r],\nonumber \\
&&{}\nonumber \\
&&p^{\mu}_s=p_s^{\mu},\nonumber \\
&&{}\nonumber \\
&&\alpha^i=\alpha^i,\nonumber \\
&&\Pi_i=\Pi_i,\nonumber \\
&&{}\nonumber \\
&&b^A_{\check r}=\epsilon^A_{\mu}(u(p_s))b^{\mu}_{\check r},\nonumber \\
&&{\tilde S}_s^{\mu\nu}=S^{\mu\nu}_s-{1\over 2}\epsilon^A_{\rho}(u(p_s))\eta
_{AB}({ {\partial \epsilon^B_{\sigma}(u(p_s))}\over {\partial p_{s\mu}} }\,
p^{\nu}_s-{ {\partial \epsilon^B_{\sigma}(u(p_s))}\over {\partial p_{s\nu}} }\,
 p_s^{\mu})S^{\rho\sigma}_s=\nonumber \\
&&=S^{\mu\nu}_s+{1\over {\sqrt{\epsilon p_s^2}(p_s^o+
\sqrt{\epsilon p_s^2})} }
[p_{s\beta}(S^{\beta\mu}_sp_s^{\nu}-S_s^{\beta\nu}p_s^{\mu})+
\sqrt{\epsilon p_s^2} (S_s^{o\mu}p_s^{\nu}-S^{o\nu}_sp_s^{\mu})],\nonumber \\
&&{}\nonumber \\ &&J^{\mu\nu}_s={\tilde x}_s^{\mu}p_s^{\nu}-{\tilde
x}^{\nu}_sp_s^{\mu}+{\tilde S}_s^{\mu\nu}.
\label{II31}
\end{eqnarray}

We have

\begin{eqnarray}
&&\lbrace {\tilde x}^{\mu}_s,p^{\nu}_s\rbrace {}^{*}=0,\nonumber \\
&&\lbrace {\tilde S}^{oi}_s,b^r_{\check A}\rbrace {}^{*}={
{\delta^{is}(p^r_s b^s_{\check A}-p^s_sb^r_{\check A})}\over
{p^o_s+\sqrt{\epsilon p^2_s}} },
\nonumber \\
&&\lbrace {\tilde S}_s^{ij},b^r_{\check A}\rbrace {}^{*}=(\delta^{ir}\delta
^{js}-\delta^{is}\delta^{jr})b^s_{\check A},\nonumber \\
&&\lbrace {\tilde S}_s^{\mu\nu},{\tilde S}_s^{\alpha\beta}\rbrace {}^{*}=
C^{\mu\nu\alpha\beta}_{\gamma\delta}{\tilde S}_s^{\gamma\delta},
\label{II32}
\end{eqnarray}

\noindent and we can define

\begin{eqnarray}
{\bar S}_s^{AB}&=&\epsilon^A_{\mu}(u(p_s))\epsilon^B_{\nu}(u(p_s))S_s^{\mu\nu}
\approx [b^A_{\check r}(\tau )b^B_{\tau}-b^B_{\check r}(\tau )b^A_{\tau}]
\nonumber \\
&&\int d^{3}{\sigma}\,  {\sigma}^{\check r} \Big( J^{{\tau}}
\sqrt{\mu^2+ \delta_{\check u\check v}
(T^{-1})^{\check ui}{\Pi}_{i}(T^{-1})^{\check vj}{\Pi}_{j}}\Big)
(\tau ,\vec \sigma )-\nonumber \\
&-&[b^A_{\check r}(\tau )b^B_{\check s}(\tau )-b^B_{\check r}(\tau )b^A
_{\check s}(\tau )]
\int d^{3}{\sigma}\, \sigma^{\check r}\Big( J^{{\tau}}(T^{-1})
^{\check sl}{\Pi}_{l}\Big)(\tau ,\vec \sigma ).
\label{II33}
\end{eqnarray}

Let us now add six more gauge-fixings by selecting the special family
of spacelike hyperplanes  $\Sigma_{\tau W}$ orthogonal to $p^{\mu}_s$
(this is possible for $\epsilon p^2_s > 0$), which can be called the
`Wigner foliation' of Minkowski spacetime. This can be done by
requiring (only six conditions are independent)

\begin{eqnarray}
T^{\mu}_{\check A}(\tau )&=&b^{\mu}_{\check A}(\tau )-\epsilon^{\mu}
_{A={\check A}}(u(p_s))\approx 0\nonumber \\
&&\Rightarrow \quad b^A_{\check A}(\tau )=\epsilon^A_{\mu}(u(p_s))b^{\mu}
_{\check A}(\tau )\approx \eta^A_{\check A}.
\label{II34}
\end{eqnarray}

Now the inverse tetrad $b^{\mu}_{\check A}$ is equal to the
polarization vectors $\epsilon^{\mu}_A(u(p_s))$ [see Appendix C] and
the indices `${\check r}$' are forced to coincide with the Wigner
spin-1 indices `r', while $\bar o=\tau$ is a Lorentz-scalar index. One
has

\begin{eqnarray}
{\bar S}_s^{AB}&\approx& (\eta^A_{r}\eta^B_{\tau}-\eta^B_{r}
\eta^A_{\tau}) {\bar S}_s^{\tau r}-\nonumber \\
&-&(\eta^A_{r}\eta^B_{s}-\eta^B_{r}\eta^A_{s}) {\bar S}_s^{rs},
\nonumber \\
 &&{}\nonumber \\
{\bar S}_s^{rs}&\approx&
\int d^{3}{\sigma}\Big( J^{{\tau}}\, [\sigma^r\, (T^{-1})^{sl}
{\Pi}_{l}-\sigma^s\, (T^{-1})^{rl}{\Pi}_{l}]\Big) (\tau ,\vec \sigma
),\nonumber \\
 {\bar S}_s^{\tau r}&\approx& -{\bar S}_s^{r\tau}=
-\int d^{3}{\sigma} \Big( J^{{\tau}}\, {\sigma}^r\,
\sqrt{\mu^2+ \delta_{uv}(T^{-1})^{ui}{\Pi}_{i}(T^{-1})^{vj}{\Pi}_{j}}\Big)
(\tau ,\vec \sigma ).
\label{II35}
\end{eqnarray}

\noindent The comparison of $\bar{S}^{AB}_{s}$ with $\tilde{S}^{\mu \nu}
_{s} $ yields

\bea
\tilde{S}^{uv}_{s} & = & \delta ^{ur} \delta^{vt} \bar{S}^{rt}_{s}  \nonumber \\
\tilde{S}^{ov}_{s} & = & - \frac{ \delta^{vr}\bar{S}^{rt}_{s} p_{s}{t}}
{p_{s}^{0}+\sqrt{\epsilon p^2_s} }.
\label{II36}
\eea

The time constancy of $T^{\mu}_{\check A}\approx 0$ with respect to
the Dirac Hamiltonian of Eq.(\ref{II24}) gives

\begin{eqnarray}
{d\over {d\tau}}[b^{\mu}_{\check r}(\tau )-\epsilon^{\mu}_r(u(p_s))]&=&
\lbrace b^{\mu}_{\check r}(\tau )-\epsilon^{\mu}_r(u(p_s)),H_D\rbrace {}^{*}=
\nonumber \\
&=&{1\over 2}{\tilde \lambda}^{\alpha\beta}(\tau )\lbrace b^{\mu}_{\check r}
(\tau ),S_{s\alpha\beta}(\tau )\rbrace {}^{*}={\tilde \lambda}^{\mu\alpha}
(\tau )b_{\check r\alpha}(\tau )\approx 0\nonumber \\
&\Rightarrow& {\tilde \lambda}^{\mu\nu}(\tau )\approx 0,
\label{II37}
\end{eqnarray}

\noindent so that the independent gauge-fixings contained in Eqs.(\ref{II34}) and
the constraints ${\tilde {\cal H}}^{\mu\nu}(\tau )\approx 0$ form six pairs of
second class constraints.

Besides Eqs.(\ref{II19}), now we have [remember that ${\dot
x}_s^{\mu}(\tau )=-{\tilde \lambda}^{\mu}(\tau )$]

\begin{eqnarray}
&&l^{\mu}=b^{\mu}_{\tau}=u^{\mu}(p_s),\nonumber \\
&&z^{\mu}_{\tau}(\tau )={\dot x}^{\mu}_s(\tau )=\sqrt{g(\tau )}u^{\mu}(p_s)-
{\dot x}_{s\nu}(\tau )\epsilon^{\mu}_r(u(p_s))\epsilon^{\nu}_r(u(p_s)),
\nonumber \\
 &&{}\nonumber \\
 &&N(\tau )=\sqrt{g(\tau )}=[{\dot x}_{s\mu}(\tau )u^{\mu}(p_s))],\quad
\quad \sqrt{\gamma}=1,\nonumber \\
 &&g_{\tau\tau}={\dot x}^2_s,\quad\quad g_{rs}=-\epsilon \, {}^3g_{rs}
 =-\epsilon \delta_{rs},\nonumber \\
 &&g_{\tau r}=-\epsilon {\dot x}_{s \mu} \epsilon^{\mu}_r(u(p_s))=-
 \epsilon \delta_{rs}N^s,\quad\quad
 N^r=\delta^{ru} {\dot x}_{s \mu} \epsilon^{\mu}_u(u(p_s)),\nonumber \\
 &&g^{\tau\tau}={1\over g}={{\epsilon}\over {N^2}},\quad\quad g^{\tau r}=-
 {{\epsilon}\over g}{\dot x}_{s\mu}\delta^{ru}\epsilon^{\mu}
_r(u(p_s))=-\epsilon {{N^r}\over {N^2}},\nonumber \\
&& g^{rs}=-\epsilon (\delta^{rs}- \delta^{ ru}\delta^{sv} {{{\dot
x}_{s\mu}\epsilon^{\mu}_u(u(p_s)) {\dot
x}_{s\nu}\epsilon^{\nu}_v(u(p_s))}\over {[{\dot x}_s\cdot u(p_s)]^2}}
)=-\epsilon (\delta^{rs}-{{N^rN^s}\over {N^2}}).
\label{II38}
\end{eqnarray}

On the hyperplane $\Sigma_{\tau W}$ all the degrees of freedom
$z^{\mu}(\tau ,
\vec \sigma )$ are reduced to the four degrees of freedom ${\tilde x}^{\mu}_s
(\tau )$, which replace $x^{\mu}_s$. The Dirac
Hamiltonian is now $H_D={\tilde \lambda}^{\mu}(\tau ){\tilde {\cal H}}_{\mu}
(\tau )$ with

\begin{eqnarray}
{\tilde {\cal H}}^{\mu}(\tau )=p_s^{\mu}&-&\nonumber \\
 &-&\int d^3\sigma  \Big( J^{\tau}
\Big[ u^{\mu}(p_s)\sqrt{\mu^2+ \delta_{uv}(T^{-1})^{ui}{\Pi}_{i}(T^{-1})^{vj}
{\Pi}_{j}}- \nonumber \\
 &-&\epsilon^{{\mu}}_{r}(u(p_s))\, \mu
(T^{-1})^{rl} {\Pi}_{l}\Big]
\Big) (\tau ,\vec \sigma )\approx 0.
\label{II39}
\end{eqnarray}

To find the new Dirac brackets, one needs to evaluate the matrix of the
old Dirac brackets of the second class constraints (without extracting the
independent ones)

\begin{eqnarray}
C=\left(
\begin{array}{cccc}
\lbrace {\tilde {\cal H}}^{\alpha\beta},{\tilde {\cal H}}^{\gamma\delta}
\rbrace {}^{*}\approx 0 & \lbrace {\tilde {\cal H}}^{\alpha\beta},T^{\sigma}
_{\check B}\rbrace {}^{*}=\\
{} & =\delta_{{\check B}B}[\eta^{\sigma\beta}\epsilon
^{\alpha}_B(u(p_s))-\eta^{\sigma\alpha}\epsilon_B^{\beta}(u(p_s))] \\
\lbrace T^{\rho}_{\check A},{\tilde {\cal H}}^{\gamma\delta}\rbrace {}^{*}
= & \lbrace T^{\rho}_{\check A},T^{\sigma}_{\check B}\rbrace {}^{*}=0\\
=\delta_{{\check A}A}[\eta^{\rho\gamma}\epsilon^{\delta}_A(u(p_s))-\eta
^{\rho\delta}\epsilon^{\gamma}_A(u(p_s))] & {}.
\end{array} \right)
\label{II40}
\end{eqnarray}

Since the constraints are redundant, this matrix has the following left and
right null eigenvectors: $\left( \begin{array}{c}  a_{\alpha\beta}=a_{\beta
\alpha} \\ 0 \end{array} \right)$  [$a_{\alpha\beta}$ arbitrary],
$\left( \begin{array}{c}   0 \\ \epsilon^B_{\sigma}(u(p_s))
\end{array} \right)$. Therefore, one has to find
a left and right quasi-inverse
$\bar C$, $\bar CC=C\bar C=D$, such that $\bar C$ and D have the same left and
right null eigenvectors. One finds

\begin{eqnarray}
\bar C&=&\left( \begin{array}{cc}
0_{\gamma\delta\mu\nu} & {1\over 4}[\eta_{\gamma\tau}\epsilon^D_{\delta}(u(p_s))
-\eta_{\delta\tau}\epsilon^D_{\gamma}(u(p_s))] \\ {1\over 4}[\eta_{\sigma\nu}
\epsilon^B_{\mu}(u(p_s))-\eta_{\sigma\mu}\epsilon^B_{\nu}(u(p_s))] &
0^{BD}_{\sigma\tau} \end{array} \right) \nonumber \\
&&{}\nonumber \\
\bar CC=C\bar C=D&=&\left( \begin{array}{cc}
{1\over 2}(\eta^{\alpha}_{\mu}\eta^{\beta}_{\nu}-\eta^{\alpha}_{\nu}\eta_{\mu}
^{\beta}) & 0^{\alpha\beta D}_{\tau} \\
0^{\rho}_{A\mu\nu} & {1\over 2}(\eta^{\rho}_{\tau}\eta^D_A-\epsilon^{D\rho}
(u(p_s))\epsilon_{A\tau}(u(p_s)) \end{array} \right)
\label{II41}
\end{eqnarray}

\noindent and the new Dirac brackets are

\begin{eqnarray}
\lbrace A,B\rbrace {}^{**}&=&\lbrace A,B\rbrace {}^{*}-{1\over 4}[\lbrace
A,{\tilde {\cal H}}^{\gamma\delta}\rbrace {}^{*}[\eta_{\gamma\tau}\epsilon^D
_{\delta}(u(p_s))-\eta_{\delta\tau}\epsilon^D_{\gamma}(u(p_s))]\lbrace
T^{\tau}_D,B\rbrace {}^{*}+\nonumber \\
&+&\lbrace A,T^{\sigma}_B\rbrace {}^{*}[\eta_{\sigma\nu}\epsilon^B_{\mu}(u(p
_s))-\eta_{\sigma\mu}\epsilon^B_{\nu}(u(p_s))]\lbrace {\tilde {\cal H}}^{\mu\nu}
,B\rbrace {}^{*}].
\label{II42}
\end{eqnarray}

\noindent While the check of $\lbrace {\tilde {\cal H}}^{\alpha\beta},B\rbrace
{}^{**}=0$ is immediate, we must use the relation $b_{{\check A}\mu}T^{\mu}
_D\epsilon^{D\rho}=-T^{\rho}_{\check A}$ [at this level we have $T^{\mu}_{\check
A}=T^{\mu}_A$] to check $\lbrace T^{\rho}_A,B\rbrace {}^{**}=0$.

Then, we find the following brackets for the remaining variables
${\tilde x}^{\mu}_s, p_s^{\mu}, \alpha^i, \Pi_i$

\begin{eqnarray}
&&\lbrace {\tilde x}^{\mu}_s,p^{\nu}_s\rbrace {}^{**}=-\eta^{\mu\nu},
\nonumber \\
&&\lbrace \alpha^i(\tau ,\vec \sigma ),\Pi_j(\tau ,{\vec \sigma}^{'})\rbrace
{}^{**}=\delta^i_j\delta^3(\vec \sigma -{\vec \sigma}^{'}),
\label{II43}
\end{eqnarray}

\noindent and the following form of the generators of the ``external" Poincar\'e
group

\begin{eqnarray}
p^{\mu}_s,&&{}\nonumber \\ J^{\mu\nu}_s&=&{\tilde
x}^{\mu}_sp^{\nu}_s-{\tilde x}^{\nu}_sp_s^{\mu}+{\tilde
S}_s^{\mu\nu},\nonumber \\ &&{\tilde S}_s^{oi}=-{ {\delta^{ir}{\bar
S}_s^{rs}p_s^s}\over {p^o_s+
\sqrt{\epsilon p_s^2}} },\nonumber \\
&&{\tilde S}_s^{ij}=\delta^{ir}\delta^{js}{\bar S}_s^{rs}.
\label{II44}
\end{eqnarray}

Let us come back to the four first class constraints ${\tilde {\cal H}}^{\mu}
(\tau )\approx 0$, $\lbrace {\tilde {\cal H}}^{\mu},{\tilde {\cal H}}^{\nu}
\rbrace {}^{**}=0$, of Eq.(\ref{II25}). They can be rewritten in the following
form  [from Eqs.(\ref{I9}), (\ref{II6}) we have $J^{\tau}=-det\,
(\partial_r\alpha^i)$,
$(T^{-1})^{ri}=\delta^{rs}\partial_s\alpha^i/det\,
(\partial_u\alpha^k)$]

\begin{eqnarray}
{\cal H}(\tau )&=&u^{\mu}(p_s){\tilde {\cal H}}_{\mu}(\tau
)=\epsilon_s- M_{sys}\approx 0,\nonumber \\ &&{}\nonumber \\
&&M_{sys}= \int d^3\sigma {\cal M}(\tau ,\vec \sigma )= \nonumber \\
 &=&\int d^{3}{\sigma} \Big( J^{{\tau}}
\sqrt{\mu^2+{\delta}_{uv}(T^{-1})^{ui}{\Pi}_{i}(T^{-1})^{vj}{\Pi}_{j}} \Big)
(\tau ,\vec \sigma )=\nonumber \\
 &=& -\int d^3\sigma \Big[ det\, (\partial_r\alpha^k)\, \sqrt{\mu^2+
  \delta^{uv}{{\partial_u\alpha^i\partial_v\alpha^j}\over
  {[det\, (\partial_r\alpha^k)]^2}}\Pi_i\Pi_j}\Big] (\tau ,\vec \sigma ),\nonumber \\
 &&{}\nonumber \\
 {\vec {\cal
H}}_p(\tau )\, &{\buildrel {def} \over =}\,& {\vec P}_{sys}= \int
d^3\sigma {\cal M}^r(\tau ,\vec \sigma )=\nonumber \\
 &=&\int d^{3}{\sigma}  \mu \Big( J^{{\tau}}
(T^{-1})^{rl}{\Pi}_{l}\Big) (\tau ,\vec \sigma )=-\int d^3\sigma
\, \mu \, \Big[ \delta^{rs}\partial_s\alpha^i \Pi_i\Big] (\tau ,\vec \sigma )\approx 0,
\label{II45}
\end{eqnarray}

\noindent where $M_{sys}$ is the invariant mass of the fluid.
The first one gives the mass spectrum of the isolated system, while
the other three say that the total 3-momentum of the N particles on
the hyperplane $\Sigma_{\tau W}$ vanishes.

There is no more a restriction on $p_s^{\mu}$ in this special gauge,
because $u^{\mu }(p_s)=p^{\mu}
_s/\sqrt{\epsilon p^2_s}$ gives the orientation of the Wigner
hyperplanes containing the isolated system with respect to an
arbitrary given external observer. Now the lapse and shift functions
are

\bea
N&=&N_{[z](flat)}=N_{(flat)}=-\lambda (\tau ) ={\dot x}^{\mu}_s(\tau
)u_{\mu}(p_s),\nonumber \\
N_r&=&N_{[z](flat)r}=N_{(flat)r}=-\lambda_r(\tau)=- {\dot
x}^{\mu}_s(\tau )\epsilon_{r\mu}(u(p_s)),
\label{II46}
\eea

\noindent
so that the velocity of the origin of the coordinates on the Wigner
hyperplane is

\beq
{\dot x}^{\mu}_s(\tau )=\epsilon [-\lambda (\tau )
u^{\mu}(p_s)+\lambda_r(\tau ) \epsilon^{\mu}_r(u(p_s)),\quad
[u^2(p_s)=\epsilon, \epsilon^2_r(u(p_s))=-\epsilon ].
\label{II47}
\eeq

\noindent The Dirac Hamiltonian is now

\beq
H_D=
\lambda (\tau ){\cal H}(\tau )-\vec \lambda (\tau )\cdot {\vec {\cal H}}_p
(\tau ),
\label{II48}
\eeq

\noindent
and we have ${\dot {\tilde x}}_s^{\mu}=\lbrace {\tilde x}_s^{\mu},
H_D\rbrace {}^{**}=-\lambda (\tau )u^{\mu}(p_s)$. Therefore, while the
old $x^{\mu}_s$ had a velocity ${\dot x}_s^{\mu}$ not parallel to the
normal $l^{\mu}=u^{\mu}(p_s)$ to the hyperplane as shown by
Eqs.(\ref{II47}), the new ${\tilde x}_s^{\mu}$ has ${\dot {\tilde
x}}^{\mu}_s \| l^{\mu}$ and no classical zitterbewegung. Moreover, we
have that $T_s=l\cdot {\tilde x}_s= l\cdot x_s$ is the
Lorentz-invariant rest frame time.

The canonical variables ${\tilde x}^{\mu}_s$, $p^{\mu}_s$, may be
replaced by the canonical pairs $\epsilon_s=
\sqrt{p^2_s}$, $T_s=p_s\cdot {\tilde x}_s/
\epsilon_s$ [to be gauge fixed with $T_s-\tau \approx 0$]; ${\vec k}_s={\vec
p}_s/\epsilon_s=\vec u(p_s)$, ${\vec z}_s= \epsilon_s ({\vec {\tilde x}}_s-
{{{\vec p}_s}\over {p^o_s}} {\tilde x}^o_s)\equiv \epsilon_s {\vec q}_s$.

One obtains in this way a new kind of instant form of the dynamics,
the  ``Wigner-covariant 1-time rest-frame instant form" with a
universal breaking of Lorentz covariance. It is the special
relativistic generalization of the non-relativistic separation of the
center of mass from the relative motion [$H={{ {\vec P}^2}\over
{2M}}+H_{rel}$]. The role of the  ``external" center of mass is taken
by the Wigner hyperplane, identified by the point ${\tilde x}_s^{\mu}
(\tau )$ and by its normal $p^{\mu}_s$. The invariant mass $M_{sys}$
of the system replaces the non-relativistic Hamiltonian $H_{rel}$ for
the relative degrees of freedom, after the addition of the
gauge-fixing $T_s-\tau \approx 0$ [identifying the time parameter
$\tau$, labelling the leaves of the foliation,  with the Lorentz
scalar time of the ``external" center of mass in the rest frame,
$T_s=p_s\cdot {\tilde x}_s/M_{sys}$ and implying $\lambda (\tau
)=-\epsilon$].  After this gauge fixing the Dirac Hamiltonian would be
pure gauge: $H_D=-\vec \lambda (\tau )\cdot {\vec {\cal H}}_p(\tau )$.
However, if we wish to reintroduce the evolution in the time $\tau
\equiv T_s$ in this frozen phase space we must use the Hamiltonian
[in it the time evolution is generated by $M_{sys}$: it is like in the
frozen Hamilton-Jacobi theory, in which the evolution can be
reintroduced by using the energy generator of the Poincar\'e group as
Hamiltonian]

\beq
H_D = M_{sys} - \vec \lambda (\tau )\cdot {\vec {\cal H}}_p(\tau ).
\label{II49}
\eeq

The Hamilton equations for $\alpha^i(\tau ,\vec \sigma )$ in the
Wigner covariant rest-frame instant form are equivalent to the
hydrodynamical Euler equations:

\bea
\partial_{\tau}\alpha^i(\tau ,\vec \sigma ) &=&
 \{ \alpha^i(\tau ,\vec \sigma ), H_D \} =\nonumber \\
 &=&- \Big( {{\delta^{uv} \partial_u\alpha^i \partial_v\alpha^j \Pi_j}\over
 {det\, (\partial_r\alpha^k)\, \sqrt{\mu^2+\delta^{uv}{{\partial_u\alpha^m\partial_v\alpha^n}
 \over {[det\, (\partial_r\alpha^k)]^2}}\Pi_m\Pi_n} }}\Big) (\tau ,\vec \sigma )+\nonumber \\
 &+& \lambda^r(\tau ) \partial_r\alpha^i(\tau ),\nonumber \\
 &&{}\nonumber \\
 \partial_{\tau}\Pi_i(\tau ,\vec \sigma ) &=&
 \{ \Pi_i(\tau ,\vec \sigma ) ,H_d \} = \quad\quad (ijk\, cyclic)\nonumber \\
 &=& {{\partial}\over {\partial \sigma^s}}
 \Big[ \epsilon^{suv}\partial_u\alpha^j\partial_v\alpha^k\,
\sqrt{\mu^2+\delta^{uv}{{\partial_u\alpha^m\partial_v\alpha^n}
 \over {[det\, (\partial_r\alpha^k)]^2}}\Pi_m\Pi_n} +\nonumber \\
 &+&{{ (\delta^m_i\delta^u_s-{{\delta^{uv}\partial_v\alpha^m}\over {det\, (\partial_r\alpha^l)}}
 \epsilon^{slt}\epsilon_{ipq}\partial_l\alpha^p\partial_t\alpha^q)
 {{\partial_u\alpha^n}\over {det\, (\partial_r\alpha^l)}}\Pi_m\Pi_n}\over
 { \sqrt{\mu^2+\delta^{uv}{{\partial_u\alpha^m\partial_v\alpha^n}
 \over {[det\, (\partial_r\alpha^k)]^2}}\Pi_m\Pi_n} }} \Big] (\tau ,\vec \sigma )
 +\nonumber \\
 &+&\lambda^r(\tau ) {{\partial}\over {\partial \sigma^s}}
 \Big[ \epsilon^{suv}\partial_u\alpha^j\partial_v\alpha^k
 {{\partial_r\alpha^m}\over {det\, (\partial_t\alpha^l)}} \Pi_m+\nonumber \\
 &+&(\delta^m_i\delta_{rs}-{{\partial_r\alpha^m}\over {det\, (\partial_t\alpha^l)}}
 \epsilon^{slt}\epsilon_{ipq}\partial_l\alpha^p\partial_t\alpha^q) \Pi_m\Big] (\tau ,\vec \sigma ).
\label{II50}
\eea

In this special gauge we have $b^{\mu}_A\equiv L^{\mu}{}_A(p_s,{\buildrel
\circ \over p}_s)$ (the standard Wigner boost for timelike Poincar\'e orbits),
$S_s^{\mu\nu}\equiv S_{sys}^{\mu\nu}$ [$S^r_{sys}= \epsilon^{ruv}
\int d^3\sigma \, \sigma^u\, \Big( J^{\tau}\, (T^{-1})^{vs} \Pi_s\Big)
(\tau ,\vec \sigma )$], and the only remaining canonical variables are
the non-covariant Newton-Wigner-like canonical ``external"
3-center-of-mass coordinate ${\vec z}_s$ (living on the Wigner
hyperplanes) and ${\vec k}_s$. Now 3 degrees of freedom of the
isolated system [an ``internal" center-of-mass 3-variable ${\vec
q}_{sys}$ defined inside the Wigner hyperplane and conjugate to ${\vec
P}_{sys}$] become gauge variables [the natural gauge fixing to the
rest-frame condition ${\vec P}_{sys}\approx 0$ is ${\vec
X}_{sys}\approx 0$, implying $\lambda
_r(\tau )=0$, so that it coincides
with the origin $x^{\mu}_s(\tau )=z^{\mu}(\tau ,\vec \sigma =0)$ of
the Wigner hyperplane]. The variable ${\tilde x}^{\mu}_s$ is playing
the role of a kinematical ``external" center of mass for the isolated
system and may be interpreted as a decoupled observer with his
parametrized clock (point particle clock). All the fields living on
the Wigner hyperplane are now either Lorentz scalar or with their
3-indices transformaing under Wigner rotations (induced by Lorentz
transformations in Minkowski spacetime) as any Wigner spin 1 index.

\vfill\eject

\section{External and internal canonical center of mass, Moller's center of
energy and Fokker-Pryce center of inertia}

Let us now consider the problem of the definition of the relativistic
center of mass of a perfect  fluid configuration,  using the dust as
an example. Let us remark that in the approach leading to the
rest-frame instant form of dynamics on Wigner's hyperplanes there is a
splitting of this concept in an ``external" and an ``internal" one.
One can either look at the isolated system from an arbitrary Lorentz
frame or put himself inside the Wigner hyperplane.

From outside one finds after the canonical reduction to Wigner
hyperplane that there is an origin $x^{\mu}_s(\tau )$ for these
hyperplanes (a covariant non-canonical centroid) and a non-covariant
canonical coordinate ${\tilde x}
^{\mu}_s(\tau )$ describing an ``external" decoupled point particle observer
with a clock measuring the rest-frame time $T_s$. Associated with them
there is the ``external" realization (\ref{II44}) of the Poincar\'e
group.

Instead, all the degrees of freedom of the isolated system (here the
perfect fluid configuration) are described by canonical variables on
the Wigner hyperplane restricted by the rest-frame condition ${\vec
P}_{sys}\approx 0$, implying that an ``internal" collective variable
${\vec q}_{sys}$ is a gauge variable and that only relative variables
are physical degrees of freedom (a form of weak Mach principle).

Inside the Wigner hyperplane at $\tau =0$ there is another realization
of the Poincar\'e group, the ``internal" Poincar\'e group.
Its generators are built by using the invariant mass $M_{sys}$ and the
3-momentum ${\vec P}_{sys}$, determined by the constraints
(\ref{II25}), as the generators of the translations and by using the
spin tensor ${\bar S}_s^{AB}$ as the generator of the Lorentz
subalgebra

\bea
P^{\tau}&=&  M_{sys}=\int d^3\sigma \Big[ J^{\tau}
\sqrt{\mu^2+\delta_{uv}(T^{-1})^{ui} (T^{-1})^{vj} \Pi_i \Pi_j}\Big]
(\tau ,\vec \sigma ),\nonumber \\
 P^r&=& {\vec P}_{sys}= -\int d^3\sigma \Big[ J^{\tau}
(T^{-1})^{ri} \Pi_i\Big] (\tau ,\vec \sigma )
\approx 0,\nonumber \\
K^r&=&J^{\tau r}={\bar S}_s^{\tau r}\equiv \int d^3\sigma \sigma^r
\Big[ J^{\tau}
\sqrt{\mu^2+\delta_{uv}(T^{-1})^{ui} (T^{-1})^{vj} \Pi_i \Pi_j}\Big]
(\tau ,\vec \sigma ),\nonumber \\
 J^r&=& S^r_{sys} ={1\over 2}\epsilon^{ruv} {\bar S}_s^{uv}
\equiv \epsilon^{ruv} \int d^3\sigma \sigma^u \Big[ J^{\tau} (T^{-1})^{vi}
\Pi_i \Big] (\tau ,\vec \sigma ).
\label{III1}
\eea

By using the methods of Ref.\cite{pauri} (where there is a complete
discussion of many definitions of relativistic center-of-mass-like
variables) we can build the three ``internal" (that is inside the
Wigner hyperplane) Wigner 3-vectors corresponding to the 3-vectors
'canonical center of mass' ${\vec q}_{sys}$, 'Moller center of energy'
${\vec r}_{sys}$ and 'Fokker-Pryce center of inertia' ${\vec y}_{sys}$
[the analogous concepts for the Klein-Gordon field are in
Ref.\cite{mate} (based on Refs.\cite{lon}), while for the relativistic
N-body problem see Ref.\cite{iten} and for the system of N charged
scalar particles plus the electromagnetic field Ref.\cite{crater}].

The non-canonical ``internal" M\o ller 3-center of energy and the
associated spin 3-vector are

\begin{eqnarray}
{\vec r}_{sys}&=& - {{\vec K}\over {P^{\tau}}} =
-{1\over {2P^{\tau}}} \int d^3\sigma
\,\, \vec \sigma \, \Big[ J^{\tau}
\sqrt{\mu^2+\delta_{uv}(T^{-1})^{ui} (T^{-1})^{vj} \Pi_i \Pi_j}\Big]
(\tau ,\vec \sigma ),\nonumber \\
 &&{}\nonumber \\
{\vec \Omega}_{sys} &=& {\vec J} -{\vec r}_{sys}\times {\vec
P},\nonumber \\
 &&\{ r^r_{sys},P^s \} =\delta^{rs},\quad\quad
\{ r^r_{sys},P^{\tau} \} ={{P^r}\over {P^{\tau}}},\nonumber \\
&&\{ r^r_{sys},r^s_{sys} \} =-{1\over {(P^{\tau})^2}} \epsilon^{rsu}
\Omega^u_{sys},\nonumber \\
&&\{ \Omega^r_{sys},\Omega^s_{sys} \}
=\epsilon^{rsu}(\Omega^u_{sys}-{1\over {(P^{\tau})^2}}({\vec
\Omega}_{sys} \cdot {\vec P})\,\, P
^u),\quad\quad \{ \Omega^r_{sys},P^{\tau} \} =0.
\label{III2}
\end{eqnarray}

The canonical ``internal" 3-center of mass ${\vec q}_{sys}$ [$\{
q^r_{sys},q^s_{sys}
\}
=0$, $\{ q^r_{sys},P^s \} =
\delta^{rs}$, $\{ J^r,q^s_{sys} \} =\epsilon^{rsu} q^u_{sys}$] is

\begin{eqnarray}
{\vec q}_{sys}&=& {\vec r}_{sys}- {{{\vec J}\times {\vec \Omega}
_{sys}}\over {\sqrt{(P^{\tau})^2-{\vec P}^2}(P^{\tau}+
\sqrt{(P^{\tau})^2-{\vec P}^2})}}=
\nonumber \\
&=&-{{{\vec K}}\over {\sqrt{(P^{\tau})^2-{\vec P}^2}}}+ {{{\vec
J}\times {\vec P}}\over {\sqrt{(P^{\tau})^2-{\vec
P}^2}(P^{\tau}+\sqrt{(P^{\tau})^2
-{\vec P}^2})}}+\nonumber \\
 &+&{{({\vec K}\cdot {\vec P})\,\, {\vec P}}\over {P^{\tau}
\sqrt{(P^{\tau})^2-{\vec P}^2}\Big( P^{\tau}+
\sqrt{(P^{\tau})^2-{\vec P}^2}\Big) }},\nonumber \\
&&\approx {\vec r}_{sys}\quad for\quad {\vec P}\approx 0;\quad\quad
\{ {\vec q}_{sys},P^{\tau} \} ={{{\vec P}}\over {P^{\tau}
}}\approx 0,\nonumber \\
 &&{}\nonumber \\
  {\vec S}_{q\, sys} &=&{\vec J}-{\vec q}_{sys}\times {\vec P}=
\nonumber \\
 &=& {{P^{\tau}{\vec J}}\over {\sqrt{(P^{\tau})^2-{\vec P}
^2}}}+{{{\vec K}\times {\vec P}}\over {\sqrt{(P^{\tau}
)^2-{\vec P}^2}}}-{{({\vec J}\cdot {\vec P})\,\, {\vec P}}\over
{\sqrt{(P^{\tau})^2-{\vec P}^2}\Big( P^{\tau}+\sqrt{(P^{\tau})^2-{\vec
P}^2}\Big) }}\approx \nonumber \\
 &\approx& {\vec S}_{sys}, \quad
for\quad {\vec P}\approx 0,\quad
S^r_{sys}= \epsilon^{ruv}
\int d^3\sigma \, \sigma^u\, \Big( J^{\tau}\, (T^{-1})^{vs} \Pi_s\Big)
(\tau ,\vec \sigma ),\nonumber \\
  &&\{ {\vec S}_{q\, sys},{\vec P} \} =\{ {\vec S}_{q\, sys},{\vec q}_{sys}
\} =0,\quad\quad \{ S^r_{q\, sys},S^s_{q\, sys} \} =\epsilon^{rsu}S^u_{q\, sys}.
\label{III3}
\end{eqnarray}

The ``internal" non-canonical Fokker-Pryce 3-center of inertia' ${\vec
y}_{sys}$ is

\begin{eqnarray}
{\vec y}_{sys}&=& {\vec q}_{sys}+{{{\vec S}_{sys}\times {\vec P}}
\over {\sqrt{(P^{\tau})^2-{\vec P}^2} (P^{\tau}+
\sqrt{(P^{\tau})^2-{\vec P}^2})}}
={\vec r}_{sys}+{{{\vec S}_{sys}\times {\vec P}}\over {P^{\tau}
\sqrt{(P^{\tau})^2-{\vec P}^2}}},\nonumber \\
&&{}\nonumber \\
 {\vec q}_{sys}&=&{\vec r}_{sys}+{{{\vec S}_{sys}\times {\vec P}}
\over {P^{\tau}(P^{\tau}+\sqrt{(P^{\tau})^2-{\vec P}
^2})}} = {{P^{\tau} {\vec r}_{sys}+\sqrt{(P^{\tau})^2-
{\vec P}^2} {\vec y}_{sys}}\over {P^{\tau}+\sqrt{(P^{\tau} )^2-{\vec
P}^2}}},\nonumber \\
 &&\{ y^r_{sys},y^s_{sys} \} ={1\over
{P^{\tau}\sqrt{(P^{\tau})
^2-{\vec P}^2} }}\epsilon^{rsu}\Big[ S^u_{sys}+{{ ({\vec S}_{sys}
\cdot {\vec P})\, P^u}\over {\sqrt{(P^{\tau})^2-{\vec P}
^2}(P^{\tau}+\sqrt{(P^{\tau})^2-{\vec P}^2})}}
\Big] ,\nonumber \\
&&{}\nonumber \\
 {\vec P}\approx 0 &\Rightarrow& {\vec q}_{sys}\approx {\vec r}_{sys}
\approx {\vec y}_{sys}.
\label{III4}
\end{eqnarray}

The Wigner 3-vector ${\vec q}_{sys}$ is therefore the canonical
3-center of mass of the perfect fluid configuration [since ${\vec
q}_{sys}\approx {\vec r}_{sys}$, it also describe that point
$z^{\mu}(\tau ,{\vec q}_{sys})=x^{\mu}_s(\tau )+ q^r_{sys}
\epsilon^{\mu}_r(u(p_s))$ where the energy of the  configuration
is concentrated].

There should exist a canonical transformation from the canonical basis
$\alpha^i(\tau ,\vec \sigma )$, $\Pi_i(\tau ,\vec \sigma )$, to a new
basis ${\vec q}_{sys}$, $\vec P={\vec P}_{\phi}$, $\alpha_{rel}^i(\tau
,\vec \sigma )$, $\Pi_{rel\, i}(\tau ,\vec \sigma )$ containing
relative variables $\alpha_{rel}^i(\tau ,\vec \sigma )$, $\Pi_{rel\,
i}(\tau ,\vec \sigma )$ with respect to the true center of mass of the
perfect fluid configuration. To identify this final canonical basis
one shall need the methods of Ref.\cite{iten}.

The gauge fixing ${\vec q}_{sys}\approx 0$ [it implies $\vec
\lambda (\tau )=0$] forces all three
internal center-of-mass variables to coincide with the origin
$x^{\mu}_s$ of the Wigner hyperplane. We shall denote $x_s^{({\vec
q}_{sys})\mu}(\tau )=x^{\mu}_s(0)+\tau u^{\mu}(p_s)$ the origin in
this gauge (it is a special centroid among the many possible ones;
$x^{\mu}_s(0)$ is arbitrary).

As we shall  see in the next Section, by adding the gauge fixings
${\vec X}_{sys}={\vec q}_{sys}\approx 0$ one can show that the origin
$x_s^{\mu}(\tau )$ becomes simultaneously the Dixon center of mass of
an extended object and both the Pirani and Tulczyjew centroids (see
Ref.\cite{bini} for a review of these concepts in relation with the
Papapetrou-Dixon-Souriau pole-dipole approximation of an extended
body). The worldline $x^{({\vec q}_{sys})\mu}
_s$ is the unique center-of-mass worldline of special relativity in the sense
of Refs.\cite{bei}.

With similar methods from the rest-frame instant form ``external"
realization of the Poincar\'e algebra of Eq. (\ref{II44}) with the
generators $p^{\mu}_s$, $J^{ij}_s={\tilde x}^i_sp^j_s-{\tilde
x}^j_sp^i_s+\delta^{ir}
\delta^{js} S^{rs}_{\phi}$, $K^i_s=J^{oi}_s={\tilde x}^o_sp^i_s-
{\tilde x}^i_sp^o_s-{{\delta^{ir} S^{rs}_{\phi}\, p^s_s}\over {p^o_s+
\epsilon_s}}={\tilde x}^o_sp^i_s-{\tilde x}^i_sp^o_s+\delta^{ir}{{({\vec S}
_{\phi}\times {\vec p}_s)^r}\over {p^o_s+\epsilon_s}}$ [for ${\tilde
x}^o_s=0$ this is the Newton-Wigner decomposition of $J^{\mu\nu}_s$]
we can build three ``external" collective 3-positions (all located on
the Wigner hyperplane): i) the ``external canonical 3-center of mass
${\vec Q}_s$ connected with the ``external" canonical non-covariant
center of mass ${\tilde x}_s^{\mu}$; ii) the ``external" M\o ller
3-center of energy ${\vec R}_s$ connected with the ``external"
non-canonical and non-covariant M\o ller center of energy $R^{\mu}_s$;
iii) the ``external" Fokker-Pryce 3-center of inertia connected with
the ``external" covariant non-canonical Fokker-Price center of inertia
$Y^{\mu}_s$ (when there are the gauge fixings ${\vec
\sigma}_{sys}\approx 0$ it coincides with the origin $x^{\mu}_s$).
It turns out that the Wigner hyperplane is the natural setting for the
study of the Dixon multipoles of extended relativistic
systems\cite{dixon} (see next Section)  and for defining the canonical
relative variables with respect to the center of mass.

The three ``external" 3-variables, the canonical ${\vec Q}_s$, the M\o
ller ${\vec R}_s$ and the Fokker-Pryce ${\vec Y}
_s$ built by using the rest-frame ``external" realization of the Poincar\'e algebra
are

\begin{eqnarray}
{\vec R}_s&=& -{1\over {p^o_s}}{\vec K}_s=({\vec {\tilde x}}_s-{{{\vec
p}_s}\over {p^o_s}} {\tilde x}^o_s)-{{{\vec S}_{sys}\times {\vec p}_s}
\over {p^o_s(p^o_s+\epsilon_s)}},\nonumber \\
{\vec Q}_s&=&{\vec {\tilde x}}_s-{{{\vec p}_s}\over {p^o_s}}{\tilde
x}^o_s= {{{\vec z}_s}\over {\epsilon_s}}= {\vec R}_s+{{ {\vec
S}_{sys}\times {\vec p}_s}\over {p^o_s(p^o_s+
\epsilon_s)}}={{p^o_s {\vec R}_s+\epsilon_s {\vec Y}_s}\over {p^o_s+\epsilon_s}}
,\nonumber \\
 {\vec Y}_s&=&{\vec Q}_s+{{ {\vec S}_{sys}\times {\vec
p}_s}\over {\epsilon_s (p^o_s+\epsilon_s)}}={\vec R}_s+{{ {\vec
S}_{sys}\times {\vec p}_s}\over {p^o_s\epsilon_s}},\nonumber \\
 &&\{R^r_s,R^s_s \} =-{1\over {(p^o_s)^2}}\epsilon^{rsu}\Omega^u_s,
\quad\quad {\vec \Omega}_s={\vec J}_s-{\vec R}_s\times {\vec p}_s,\nonumber \\
&&\{ Y^r_s,Y^s_s \} ={1\over {\epsilon_sp^o_s}}\epsilon^{rsu}\Big[
S^u_{sys} +{{ ({\vec S}_{sys}\cdot {\vec p}_s)\, p^u_s}\over
{\epsilon_s(p^o_s+
\epsilon_s)}}\Big] ,\nonumber \\
&&{}\nonumber \\
&&{\vec p}_s\cdot {\vec Q}_s={\vec p}_s\cdot {\vec R}_s={\vec p}_s\cdot {\vec
Y}_s={\vec k}_s\cdot {\vec z}_s,\nonumber \\
 &&{}\nonumber \\
{\vec p}_s=0 &\Rightarrow& {\vec Q}_s={\vec Y}_s={\vec R}_s,
\label{III5}
\end{eqnarray}

\noindent with the same velocity and coinciding in the Lorentz rest frame where
${\buildrel \circ \over p}^{\mu}_s=\epsilon_s (1;\vec 0)$

In Ref.\cite{pauri} in a one-time framework without constraints and at a fixed
time, it is shown that the 3-vector ${\vec Y}_s$ [but not ${\vec Q}_s$ and
${\vec R}_s$] satisfies the condition $\{ K^r_s,Y^s_s \} =
Y^r_s\, \{ Y^s_s,p^o_s \}$ for being the space component of a 4-vector
$Y^{\mu}_s$. In the enlarged canonical treatment including time variables, it is
not clear which are the time components to be added to ${\vec Q}_s$, ${\vec R}
_s$, ${\vec Y}_s$, to rebuild 4-dimesnional quantities ${\tilde x}^{\mu}_s$,
$R^{\mu}_s$, $Y^{\mu}_s$, in an arbitrary Lorentz frame $\Gamma$, in which the
origin of the Wigner hyperplane is the 4-vector $x^{\mu}_s = (x^o_s; {\vec x}
_s)$. We have

\begin{eqnarray}
{\tilde x}^{\mu}_s(\tau )&=&
({\tilde x}^o_s(\tau ); {\vec {\tilde x}}_s(\tau ) )=
x^{\mu}_s-{1\over {\epsilon_s(p^o_s+\epsilon_s)}}\Big[
p_{s\nu}S_s^{\nu\mu}+\epsilon_s(S^{o\mu}_s-S^{o\nu}_s{{p_{s\nu}p_s^{\mu}}\over
{\epsilon^2_s}}) \Big],\quad\quad p^{\mu}_s,\nonumber \\
{\tilde x}^o_s&=&\sqrt{1+{\vec k}_s^2} (T_s+{{{\vec k}_s\cdot {\vec z}_s}\over
{\epsilon_s}})=\sqrt{1+{\vec k}^2_s}(T_s+{\vec k}_s\cdot {\vec q}_s)\not=
x^0_s,\quad\quad p^o_s=\epsilon_s\sqrt{1+{\vec k}_s^2},\nonumber \\
{\vec {\tilde x}}_s&=&{{ {\vec z}_s}\over {\epsilon_s}}+(T_s+{{{\vec k}_s\cdot
{\vec z}_s}\over {\epsilon_s}}) {\vec k}_s={\vec q}_s+(T_s+{\vec k}_s\cdot
{\vec q}_s){\vec k}_s,\quad\quad {\vec p}_s=\epsilon_s {\vec k}_s.
\label{III6}
\end{eqnarray}

\noindent for the non-covariant (frame-dependent) canonical center of mass and
its conjugate momentum.

Each Wigner hyperplane intersects the worldline of the arbitrary origin
4-vector $x^{\mu}_s(\tau )=z^{\mu}(\tau ,\vec 0)$ in $\vec \sigma =0$, the
pseudo worldline of ${\tilde x}^{\mu}_s(\tau )=z^{\mu}(\tau ,{\tilde {\vec
\sigma}})$ in some ${\tilde {\vec \sigma}}$ and the worldline of the
 Fokker-Pryce 4-vector $Y^{\mu}_s(\tau )=z^{\mu}(\tau ,{\vec \sigma}_Y)$ in
some ${\vec \sigma}_Y$ [on this worldline one can put the ``internal
center of mass" with the gauge fixing ${\vec q}_{\phi}\approx 0$
(${\vec q}_{\phi}\approx {\vec r}_{\phi}\approx {\vec y}_{\phi}$ due
to ${\vec P}_{\phi}\approx 0$)]; one also has $R^{\mu}_s=z^{\mu}(\tau
,{\vec \sigma}_R)$. Since we have $T_s=u(p_s)\cdot x_s=u(p_s)\cdot
{\tilde x}_s\equiv \tau$ on the Wigner hyperplane labelled by $\tau$,
we require that also $Y^{\mu}_s$, $R^{\mu}_s$ have time components
such that they too satisfy $u(p_s)\cdot Y_s=u(p_s)\cdot R_s=T_s\equiv
\tau$. Therefore, it is reasonable to assume that ${\tilde
x}^{\mu}_s$, $Y^{\mu}_s$ and $R^{\mu}_s$ satisfy the following
equations consistently with Eqs.(\ref{III2}), (\ref{III3}) when
$T_s\equiv
\tau$ and ${\vec q}_{sys}\approx 0$

\begin{eqnarray}
{\tilde x}^{\mu}_s&=&( {\tilde x}^o_s; {\vec {\tilde x}}_s)=( {\tilde
x}^o_s; {\vec Q}_s+{{{\vec p}_s}\over {p^o_s}} {\tilde x}^o_s
)=\nonumber \\
 &=&({\tilde x}^o_s;
{{{\vec z}_s}\over {\epsilon_s}}+(T_s+{{{\vec k}_s\cdot {\vec z}_s}\over
{\epsilon_s}}){\vec k}_s )
=x^{({\vec q}_{sys})\mu}_s+\epsilon^{\mu}_u(u(p_s)) {\tilde \sigma}^u,
\nonumber \\
 Y^{\mu}_s&=& ({\tilde x}^o_s; {\vec Y}_s)=\nonumber \\
  &=&({\tilde x}^o_s;\, {1\over {\epsilon_s}}[{\vec z}_s+{{{\vec
S}_{sys}\times {\vec p}_s}\over {\epsilon_s[1+u^o(p_s)]}}]+(T_s+
{{{\vec k}_s\cdot {\vec z}_s}\over {\epsilon_s}}){\vec k}_s\,
)=\nonumber \\
 &=&{\tilde x}^{\mu}_s+\eta^{\mu}_r{{({\vec
S}_{sys}\times {\vec p}_s)^r}\over {\epsilon_s[1+u^o(p_s)]}}=\nonumber
\\
 &=&x^{({\vec q}_{sys})\mu}_s+\epsilon^{\mu}_u(u(p_s))
\sigma^u_Y,\nonumber \\
 R^{\mu}_s&=& ( {\tilde x}^o_s; {\vec R}_s)=\nonumber \\
  &=&( {\tilde x}^o_s;\, {1\over
{\epsilon_s}}[{\vec z}_s- {{{\vec S}_{sys}\times {\vec p}_s}\over
{\epsilon_s u^o(p_s) [1+u^o(p_s)]}}]+(T_s+ {{{\vec k}_s\cdot {\vec
z}_s}\over {\epsilon_s}}){\vec k}_s\, )=\nonumber \\
 &=&{\tilde
x}^{\mu}_s-\eta^{\mu}_r{{({\vec S}_{sys}\times {\vec p}_s)^r}
\over {\epsilon_su^o(p_s)[1+u^o(p_s)]}}=\nonumber \\
&=&x^{({\vec q}_{sys})\mu}_s+\epsilon^{\mu}_u(u(p_s))
\sigma^u_R,\nonumber \\
 &&{}\nonumber \\
 T_s&=&u(p_s)\cdot x_s^{({\vec
q}_{sys})}=u(p_s)\cdot {\tilde x}_s=u(p_s)\cdot Y_s=u(p_s)\cdot
R_s,\nonumber \\
 &&{}\nonumber \\
 {\tilde \sigma}^r&=&\epsilon_{r\mu}(u(p_s)) [x^{({\vec
q}_{sys})\mu}_s-{\tilde x}^{\mu}_s]= {{ \epsilon_{r\mu}(u(p_s))
[u_{\nu}(p_s)S^{\nu\mu}_s+S^{o\mu}_s]}\over {[1+u^o(p_s)]}}=\nonumber
\\
 &=&-S_{sys}^{\tau r}+{{S_{sys}^{rs}p^s_s}\over
{\epsilon_s[1+u^o(p_s)]}}
=\epsilon_s r^r_{\phi}+{{S_{sys}^{rs}u^s(p_s)}\over {1+u^o(p_s)}}
\approx \nonumber \\
 &\approx& \epsilon_s q^r_{sys}+{{S_{sys}^{rs}u^s(p_s)}\over {1+
u^o(p_s)}}\approx {{S_{sys}^{rs}u^s(p_s)}\over {1+u^o(p_s)}}
,\nonumber
\\
 \sigma^r_Y&=&\epsilon_{r\mu}(u(p_s))[x^{({\vec q}_{sys})\mu}_s-Y^{\mu}_s]=
{\tilde \sigma}^r-\epsilon_{ru}(u(p_s)){{({\vec S}_{sys}\times {\vec
p}_s)^u}\over {\epsilon_s[1+u^o(p_s)]}}=\nonumber \\
 &=&{\tilde \sigma}^r+{{S^{rs}_{sys}u^s(p_s)}\over {1+u^o(p_s)}}=
\epsilon_s r^r_{sys} \approx \epsilon_s q^r_{sys} \approx 0,\nonumber \\
\sigma^r_R&=&\epsilon_{r\mu}(u(p_s))
[x^{({\vec q}_{sys})\mu}_s-R^{\mu}_s]={\tilde \sigma}^r+
\epsilon_{ru}(u(p_s)) {{({\vec S}_{sys}\times {\vec p}_s)^u}
\over {\epsilon_su^o(p_s)[1+u^o(p_s)]}}=\nonumber \\
&=&{\tilde \sigma}^r-{{S_{sys}^{rs}u^s(p_s)}\over {u^o(p_s)[1+
u^o(p_s)]}}=\epsilon_sr^r_{sys}+{{[1-u^o(p_s)]S^{rs}_{sys}u^s(p_s)}\over
{u^o(p_s)[1+u^o(p_s)]}}\approx \nonumber \\
 &\approx& {{[1-u^o(p_s)]S^{rs}_{sys}u^s(p_s)}\over
{u^o(p_s)[1+u^o(p_s)]}},\nonumber \\
 &&{}\nonumber \\
  &\Rightarrow&
x^{({\vec q}_{sys})\mu}_s(\tau ) = Y^{\mu}_s,\quad for\quad {\vec
q}_{sys}\approx 0,
\label{III7}
\end{eqnarray}

\noindent namely in the gauge ${\vec q}_{sys}\approx 0$
the external Fokker-Pryce non-canonical center of inertia coincides
with the origin $x^{({\vec q}_{sys})\mu}_s(\tau )$ carrying the
``internal" center of mass (coinciding with the ``internal" M\"oller
center of energy and with the ``internal" Fokker-Pryce center of
inertia) and also being the Pirani centroid and the Tulczyjew
centroid.

Therefore, if we would find the center-of-mass canonical basis, then,
in the gauge ${\vec q}_{sys}\approx 0$ and $T_s
\approx \tau$, the perfect fluid configurations would have the
four-momentum density peaked on the worldline $x^{({\vec
q}_{sys})\mu}_s(T_s)$; the canonical variables $\alpha^i_{rel}(\tau
,\vec \sigma )$, $\Pi_{rel\, i}(\tau ,\vec \sigma )$ would
characterize the relative motions with respect to the ``monopole"
configuration describing the center of mass of the fluid
configuration. The ``monopole" configurations would be identified by
the vanishing of the relative variables.

Remember that the canonical center of mass  lies in between the Moller
center of energy and the Fokker-Pryce center of inertia and that the
non-covariance region around the Fokker-Pryce 4-vector extends to a
worldtube with radius (the Moller radius) $|{\vec S}_{sys}| /
P^{\tau}$.

\vfill\eject

\section{Dixon's Multipoles in Minkowski Spacetime.}

Let us now look at other properties of a perfect fluid configuration
on the Wigner hyperplanes, always using the dust as an explicit
example. To identify which kind of collective variables describe the
center of mass of a fluid configuration let us consider  it as a
relativistic extended body and let us study its energy-momentum tensor
and its Dixon multipoles\cite{dixon} in Minkowski spacetime.

The Euler-Lagrange equations from the action (\ref{I10}) [(\ref{II1})
for the dust] are

\begin{eqnarray}
&&\Big( {{\partial {\cal L}}\over {\partial z^{\mu}}}-\partial_{\check
A} {{\partial {\cal L}}\over {\partial z^{\mu}_{\check A}}}\Big) (\tau
,\vec
\sigma )=\eta_{\mu\nu}\partial_{\check A}[\sqrt{g} T^{\check A\check B}[\alpha ]
z_{\check B}^{\nu}](\tau ,\vec
\sigma )\, {\buildrel \circ \over =}\, 0,\nonumber \\
&&\Big( {{\partial {\cal L}}\over {\partial
\phi}}-\partial_{\check A}{{\partial {\cal L}}\over {\partial \partial_{\check A}\phi}}\Big)
(\tau ,\vec \sigma )\, {\buildrel \circ \over =}\, 0,
\label{IV1}
\end{eqnarray}

\noindent where we introduced the energy-momentum tensor [with a different
sign with respect to the standard convention to conform with
Ref.\cite{brown}]

\begin{eqnarray}
T^{\check A\check B}(\tau ,\vec \sigma )[\alpha ]&=&\Big[ {2\over
{\sqrt{g}}}{{\delta S}\over {\delta g_{\check A\check B}}}\Big] (\tau
,\vec \sigma )=\nonumber \\
 &=&\Big[ - \rho g^{\check A\check B} +n {{\partial \rho}\over
 {\partial n}}{|}_s\, (g^{\check A\check B}-{{J^{\check A} J^{\check B}}\over
 {g_{\check C\check D} J^{\check C} J^{\check D}}})\Big] (\tau ,\vec \sigma )=
 \nonumber \\
 &=&\Big[ -\epsilon \rho U^{\check A} U^{\check B} +p (g^{\check A\check B}-\epsilon
 U^{\check A} U^{\check B})\Big] (\tau ,\vec \sigma )\nonumber \\
 &&{}\nonumber \\
 &{\buildrel {dust} \over =}\,& -\epsilon \mu n U^{\check A} U^{\check B}
 =-\epsilon  {{J^{\check A} J^{\check B}}\over {N^2 \sqrt{\gamma}\, J^{\tau}}}
 \sqrt{\mu^2+{}^3g_{\check u\check v}(T^{-1})^{\check ui}
 (T^{-1})^{\check vj} \Pi_i\Pi_j}.
\label{IV2}
\end{eqnarray}

When $\partial_{\check A}[\sqrt{g} z^{\mu}_{\check B}]=0$, as it
happens on the Wigner hyperplanes in the gauge $T_s-\tau \approx 0$,
$\vec \lambda (\tau )=0$, we get the conservation of the
energy-momentum tensor $T^{\check A\check B}$, i.e. $\partial_{\check
A}T^{\check A\check B}\, {\buildrel \circ \over
=}\, 0$. Otherwise, there is compensation coming from the dynamics of
the surface.

As shown in Eq.(\ref{a9}) the conserved, manifestly Lorentz covariant
energy-momentum tensor of the perfect fluid with equation of state
$\rho =\rho (n,s)$ [so that $p=n {{\partial \rho}\over {\partial
n}}{|}_s-\rho)$] is

\bea
T^{\mu\nu}(x)[\tilde \alpha ]&=& \Big[ -\epsilon \rho \,
{}^4g^{\mu\nu} +n {{\partial
\rho}\over {\partial n}}{|}_s\, ({}^4g^{\mu\nu} -{{J^{\mu}
J^{\nu}}\over {{}^4g_{\alpha\beta}J^{\alpha} J^{\beta}}})\Big]
(x)=\nonumber \\
 &=&\Big[ -\epsilon \rho U^{\mu} U^{\nu} + p ({}^4g^{\mu\nu}-
 \epsilon U^{\mu} U^{\nu})\Big] (x)=\nonumber \\
 &=& \Big[ -\epsilon (\rho +p) U^{\mu}U^{\nu} + p \, {}^4g^{\mu\nu}\Big] (x)
 \nonumber \\
 &{\buildrel {dust} \over =}\,& -\epsilon \mu \Big[ n\, U^{\mu} U^{\nu}\Big] (x),
 \nonumber \\
 &&{}\nonumber \\
 n U^{\mu}&=& J^{\mu}=
-\epsilon^{\mu\nu\rho\sigma}\partial_{\nu}{\tilde \alpha}
^1\partial_{\rho}{\tilde \alpha}^2\partial_{\sigma}{\tilde \alpha}^3=n z^{\mu}
_{\check A} U^{\check A}=\nonumber \\
&=&z^{\mu}_{\check A} J^{\check A}= z^{\mu}_{\tau} J^{\tau}+z^{\mu}_{\check r}
J^{\check r}.
\label{IV3}
\eea

Therefore, in $\Sigma_{\tau}$-adapted coordinates on each
$\Sigma_{\tau}$ we get

\begin{eqnarray}
T^{\check A\check B}(\tau ,\vec \sigma )[\alpha ]&=&z^{\check
A}_{\mu}(\tau ,\vec \sigma )z^{\check B}_{\nu}(\tau ,\vec \sigma
)T^{\mu\nu}(x=z(\tau ,\vec \sigma ))[\tilde \alpha ]=\nonumber \\
&=&z^{\check A}_{\mu}(\tau ,\vec \sigma )z^{\check B}_{\nu}(\tau ,\vec
\sigma )T^{\mu\nu}(\tau ,\vec \sigma )[\alpha =\tilde \alpha \circ z] ,
\label{IV4}
\end{eqnarray}

 On Wigner hyperplanes, where Eqs.(\ref{II38}) hold and where we have

\begin{eqnarray}
z^{\mu}(\tau ,\vec \sigma )&=&x^{\mu}_s(\tau
)+\epsilon^{\mu}_u(u(p_s))\sigma^u,\nonumber \\
 &&{}\nonumber \\
{}^4\eta^{\mu\nu}&=&\epsilon \Big[ u^{\mu}(p_s)
u^{\nu}(p_s)-\sum_{r=1}^3 \epsilon^{\mu}_r(u(p_s))
\epsilon^{\nu}_r(u(p_s))\Big] ,\nonumber \\
 &&{}\nonumber \\
 N&=& {\dot x}_s\cdot u(p_s),\quad\quad N^r=\delta^{ru} {\dot x}_{s \mu}
 \epsilon^{\mu}_u(u(p_s)),\nonumber \\
 &&{}\nonumber \\
 Y^r&=& {{J^r+N^r J^{\tau}}\over N}={{J^r+\delta^{ru}
 {\dot x}_{s \mu}\epsilon^{\mu}_u(u(p_s)) J^{\tau}}\over {{\dot x}_s\cdot u(p_s)}},
 \nonumber \\
 n&=&{{\sqrt{\epsilon g_{AB}J^AJ^B}}\over N}=\sqrt{(J^{\tau})^2-\delta_{rs}Y^rY^s}
 \nonumber \\
 &{\buildrel {dust} \over =}\,& {{\mu J^{\tau}}\over { \sqrt{\mu^2+
 \delta_{uv}(T^{-1})^{ui}(T^{-1})^{vj} \Pi_i\Pi_j}}},
\label{IV5}
\end{eqnarray}

\noindent we get [$\check A=A$]

\begin{eqnarray}
&&T^{\mu\nu}[x^{\beta}_s(\tau
)+\epsilon^{\beta}_u(u(p_s))\sigma^u][\alpha ]=\nonumber \\
 &&{}\nonumber \\
 &=&[\delta^{\tau}_A {\dot x}^{\mu}_s(\tau )+\delta^r_A \epsilon^{\mu}_r(u(p_s))]
 [\delta^{\tau}_B {\dot x}^{\nu}_s(\tau )+\delta^s_B \epsilon^{\nu}_s(u(p_s))]
 T^{AB}(\tau ,\vec \sigma )=\nonumber \\
 &=& {\dot x}^{\mu}_s(\tau ) {\dot x}^{\nu}_s(\tau ) T^{\tau\tau}(\tau ,\vec \sigma )+
 \epsilon^{\mu}_r(u(p_s)) \epsilon^{\nu}_s(u(p_s)) T^{rs}(\tau ,\vec \sigma )+
 \nonumber \\
 &+&[{\dot x}^{\mu}_s(\tau ) \epsilon^{\nu}_r(u(p_s))+
 {\dot x}^{\nu}_s(\tau ) \epsilon^{\mu}_r(u(p_s))] T^{r\tau}(\tau ,\vec \sigma ),
 \nonumber \\
  &&{}\nonumber \\
  T^{\tau\tau}(\tau \vec \sigma )&=&
\Big[ -{{\epsilon \rho}\over { [{\dot x}_s\cdot u(p_s)]^2}} +\nonumber \\
 &+&n {{\partial \rho}\over
 {\partial n}}{|}_s\, ({{\epsilon}\over {[{\dot x}_s\cdot u(p_s)]^2}}-
 {{(J^{\tau})^2}\over
 {[{\dot x}_s\cdot u(p_s)]^2 [(J^{\tau})^2-\delta_{uv}Y^uY^v]}})\Big] (\tau ,\vec \sigma )
 \nonumber \\
  &{\buildrel {dust} \over =}\,&
  -\epsilon \Big[ {{J^{\tau} }\over {[{\dot x}_s\cdot u(p_s)]^2 }}
 \sqrt{\mu^2+\delta_{uv}(T^{-1})^{ui}
 (T^{-1})^{vj} \Pi_i\Pi_j}\Big] (\tau ,\vec \sigma ),\nonumber \\
  T^{r\tau}(\tau ,\vec \sigma )&=&T^{\tau r}(\tau ,\vec \sigma )=
  \Big[ -\epsilon \rho {{\delta^{ru}{\dot x}_{s\mu} \epsilon^{\mu}_u(u(p_s))}
  \over {[{\dot x}_s\cdot u(p_s)]^2}} -\nonumber \\
  &-&n {{\partial \rho}\over
 {\partial n}}{|}_s\, (\epsilon {{\delta^{ru}{\dot x}_{s\mu} \epsilon^{\mu}_u(u(p_s))}
  \over {[{\dot x}_s\cdot u(p_s)]^2}}+\nonumber \\
&+&{{J^{r} J^{\tau}}\over
 {[{\dot x}_s\cdot u(p_s)]^2 [(J^{\tau})^2-\delta_{uv}Y^uY^v]}})\Big] (\tau ,\vec \sigma )
  \nonumber \\
  &{\buildrel {dust} \over =}\,&
-\epsilon \Big[ {{J^{r} }\over {[{\dot x}_s\cdot u(p_s)]^2 }}
 \sqrt{\mu^2+\delta_{uv}(T^{-1})^{ui}
 (T^{-1})^{vj} \Pi_i\Pi_j}\Big] (\tau ,\vec \sigma ),\nonumber \\
  T^{rs}(\tau ,\vec \sigma )&=&
  \Big[ \epsilon \rho (\delta^{rs}-\delta^{ru}\delta^{sv}{{{\dot x}_{s\mu}
  \epsilon^{\mu}_u(u(p_s)) {\dot x}_{s\nu} \epsilon^{\nu}_v(u(p_s))}\over
  {[{\dot x}_s\cdot u(p_s)]^2}})-\nonumber \\
  &-&n {{\partial \rho}\over
 {\partial n}}{|}_s\, \Big( \epsilon (\delta^{rs}-\delta^{ru}\delta^{sv}{{{\dot x}_{s\mu}
  \epsilon^{\mu}_u(u(p_s)) {\dot x}_{s\nu} \epsilon^{\nu}_v(u(p_s))}\over
  {[{\dot x}_s\cdot u(p_s)]^2}})+\nonumber \\
 &+&{{J^{r} J^{s}}\over[{\dot x}_s\cdot u(p_s)]^2
 {[{\dot x}_s\cdot u(p_s)]^2 [(J^{\tau})^2-\delta_{uv}Y^uY^v]}}\Big)\Big] (\tau ,\vec \sigma )
  \nonumber \\
  &{\buildrel {dust} \over =}\,&
-\epsilon \Big[ {{J^{r} J^{s}}\over {[{\dot x}_s\cdot u(p_s)]^2\, J^{\tau}}}
 \sqrt{\mu^2+\delta_{uv}(T^{-1})^{ui}
 (T^{-1})^{vj} \Pi_i\Pi_j}\Big] (\tau ,\vec \sigma ).
\label{IV6}
\end{eqnarray}

Since we have

\begin{eqnarray}
{\dot x}_s^{\mu}(\tau )&=&-\lambda^{\mu}(\tau )=\epsilon
[u^{\mu}(p_s)u^{\nu}(p_s)-\epsilon^{\mu}_r(u(p_s))\epsilon^{\nu}_r(u(p_s))]
{\dot x}_{s \nu}(\tau )=\nonumber \\
 &=&\epsilon \Big[ - u^{\mu}(p_s) \lambda (\tau
)+\epsilon^{\mu}_r(u(p_s)) \lambda_r(\tau )\Big],\nonumber \\
 {\dot x}^2_s(\tau )&=& \lambda^2(\tau )-{\vec \lambda}(\tau ) > 0,\nonumber \\
 U^{\mu}_s(\tau )&=& {{ {\dot x}^{\mu}_s(\tau )}\over { \sqrt{{\dot
x}^2_s(\tau )} }}=\epsilon {{-\lambda (\tau
)u^{\mu}(p_s)+\lambda_r(\tau )\epsilon^{\mu}_r (u(p_s))}\over
{\sqrt{\lambda^2(\tau )-{\vec
\lambda}^2(\tau )} }},
\label{IV7}
\end{eqnarray}

\noindent the timelike worldline described by the origin of the Wigner
hyperplane is arbitrary (i.e. gauge dependent): $x^{\mu}_s(\tau )$ may
be any covariant non-canonical centroid. As already said the real
``external" center of mass is the canonical non-covariant ${\tilde
x}^{\mu}_s(T_s)= x^{\mu}
_s(T_s)-{1\over {\epsilon_s(p^o_s+\epsilon_s)}}\Big[
p_{s\nu}S^{\nu\mu}_s+\epsilon_s(S^{o\mu}_s+S^{o\nu}_s
{{p_{s\nu}p^{\mu}_s}\over {\epsilon_s^2}}) \Big]$: it describes a
decoupled point particle observer.

In the gauge $T_s-\tau \approx 0$, ${\vec X}_{sys}={\vec
q}_{sys}\approx 0$, implying $\lambda (\tau )=-1$, $\vec \lambda (\tau
)=0$ [$g_{\tau\tau}=\epsilon$, $N=1$,  $N^r=g_{\tau r}=0$], we get
${\dot x}^{\mu}_s(T_s)=u^{\mu}(p_s)$. Therefore, in this gauge, we
have the centroid

\beq
x^{\mu}_s(T_s)=x_s^{({\vec q}_{sys})\mu }(T_s)= x^{\mu}_s(0)+T_s
u^{\mu}(p_s),
\label{IV8}
\eeq

\noindent which carries the fluid ``internal" collective variable ${\vec
X}_{sys}={\vec q}_{sys}\approx 0$.

In this gauge we get the following form of the energy-momentum tensor
[$Y^r= J^r$]

\begin{eqnarray}
T^{\mu\nu}[x^{({\vec q}_{sys})
\beta}_s(T_s)+\epsilon^{\beta}_u(u(p_s))\sigma^u][\alpha ]&=&
u^{\mu}(p_s) u^{\nu}(p_s) T^{\tau\tau}(T_s,\vec \sigma )+\nonumber \\
&+&[u^{\mu}(p_s) \epsilon^{\nu}_r(u(p_s))+u^{\nu}(p_s)
\epsilon^{\mu}_r(u(p_s))] T^{r\tau}(T_s,\vec \sigma )+\nonumber \\
&+&\epsilon^{\mu}_r(u(p_s)) \epsilon^{\nu}_s(u(p_s)) T^{rs}(T_s,\vec
\sigma ),\nonumber \\
 &&{}\nonumber \\
 T^{\tau\tau}(T_s,\vec \sigma )&=&\Big[ -\epsilon \rho+\nonumber \\
  &+& n {{\partial \rho}\over {\partial n}}{|}_s\, (\epsilon -{{(J^{\tau})^2}\over
  {(J^{\tau})^2-\delta_{uv}Y^uY^v}})\Big](T_s,\vec \sigma )\nonumber \\
 &{\buildrel {dust} \over =}\,&-\epsilon \Big[ J^{\tau}
 \sqrt{\mu^2+\delta_{uv}(T^{-1})^{ui} (T^{-1})^{vj} \Pi_i\Pi_j}\Big]
 (T_s,\vec \sigma ),\nonumber \\
 T^{r\tau}(T_s,\vec \sigma )&=&\Big[ -n {{\partial \rho}\over {\partial n}}{|}s\,
 {{J^r J^{\tau}}\over {(J^{\tau})^2-\delta_{uv}Y^u Y^v}}\Big] (T_s,\vec \sigma )\nonumber \\
 &{\buildrel {dust} \over =}\,&-\epsilon \Big[ J^r
 \sqrt{\mu^2+\delta_{uv}(T^{-1})^{ui} (T^{-1})^{vj} \Pi_i\Pi_j}\Big]
 (T_s,\vec \sigma ),\nonumber \\
 T^{rs}(T_s,\vec \sigma )&=&\Big[ \epsilon \rho \delta^{rs}-n
 {{\partial \rho}\over {\partial n}}{|}_s\, \Big( \epsilon \delta^{rs}+\nonumber \\
 &+&{{J^r J^s}\over {(J^{\tau})^2-\delta_{uv} Y^u Y^v}}\Big) \Big] (T_s,\vec \sigma )
 \nonumber \\
 &{\buildrel {dust} \over =}\,&-\epsilon \Big[ {{J^r J^s}\over {J^{\tau} }}
 \sqrt{\mu^2+\delta_{uv}(T^{-1})^{ui} (T^{-1})^{vj} \Pi_i\Pi_j}\Big]
 (T_s,\vec \sigma ),\nonumber \\
 &&{}\nonumber \\
 &&with\,\, total\,\, 4-momentum\nonumber \\
 &&{}\nonumber \\
  P^{\mu}_T[\alpha ]&=&\int
d^3\sigma T^{\mu\nu}[x^{\mu}_s(T_s)+\epsilon^{\mu}
_u(u(p_s))\sigma^u][\alpha ]u_{\nu}(p_s)=\nonumber \\
&=&-P^{\tau} u^{\mu}(p_s)-P^r\epsilon^{\mu}_r(u(p_s))\approx -P^{\tau}
u^{\mu}(p_s)=\nonumber \\
 &=&-M_{sys} u^{\mu}(p_s)  \approx -p^{\mu}_s,\nonumber \\
 &&{}\nonumber \\
 &&and\,\, total\,\, mass\nonumber \\
 &&{}\nonumber \\
 M[\alpha ]&=&P^{\mu}_T[\alpha ] u_{\mu}(p_s) =-P^{\tau}=-M_{sys}.
\label{IV9}
\eea

\noindent The stress tensor of the perfect fluid configuration on the Wigner
hyperplanes is $T^{rs}(T_s,\vec \sigma )$.

We can rewrite the energy-momentum tensor in such a way  that it
acquires a form reminiscent of the energy-momentum tensor of an ideal
relativistic fluid as seen from a local observer at rest (see the
Eckart decomposition in Appendix B):

\bea
T^{\mu\nu}[\tilde \alpha ]&=& \Big[ \rho [\alpha ,\Pi ]\,\,
u^{\mu}(p_s)u^{\nu}(p_s)+\nonumber \\
 &+&{\cal P}[\alpha ,\Pi ]\,\, [\eta
^{\mu\nu}-u^{\mu}(p_s)u^{\nu}(p_s)]+\nonumber \\
&+&u^{\mu}(p_s)q^{\nu}[\alpha ,\Pi ]+ u^{\nu}(p_s)q^{\mu}[\alpha ,\Pi
]+\nonumber \\
 &+&T^{rs}_{an}[\alpha ,\Pi ] \,\,
\epsilon^{\mu}_r(u(p_s))\epsilon^{\nu}
_s(u(p_s))\Big] (T_s,\vec \sigma ),\nonumber \\
 &&{}\nonumber \\
 \rho [\alpha ,\Pi ]&=& T^{\tau\tau},\nonumber \\
 {\cal P}[\alpha ,\Pi ]&=& {1\over 3} \sum_u T^{uu},\nonumber \\
 q^{\mu}[\alpha ,\Pi ]&=&\epsilon^{\mu}_r(u(p_s)) T^{r\tau},\nonumber \\
 T^{rs}_{an}[\alpha ,\Pi
]&=&T^{rs}-{1\over 3}\delta^{rs} \sum_u T^{uu},\quad\quad
 \delta_{uv}T^{uv}_{an}[\alpha ,\Pi ]=0,
\label{IV10}
\eea

\noindent where

i) the constant normal $u^{\mu}(p_s)$ to the Wigner hyperplanes
replaces the hydrodynamic velocity field of the fluid;

ii) $\rho [\alpha ,\Pi ](T_s,\vec
\sigma )$ is the energy density;

iii) ${\cal P}[\alpha ,\Pi ](T_s,\vec \sigma )$ is the analogue of the
pressure (sum of the thermodynamical pressure and of the
non-equilibrium bulk stress or viscous pressure);

iv) $q^{\mu}[\alpha ,\Pi ](T_s,
\vec \sigma )$ is the analogue of the heat flow;

v) $T^{rs}_{an}[\alpha ,\Pi ](T_s,\vec \sigma )$ is the shear (or
anisotropic) stress tensor.

${}$

We can now study the manifestly Lorentz covariant Dixon multipoles
\cite{dixon} for the perfect fluid configurationon the Wigner
hyperplanes in the gauge $\lambda (\tau )=-1$, $\vec \lambda (\tau
)=0$ [so that ${\dot x}_s^{\mu}(T_s )=u^{\mu}(p_s)$, ${\ddot
x}^{\mu}_s(T_s)=0$, $x^{\mu}_s(T_s)=x_s^{({\vec
q}_{sys})\mu}(T_s)=x_s^{\mu}(0)+u^{\mu}(p_s) T_s$] with respect to the
origin an arbitrary timelike worldline $w^{\mu}(T_s)=z^{\mu}(T_s,\vec
\eta (T_s))=x_s^{({\vec q}_{sys}) \mu}(T_s)+\epsilon^{\mu}_r(u(p_s))\eta^r(T_s)$.
Since we have  $z^{\mu}(T_s,\vec \sigma )= x_s^{({\vec
q}_{sys})\mu}(T_s)+\epsilon^{\mu}_r(u(p_s)) \sigma^r =w^{\mu}(T_s)+
\epsilon^{\mu}_r(u(p_s)) [\sigma^r-\eta^r(T_s)]\, {\buildrel {def} \over =}\,
w^{\mu}(T_s)+ \delta z^{\mu}(T_s,\vec \sigma )$ [for $\vec \eta
(T_s)=0$ we get the multipoles with respect to the origin of
coordinates], we obtain [ $(\mu_1..\mu_n)$ means symmetrization, while
$[\mu_1..\mu_n]$ means antisymmetrization;
$t_T^{\mu_1..\mu_n\mu\nu}(T_s,\vec \eta
=0)=t_T^{\mu_1..\mu_n\mu\nu}(T_s)$]

\begin{eqnarray}
t_T^{\mu_1...\mu_n\mu\nu}(T_s,\vec
\eta )&=&t_T^{(\mu_1...\mu_n)(\mu\nu)}(T_s,\vec \eta )=
\nonumber \\
&=&\int d^3\sigma  \, \delta z^{\mu_1}(T_s,\vec \sigma )...\delta
z^{\mu_n}(T_s,\vec
\sigma )\, T^{\mu\nu}[x^{({\vec q}_{sys})\beta}_s(T_s)
+\epsilon^{\beta}_u(u(p_s))\sigma^u][\alpha ]
=\nonumber \\
&=&\epsilon^{\mu_1}_{r_1}(u(p_s))...\epsilon^{\mu_n}_{r_n}(u(p_s))
\epsilon^{\mu}_A(u(p_s))\epsilon^{\nu}_B(u(p_s))\, I_T^{r_1..r_nAB}(T_s,\vec \eta )=\nonumber \\
&=&\epsilon^{\mu_1}_{r_1}(u(p_s))...\epsilon^{\mu_n}_{r_n}(u(p_s))
\Big[ u^{\mu}(p_s) u^{\nu}(p_s) I_T^{r_1...r_n\tau\tau}(T_s,\vec \eta )+\nonumber \\
&+&\epsilon^{\mu}_r(u(p_s))\epsilon^{\nu}_s(u(p_s))
I_T^{r_1...r_nrs}(T_s,\vec \eta )+\nonumber \\
&+&[u^{\mu}(p_s)\epsilon^{\nu}_r(u(p_s))+u^{\nu}(p_s)\epsilon^{\mu}_r(u(p_s))]
I_T^{r_1...r_nr\tau}(T_s,\vec \eta ) \Big] ,\nonumber \\
 I_T^{r_1..r_nAB}(T_s,\vec \eta )&=& \int d^3\sigma \,
 [\sigma^{r_1}-\eta^{r_1}(T_s)]...[\sigma^{r_n}-\eta^{r_n}(T_s)]
 T^{AB}(T_s,\vec \sigma )[\alpha ],\nonumber \\
 &&{}\nonumber \\
u_{\mu_1}(p_s)&& t_T^{\mu_1...\mu_n\mu\nu}(T_s,\vec \eta )=0,\nonumber
\\
 &&{}\nonumber \\
 For \vec \eta =0 && n=0\, (monopole)\quad I^{\tau\tau}_T(T_s)=P^{\tau},\quad\quad
 I_T^{r\tau}(T_s)=P^r,\nonumber \\
 &&{}\nonumber \\
t_T^{\mu_1...\mu_n\mu}{}_{\mu}(T_s)&=&\int d^3\sigma \delta
x^{\mu_1}_s(\vec
\sigma )...\delta x^{\mu_n}_s(\vec \sigma ) T^{\mu}{}_{\mu}
[x^{({\vec
q}_{sys})\beta}_s(T_s)+\epsilon^{\beta}_u(u(p_s))\sigma^u][\alpha ]
=\nonumber \\
&{\buildrel {def} \over
=}&\epsilon^{\mu_1}_{r_1}(u(p_s))...\epsilon^{\mu_n}
_{r_n}(u(p_s))I_T^{r_1...r_nA}{}_A(T_s)\nonumber \\
&&{}\nonumber \\
 I_T^{r_1r_2A}{}_A(T_s)&=&{\check I}_T^{r_1r_2A}{}_A(T_s)-{1\over 3}\delta^{r_1r_2}\delta
_{uv}I_T^{uvA}{}_A(T_s)=i^{r_1r_2}_T(T_s)-{1\over 2}\delta^{r_1r_2}\delta_{uv}i_T
^{uv}(T_s),\nonumber \\
{\check I}_T^{r_1r_2A}{}_A(T_s)&=&i_T^{r_1r_2}(T_s)-{1\over
3}\delta^{r_1r_2}\delta
_{uv}i_T^{uv}(T_s),\quad\quad \delta_{uv}{\check I}_T^{uvA}{}_A(T_s)=0,\nonumber \\
i^{r_1r_2}_T(T_s)&=&I_T^{r_1r_2A}{}_A(T_s)-\delta^{r_1r_2}\delta_{uv}I_T^{uvA}{}_A(T_s),
\nonumber \\
 &&{}\nonumber \\
 {\tilde t}_T^{\mu_1...\mu_n}(T_s)&=& t_T^{\mu_1...\mu_n\mu\nu}(T_s)
u_{\mu}(p_s)u_{\nu}(p_s)=\nonumber \\
&=&\epsilon^{\mu_1}_{r_1}(u(p_s))...\epsilon^{\mu_n}_{r_n}(u(p_s))
I_T^{r_1..r_n\tau\tau}(T_s),\nonumber \\
 &&{}\nonumber \\
 {\tilde t}_T^{\mu_1}(T_s)
 &=&\epsilon^{\mu_1}_{r_1}(u(p_s)) I_T^{r_1\tau\tau}(T_s)=-P^{\tau}
\epsilon^{\mu_1}_{r_1}(u(p_s)) r^{r_1}_{sys},\nonumber \\
 &&{}\nonumber \\
{\tilde t}_T^{\mu_1\mu_2}(T_s)&=&
\epsilon^{\mu_1}_{r_1}(u(p_s))\epsilon^{\mu_2}_{r_2}(u(p
_s))I_T^{r_1r_2\tau\tau}(T_s),\nonumber \\
 &&{}\nonumber \\
I^{r_1r_2\tau\tau}_T(T_s)&=&{\hat I}^{r_1r_2\tau\tau}_T(T_s)-{1\over
3}\delta^{r_1r_2}
\delta_{uv}I^{uv\tau\tau}_T(T_s)={\tilde i}^{r_1r_2}_T(T_s)-{1\over 2}\delta
^{r_1r_2}\delta_{uv}{\tilde i}^{uv}_T(T_s),\nonumber \\
{\hat I}^{r_1r_2\tau\tau}_T(T_s)&=&{\tilde i}^{r_1r_2}_T(T_s)-{1\over
3}\delta^{r_1r_2}
\delta_{uv}{\tilde i}^{uv}_T(T_s),\quad\quad \delta_{uv}{\hat I}^{uv\tau\tau}_T(T_s)=0,
\nonumber \\
{\tilde
i}_T^{r_1r_2}(T_s)&=&I_T^{r_1r_2\tau\tau}(T_s)-\delta^{r_1r_2}\delta
_{uv}I_T^{uv\tau\tau}(T_s).
\label{IV11}
\end{eqnarray}

The Wigner covariant multipoles $I_T^{r_1..r_n\tau\tau}(T_s)$,
$I_T^{r_1..r_nrs}(T_s)$, $I_T^{r_1..r_nr\tau}(T_s)$ are the mass,
stress and momentum multipoles respectively.

The quantities ${\check I}_T^{r_1r_2A}{}_A(T_s)$ and $i^{r_1r_2}_T
(T_s)$ are the traceless quadrupole moment and the inertia tensor
defined by Thorne in Ref.\cite{thorne}.

The quantities $I_T^{r_1r_2\tau\tau}(T_s)$ and ${\tilde i}_T^{r_1r_2}
(T_s)$ are Dixon's definitions of quadrupole moment and of tensor of
inertia respectively.

Moreover, Dixon's definition of ``center of mass" of an extended
object is ${\tilde t}^{\mu_1}_T(T_s)=0$ or
$I_T^{r\tau\tau}(T_s)=-P^{\tau} r^r_{sys}=0$: therefore the quantity
${\vec r}_{sys} $ defined in the previous equation is a non-canonical
[$\{ r^r_{sys},r^s_{sys}
 \} =S^{rs}_{sys}$] candidate for the ``internal" center of mass of the
field configuration: its vanishing is a gauge fixing for ${\vec
P}\approx 0$ and implies $x^{\mu}_s(T_s)=x^{({\vec
q}_{sys})\mu}_s(T_s)=x^{\mu}_s(0)+u^{\mu}(p_s) T_s$. As we have seen
in the previous Section ${\vec r}_{sys}$ is the ``internal" M\o ller
3-center of energy and we have ${\vec r}_{sys}\approx {\vec
q}_{sys}\approx {\vec y}_{sys}$.

When $I^{r\tau\tau}_T(T_s)=0$, the equations
$0={{dI^{r\tau\tau}_T(T_s)}\over {dT_s}}=-P^{\tau} {{dr^r_{sys}}\over
{dT_s}}=\, {\buildrel \circ \over =}\, -P^r$ implies the correct
momentum-velocity relation ${{{\vec P}}\over {P^{\tau}}}\, {\buildrel
\circ \over =}\, {{d{\vec r}_{sys}}\over {dT_s}} \approx 0$.

Then there are the related Dixon multipoles

\begin{eqnarray}
p^{\mu_1...\mu_n\mu}_T(T_s)&=&t_T^{\mu_1...\mu_n\mu\nu}(T_s)
u_{\nu}(p_s)= p_T^{(\mu_1...\mu_n)\mu}(T_s)=\nonumber \\
 &=&\epsilon^{\mu_1}_{r_1}(u(p_s))...\epsilon^{\mu_n}_{r_n}(u(p_s))
\epsilon^{\mu}_A(u(p_s)) I_T^{r_1..r_nA\tau}(T_s),\nonumber \\
 &&{}\nonumber \\
u_{\mu_1}(p_s)&& p_T^{\mu_1...\mu_n\mu}(T_s)=0,\nonumber \\
 &&{}\nonumber \\
 n=0&&\Rightarrow p^{\mu}_T(T_s)=P^{\mu}_T[\alpha ]=-\epsilon \epsilon^{\mu}_A(u(p_s))
 P^A\approx -\epsilon p^{\mu}_s,\nonumber \\
 &&{}\nonumber \\
 p_T^{\mu_1..\mu_n\mu}(T_s)u_{\mu}(p_s)&=&{\tilde
t}_T^{\mu_1...\mu_n}(T_s)
=\epsilon^{\mu_1}_{r_1}(u(p_s))...\epsilon^{\mu_n}_{r_n}(u(p_s))
I_T^{r_1..r_n\tau\tau}(T_s).
\label{IV12}
\end{eqnarray}

The spin dipole is defined as

\begin{eqnarray}
S^{\mu\nu}_T(T_s)[\alpha ]&=&2 p_T^{[\mu\nu ]}(T_s)=2
\epsilon^{[\mu}_r(u(p_s))\epsilon^{\nu ]}_A(u(p_s)) I_T^{rA\tau}(T_s)=\nonumber \\
&=&S^{\mu\nu}_s=\nonumber \\
 &=&\epsilon^{\mu}_r(u(p_s))\epsilon^{\nu}_s(u(p_s))S^{rs}_{sys}+
[\epsilon^{\mu}_r(u(p_s)) u^{\nu}(p_s)-\epsilon^{\nu}_r(u(p_s))
u^{\mu} (p_s)]S^{\tau r}_{sys},\nonumber \\
 &&{}\nonumber \\
 &&u_{\mu}(p_s) S^{\mu\nu}_T(T_s)[\alpha ]=-\epsilon^{\nu}_r(u(p_s))
S^{\tau r}_{sys}=-{\tilde
t}^{\nu}_T(T_s)=P^{\tau}\epsilon^{\nu}_r(u(p_s)) r^r_{sys},
\label{IV13}
\end{eqnarray}

\noindent with $u_{\mu}(p_s)S^{\mu\nu}_T(T_s)[\alpha ]=0$
when ${\tilde t}_T^{\mu_1}(T_s)=0$ and this condition can be taken as
a definition of center of mass equivalent to Dixon's one. When this
condition holds, the barycentric spin dipole is
$S^{\mu\nu}_T(T_s)[\alpha ]=2 \epsilon^{[\mu}_r(u(p_s)) \epsilon^{\nu
]}_s(u(p_s)) I_T^{rs\tau}(T_s)$, so that
$I^{[rs]\tau}_T(T_s)=\epsilon^r_{\mu}(u(p_s)) \epsilon^s_{\nu}(u(p_s))
S^{\mu\nu}_T(T_s)[\alpha ]$.

As shown in Ref.\cite{dixon}, if the fluid configuration has a compact
support W on the Wigner hyperplanes $\Sigma_{W\tau}$ and if $f(x)$ is
a $C^{\infty}$ complex-valued scalar function on Minkowski spacetime
with compact support [so that its Fourier transform $\tilde f(k)=\int
d^4x f(x) e^{ik\cdot x}$ is a slowly increasing entire analytic
function on Minkowski spacetime ($|(x^o+iy^o)
^{q_o}...(x^3+iy^3)^{q_3}f(x^{\mu}+iy^{\mu})| < C_{q_o...q_3} e^{a_o|y^o|+...+
a_3|y^3|}$, $a_{\mu} > 0$, $q_{\mu}$ positive integers for every $\mu$ and
$C_{q_o...q_3} > 0$), whose inverse is $f(x)=\int {{d^4k}\over {(2\pi )^4}}
\tilde f(k)e^{-ik\cdot x}$], we have [we consider $\vec \eta =0$ with
$\delta z^{\mu}=\delta x_s^{\mu}$]

\begin{eqnarray}
< T^{\mu\nu},f >&=& \int d^4x T^{\mu\nu}(x) f(x)=\nonumber \\
&=&\int dT_s\int d^3\sigma f(x_s+\delta x_s)T^{\mu\nu}[x_s(T_s)+\delta x_s(\vec
\sigma )][\alpha ]=\nonumber \\
&=&\int dT_s \int d^3\sigma \int {{d^4k}\over {(2\pi )^4}} \tilde f(k)
e^{-ik\cdot [x_s(T_s)+\delta x_s(\vec \sigma )]}T^{\mu\nu}
[x_s(T_s)+\delta x_s(\vec \sigma )][\alpha ]=\nonumber \\ &=&\int
dT_s\int {{d^4k}\over {(2\pi )^4}} \tilde f(k) e^{-ik\cdot x_s(T_s)}
\int d^3\sigma T^{\mu\nu}[x_s(T_s)+\delta x_s(\vec \sigma )][\alpha ]\nonumber \\
&&\sum_{n=0}^{\infty} {{(-i)^n}\over {n!}} [k_{\mu}\epsilon^{\mu}_u(u(p_s))
\sigma^u]^n=\nonumber \\
&=&\int dT_s\int {{d^4k}\over {(2\pi )^4}} \tilde f(k) e^{-ik\cdot
x_s(T_s)}
\sum_{n=0}^{\infty} {{(-i)^n}\over {n!}} k_{\mu_1}...k_{\mu_n} t_T^{\mu_1...
\mu_n\mu\nu}(T_s),
\label{IV14}
\end{eqnarray}

\noindent and, but only for $f(x)$ analytic in W \cite{dixon}, we get

\begin{eqnarray}
< T^{\mu\nu},f >&=& \int dT_s \sum_{n=0}^{\infty}{1\over {n!}} t_T^{\mu_1...
\mu_n\mu\nu}(T_s) {{\partial^nf(x)}\over {\partial x^{\mu_1}...\partial x^{\mu
_n}}}{|}_{x=x_s(T_s)},\nonumber \\
&&\Downarrow \nonumber \\ T^{\mu\nu}(x)[\alpha ]&=&
\sum_{n=0}^{\infty}{{(-1)^n}\over {n!}} {{\partial^n}\over {\partial
x^{\mu_1}...\partial x^{\mu_n}}} \int dT_s
\delta^4(x-x_s(T_s)) t_T^{\mu_1...\mu_n\mu\nu}(T_s)=\nonumber \\
 &=& \epsilon^{\mu}_A(u(p_s)) \epsilon^{\nu}_B(u(p_s)) T^{AB}(T_s,\vec \sigma )[\alpha ]=
 \nonumber \\
 &=&\epsilon^{\mu}_A(u(p_s)) \epsilon^{\nu}_B(u(p_s)) \sum_{n=0}^{\infty}
 {{(-)^n}\over {n!}} I_T^{r_1...r_nAB}(T_s,\vec \eta ){{\partial^n\delta^3(\vec \sigma -
 \vec \eta (T_s))}\over {\partial \sigma^{r_1}...\partial \sigma^{r_n}}}
 {|}_{\vec \eta =0}.
\label{IV15}
\end{eqnarray}

For a non analytic $f(x)$ we have

\begin{eqnarray}
< T^{\mu\nu},f > &=& \int dT_s \sum^N_{n=0} {1\over {n!}} t_T^{\mu_1...\mu_n
\mu\nu}(T_s) {{\partial^nf(x)}\over {\partial x^{\mu_1}...\partial x^{\mu
_n}}}{|}_{x=x_s(T_s)}+\nonumber \\
&+&\int dT_s \int {{d^4k}\over {(2\pi )^4}} \tilde f(k) e^{-ik\cdot x_s(T_s)}
\sum_{n=N+1}^{\infty} {{(-i)^n}\over {n!}} k_{\mu_1}...k_{\mu_n} t_T^{\mu_1...
\mu_n\mu\nu}(T_s),
\label{IV16}
\end{eqnarray}

\noindent and, as shown in Ref.\cite{dixon}, from the knowledge of the moments
$t_T^{\mu_1...\mu_n\mu}(T_s)$ for all $n > N$ we can get $T^{\mu\nu}(x)$ and,
thus, all the moments with $n\leq N$.

In Appendix D  other types of Dixon's multipoles are analyzed. From
this study it turns out that the multipolar expansion(\ref{IV15}) may
be rearranged with the help of the Hamilton equations implying
$\partial_{\mu}T^{\mu\nu}\, {\buildrel \circ \over =}\, 0$, so that
for analytic fluid configurations from Eq.(\ref{d5}) we get

\begin{eqnarray}
T^{\mu\nu}(x)[\alpha ]\, &{\buildrel \circ \over =}\,&
-\epsilon u^{(\mu}(p_s)
\epsilon^{\nu )}_A(u(p_s)) \int dT_s\, \delta^4(x-x_s(T_s))\,\,
P^A+\nonumber \\
 &+&{1\over 2} {{\partial}\over {\partial x^{\rho}}} \int dT_s\,
 \delta^4(x-x_s(T_s))\, S_T^{\rho (\mu}(T_s)[\alpha ]\,\, u^{\nu )}(p_s)+\nonumber \\
 &+&\sum_{n=2}^{\infty}{{(-1)^n}\over {n!}} {{\partial^n}\over
 {\partial x^{\mu_1}...\partial x^{\mu_n}}} \int dT_s\, \delta^4(x-x_s(T_s))\,\,
 {\cal I}_T^{\mu_1..\mu_n\mu\nu}(T_s),\nonumber \\
 &&{}\nonumber \\
 T^{\mu\nu}(w+\delta z)&=& -\epsilon u^{(\mu}(p_s)\, \epsilon^{\nu )}_A(u(p_s))
 P^A \delta^3(\vec \sigma -\vec \eta (T_s))+ \nonumber \\
 &+&{1\over 2} S^{\rho (\mu}_T(T_s,\vec \eta )[\alpha ] u^{\nu )}(p_s)
 \epsilon^r_{\rho}(u(p_s)) {{\partial \delta^3(\vec \sigma -\vec \eta (t_s))}\over
 {\partial \sigma^r}}+\nonumber \\
 &+&\sum_{n=2}^{\infty} {{(-)^n}\over {n!}} \Big[ {{n+3}\over {n+1}} u^{\mu}(p_s)u^{\nu}(p_s)
 I_T^{r_1...r_n\tau\tau}(T_s,\vec \eta )+\nonumber \\
 &+&{1\over n} [u^{\mu}(p_s)\epsilon^{\nu}_r(u(p_s))+u^{\nu}(p_s)\epsilon^{\mu}_r(u(p_s))]
 I_T^{r_1...r_nr\tau}(T_s,\vec \eta )+\nonumber \\
 &+&\epsilon^{\mu}_{s_1}(u(p_s))\epsilon^{\nu}_{s_2}(u(p_s))[I_T^{r_1...r_ns_1s_2}(T_s,\vec \eta )
 -\nonumber \\
 &-&{{n+1}\over n} (I_T^{(r_1...r_ns_1)s_2}(T_s,\vec \eta ) +
 I_T^{(r_1...r_ns_2)s_1}(T_s,\vec \eta )) +\nonumber \\
 &+& I_T^{(r_1...r_ns_1s_2)}(T_s,\vec \eta ) ] \Big] ,
\label{IV17}
\end{eqnarray}

\noindent where for $n\geq 2$ and $\vec \eta =0$
$\quad {\cal I}_T^{\mu_1..\mu_n\mu\nu}(T_s)
={{4(n-1)}\over {n+1}} J_T^{(\mu_1..\mu_{n-1} | \mu | \mu_n)\nu}(T_s)$,
with the quantities $J_T^{\mu_1..\mu_n\mu\nu\rho\sigma}(T_s)$ being
the Dixon $2^{2+n}$-pole inertial moment tensors given in
Eqs.(\ref{d7}) [the quadrupole and related inertia tensor are
proportional to $I_T^{r_1r_2\tau\tau}(T_s)$].

The equations $\partial_{\mu}T^{\mu\nu}\, {\buildrel \circ \over =}\,
0$ imply the Papapetrou-Dixon-Souriau equations for the `pole-dipole'
system $P_T^{\mu}(T_s)$ and $S^{\mu\nu}_T(T_s)[\alpha ]$ [see
Eqs.(\ref{d1}) and (\ref{d4}); here $\vec \eta =0$]

\begin{eqnarray}
{{d P^{\mu}_T(T_s)}\over {dT_s}}\, &{\buildrel \circ \over =}\,&
0,\nonumber \\
 {{d S^{\mu\nu}_T(T_s)[\alpha ]}\over {dT_s}}\,
&{\buildrel \circ \over =}\,& 2 P^{[\mu}_T(T_s)\,\, u^{\nu
]}(p_s)=-2\epsilon P^r \epsilon^{[\mu}_r(u(p_s))\,\, u^{\nu ]}(p_s)
\approx 0.
\label{IV18}
\end{eqnarray}

The Cartesian Dixon's multipoles could be re-expressed in terms of
either spherical or STF (symmetric tracefree) multipoles \cite{thorne}
[both kinds of tensors are associated with the irreducible
representations of the rotation group: one such multipole of order $l$
has exactly $2l+1$ independent components].

\vfill\eject

\section{Isentropic and Non-Isentropic Fluids.}

Let us now consider isentropic ($s=const.$) perfect fluids. For them
we have from Eqs.(\ref{I9}), (\ref{I10}) and (\ref{II1}) [in general
$\mu$ is not the chemical potential but only a parameter]

\bea
n&=&{{|J|}\over {N\sqrt{\gamma}}}={X\over {\sqrt{\gamma}}},
\nonumber \\
 &&{}\nonumber \\
 \rho &=& \rho (n)=\rho ({{|J|}\over {N\sqrt{\gamma}}})=\mu f({X\over
{\sqrt{\gamma}}}),\nonumber \\
 L&=&-\mu N \sqrt{\gamma} f({X\over {\sqrt{\gamma}}}).
\label{V1}
\eea

Some possible equations of state for such fluids are (see also
Appendix A):\hfill\break
\hfill\break
1) $p=0$, dust: this implies

\bea
\rho (n)&=&\mu n =\mu {X\over {\sqrt{\gamma}}},\quad\quad i.e. \nonumber \\
 &&{}\nonumber \\
f({X\over {\sqrt{\gamma}}})&=&{X\over {\sqrt{\gamma}}},
\quad\quad {{\partial f({X\over {\sqrt{\gamma}}})}\over {\partial
X}}={1\over {\sqrt{\gamma}}}.
\label{V2}
\eea

2) $p=k \rho (n)=n {{\partial \rho (n)}\over {\partial n}}-\rho (n)$
($k\not= -1$ because otherwise $\rho = const.$, $\mu =0$, $f=-Ts$).
For $k={1\over 3}$ one has the photon gas. The previous differential
equation for $\rho (n)$ implies

\bea
\rho (n) &=& (an)^{k+1}=\mu n^{k+1}=\mu ({X\over {\sqrt{\gamma}}})^{k+1},
\quad\quad (\mu =a^{k+1}),\quad i.e.\nonumber \\
 &&{}\nonumber \\
f({X\over {\sqrt{\gamma}}})&=&({X\over
{\sqrt{\gamma}}})^{k+1},\quad\quad {{\partial f({X\over
{\sqrt{\gamma}}})}\over {\partial X}}= {{k+1}\over
{\sqrt{\gamma}}}({X\over {\sqrt{\gamma}}})^k,
\label{V3}
\eea

\noindent
[for $k\rightarrow 0$ we recover 1)].

More in general one can have $k=k(s)$: this is a non-isentropic
perfect fluid with $\rho
=\rho (n,s)$.

 3) $p=k \rho^{\gamma}(n)=n {{\partial \rho
(n)}\over {\partial n}}-\rho (n)$ ($\gamma \not= 1$) \cite{ryan}. It
is an isentropic polytropic perfect fluid ($\gamma =1+{1\over n}$).
The differential equation for $\rho (n)$ implies [$a$ is an
integration constant; the chemical potential is $\mu
={{\partial
\rho}\over {\partial n}}{|}_s$]

\bea
\rho (n)&=&{{a n}\over {[1-k(a n)^{\gamma -1}]^{ {1\over {\gamma -1}} } }}=
{{a n}\over {[1-k(a n)^{{1\over n}}]^n }},\quad\quad i.e.
\nonumber \\
 &&{}\nonumber \\
  f({X\over {\sqrt{\gamma}}})&=&{{X\over
{\sqrt{\gamma}}}\over {[1-k(a {X\over {\sqrt{\gamma}}})^{\gamma
-1}]^{ {1\over {\gamma
-1}} } }}= {{X\over {\sqrt{\gamma}}}\over {[1-k(a {X\over
{\sqrt{\gamma}}})
^{{1\over n}}]^n }},\nonumber \\
 &&{}\nonumber \\
{{\partial f({X\over {\sqrt{\gamma}}})}\over {\partial X}}&=&{1\over
{\sqrt{\gamma}}}[1-k(a {X\over {\sqrt{\gamma}}})^{\gamma
-1}]^{-{{\gamma}\over {\gamma -1}} }={1\over {\sqrt{\gamma}}} [1-k(a
{X\over {\sqrt{\gamma}}})^{{1\over n}}]^{-(n+1)}.
\label{V4}
\eea

Instead in Ref.\cite{damour,an} a polytropic perfect fluid is defined
by the equation of state [see the last of Eqs.(\ref{b4}); $m$ is a
mass]

\beq
\rho (n,ns)=m n +{{k(s)}\over {\gamma -1}} (m n)^{\gamma},
\label{V5}
\eeq

\noindent
and has pressure $p=k(s) (m n)^{\gamma}=(\gamma -1)(\rho -m n)$ and
chemical potential or specific enthalpy $\mu^{'} =mc^2+mk(s)
{{\gamma}\over {\gamma -1}} (m n)^{\gamma -1}$ of Eq.(\ref{b3}).

4) $p=p(\rho )$, barotropic perfect fluid. In the isentropic case  one
gets $\rho =\rho (n)$ by solving $p(\rho (n))=n {{\partial \rho
(n)}\over {\partial n}}-\rho (n)$.

5)Relativistic ideal (Boltzmann) gas\cite{israel} (this is a
non-isentropic case):

\beq
p=n k_B T\quad and\quad \rho =m c^2 n \Gamma (\beta )-p,\quad\quad \mu
={{\rho +p}\over n}=mc^2 \Gamma (\beta ),
\label{V6}
\eeq

\noindent
with $\beta ={{m c^2}\over {k_BT}}$, $\Gamma (\beta )={{K_3(\beta
)}\over {K_2(\beta )}}$ ($K_i$ are modified Bessel functions). One
gets the equation of state $\rho =\rho (n,s)$ by solving the
differential equation

\beq
{{\partial \rho}\over {\partial n}}{|}_s =m c^2 \Gamma ({{m c^2
n}\over {n{{\partial \rho}\over {\partial n}}{|}_s-\rho}} ).
\label{V7}
\eeq

5a) Ultrarelativistic case $\beta << 1$ ($mc^2 << k_BT$): since we
have $\Gamma (\beta )\approx {4\over {\beta}}+{{\beta}\over
2}+O(\beta^3)$, we get $\rho = 3n k_B T + {{m^2c^4 n}\over
{2k_BT}}+O(k_BT \beta^4)$, namely $p\approx {1\over 3}\rho$ and $\rho
\approx \rho (n)=\mu n^{4/3}$.

5b) Non-relativistic case $\beta >> 1$ ($k_BT << mc^2$):
since we have $\Gamma (\beta )\approx 1
+{5\over {2\beta}}+O(\beta^{-2})$, we get $\rho \approx m c^2 n+{3\over 2}p$,
so that we have to solve the differential equation $3n {{\partial \rho}\over
{\partial n}}{|}_s-5\rho+2m c^2 n\approx 0$. Its solution is $\rho (n,s)
\approx m c^2 n+k(s) n^{5/3}$. To find $k(s)$ let us use the definition of
temperature: $p=n k_BT=k_B {{\partial \rho}\over {\partial s}}{|}_n=k_B
n^{5/3} {{\partial k(s)}\over {\partial s}}=n{{\partial \rho}\over {\partial
n}}{|}_s-\rho ={2\over 3}n^{5/3}k(s)$. This leads to the equation $d ln\, k(s)
={{2 ds}\over {3k_B}}$, whose solution is $k(s)=he^{{{2s}\over {3k_B}} }=
e^{{{2(s-s_o)}\over {3k_B}} }$ with $h=e^{-{{2s_o}\over {3k_B}}
}=const.$. Therefore, in this case we get [it is a polytropic like in
Eq.(\ref{V5}) with $\gamma =5/3$ and $k(s)={2\over 3} m^{-5/3} e^{
{{2(s-s_o)}\over {3k_B}} }$]

\beq
\rho(n,s)\approx m c^2 n +  n^{{5\over 3}} e^{ {{2(s-s_o)}\over
{3k_B}} }\quad and\quad T={1\over n}{{\partial \rho}\over {\partial
s}}{|}_n
\approx {2\over {3k_B}} n^{{5\over 3}} e^{ {{2(s-s_o)}\over {3k_B}} }.
\label{V8}
\eeq

The action for these fluids is [$s=s(\alpha^i)$]

\beq
S=\int d\tau d^3\sigma L(\alpha^i(\tau ,\vec \sigma ), z^{\mu}(\tau
,\vec \sigma ))
=-\int d\tau d^3\sigma N \sqrt{\gamma} \rho (n,s),
\label{V9}
\eeq

\noindent and we have  as in Section II

\bea
J^{\tau} &=&-\epsilon^{\check r\check u\check v}\partial _{\check r}\alpha^1
\partial _{\check u}\alpha^2\partial _{\check v}\alpha^3,\nonumber\\
J^{\check r} &=&  \sum_{i=1}^3 \partial _{\tau}\alpha^i \epsilon^{\check
r\check u\check v}\partial _{\check u}\alpha^j \partial _{\check v}
\alpha^k, \nonumber\\
& &i,~j,~ k, ~ = ~ \mbox{cyclic},\nonumber \\
|J|&=&\sqrt{N^{2}(J^{{\tau}})^{2}\,\, -
{}^3g_{\check r\check s}[J^{\check r}+N^{\check r}J^{{\tau}}][J^{\check v}+
N^{\check v}J^{{\tau}}]}=N X,\nonumber \\
&&{}\nonumber \\
X&=&\sqrt{(J^{{\tau}})^{2}-{}^3g_{\check r\check s}Y^{\check r}Y^{\check s}},
\nonumber \\
Y^{\check r}&=&\frac{1}{N}(J^{\check r}+N^{\check r}J^{{\tau}}),\nonumber \\
&&{}\nonumber \\
{{\partial X}\over {\partial \partial_{\tau}\alpha^i}}&=&{{Y^{\check r}
T_{\check ri}}\over {N X}},
\nonumber \\
&&T_{\check ti}=-{}^3g_{\check t\check r}\epsilon^{\check r\check u\check v}
\partial _{\check u} \alpha^j \partial _{\check v}\alpha^k,\quad
(i,j,k\quad cyclic),\nonumber \\
{{\partial X}\over {\partial N}}&=&{{{}^3g_{\check r\check s}Y^{\check
r}Y^{\check s}}\over {N X}}={{(J^{\tau})^2-X^2}\over {NX}},\nonumber \\
{{\partial X}\over {\partial N^{\check u}}}&=&- J^{\tau}
{{{}^3g_{\check u\check s}Y^{\check s}}\over {N X}}.
\label{V10}
\eea

In the cases 1), 2) and 3) [in case 3) we rename $\mu$ the constant
$a$] the canonical momenta  can be written in the form

\bea
\Pi_i(\tau ,\vec \sigma )&=&{{\partial L(\tau ,\vec \sigma )}\over
{\partial \partial_{\tau}\alpha^i(\tau ,\vec \sigma )}}=
\mu \Big[  {{\partial f(x)}\over {\partial x}}{|}_{x={X\over {\sqrt{\gamma}}}}
 {{Y^{\check r}T_{\check ri}}\over X} \Big] (\tau
,\vec \sigma ),\nonumber \\
&&{}\nonumber \\
\Rightarrow&& Y^{\check r}=-{{(T^{-1})^{\check ri}\Pi_i}\over {\mu}} X
({{\partial f(x)}\over {\partial x}}{|}_{x={X\over
{\sqrt{\gamma}}}})^{-1}= Y^{\check r}(X),\nonumber \\
\Rightarrow&& \partial_{\tau}\alpha^i =-{{J^{\check r}\partial_{\check r}
\alpha^i}\over {J^{\tau}}}={{N^{\check r}J^{\tau}-N Y^{\check r}}\over
{J^{\tau}}}\partial_{\check r}\alpha^i=\nonumber \\
&&\quad\quad =N^{\check r}\partial_{\check r}\alpha^i +N \partial_{\check r}
\alpha^i (T^{-1})^{\check rj} \Pi_j {X\over {\mu J^{\tau}}} [{{\partial f(x)}
\over {\partial x}}{|}_{x={X\over {\sqrt{\gamma}}}} ]^{-1},\nonumber \\
&&{}\nonumber \\
 &&\Downarrow \nonumber \\
\rho_{\mu}(\tau ,\vec \sigma )&=&-{{\partial L(\tau ,\vec \sigma )}\over
{\partial z_{\tau}^{\mu}(\tau ,\vec \sigma )}}=-\Big[ l_{\mu} {{\partial L}\over
{\partial N}}-\epsilon z_{\check s\mu}\,
{}^3g^{\check s\check r}{{\partial L}\over {\partial N^{\check r}}}\Big] (\tau
,\vec \sigma )=\nonumber \\
&=&\mu \sqrt{\gamma} \Big[ l_{\mu} {{\partial N f({X\over {\sqrt{\gamma}}})}
\over {\partial N}}-\epsilon z_{\check s\mu}\, {}^3g^{\check s\check r} N
{{\partial f({X\over {\sqrt{\gamma}}})}\over {\partial N^{\check
r}}}\Big] (\tau ,\vec \sigma )=\nonumber \\ &=&\Big[ {{\mu}\over X}
\Big( \sqrt{\gamma}X f({X\over {\sqrt{\gamma}}})+
[(J^{\tau})^2-X^2]{{\partial f(x)}\over {\partial x}}{|}_{x={X\over
{\sqrt{\gamma}}}}\Big) l_{\mu}+ \nonumber \\ &+&{{\mu}\over X}
{{\partial f(x)}\over {\partial x}}{|}_{x={X\over {\sqrt{\gamma}}}}
J^{\tau} Y^{\check r} z_{\check r\mu}\Big] (\tau ,\vec \sigma
)=\nonumber \\ &=&\Big[ {{\mu}\over X} \Big( \sqrt{\gamma}X f({X\over
{\sqrt{\gamma}}})+ [(J^{\tau})^2-X^2]{{\partial f(x)}\over {\partial
x}}{|}_{x={X\over {\sqrt{\gamma}}}}\Big) l_{\mu}\Big] (\tau ,\vec
\sigma )-\nonumber \\ &-&\epsilon \Big[ J^{\tau} (T^{-1})^{\check
ri}\Pi_i z_{\check r\mu}\Big] (\tau ,\vec \sigma )=\nonumber \\
&=&\Big[ \mu  \sqrt{\gamma} G(X,J^{\tau},\sqrt{\gamma}) l_{\mu}+
J^{\tau} (T^{-1})^{\check ri}
\Pi_i z_{\check r\mu}\Big] (\tau ,\vec \sigma ),\nonumber \\
&& with \nonumber \\
  G({X\over {\sqrt{\gamma}}},{{(J^{\tau})^2}\over
{\gamma}})&=& f({X\over {\sqrt{\gamma}}})+{{ {{(J^{\tau})^2}\over
{\gamma}}-({X\over {\sqrt{\gamma}}})^2}\over {X\over {\sqrt{\gamma}}}}
{{\partial f(x)}\over {\partial x}}{|}_{x={X\over
{\sqrt{\gamma}}}}=f(n)+ {{ {{(J^{\tau})^2}\over {\gamma}}-n^2}\over n}
{{\partial f(n)}\over {\partial n}}.
\label{V11}
\eea

To get the Hamiltonian expression of the constraints ${\cal
H}^{\mu}(\tau ,\vec \sigma )\approx 0$, we have to find the solution
$X$ of the equation $X^2+{}^3g_{\check r\check s}Y^{\check
r}(X)Y^{\check s}(X)=(J^{\tau})^2$ with $Y^{\check r}(X)$ given by the
second line of Eq.(\ref{V11}). This equation may be written in the
following forms

\bea
&&X^2 \Big[ \mu^2+ A^2 ({{\partial f(x)}\over {\partial x}}{|}_{x=
{X\over {\sqrt{\gamma}}}})^{-2}\Big]
=B^2,\quad or\nonumber \\
({X\over {\sqrt{\gamma}}})^2&& \Big[ 1+{{A^2}\over {\mu^2+A^2}} \Big(
({{\partial f(x)}\over {\partial x}}{|}
_{x={X\over {\sqrt{\gamma}}}})^{-2}-1\Big) \Big] ={{B^2}\over
{\gamma (\mu^2+A^2)}},\quad or\nonumber \\
n^2&&\Big[ 1+{{A^2}\over {\mu^2+A^2}} \Big(
({{\partial f(n)}\over {\partial n}})^{-2}-1\Big) \Big] ={{B^2}\over
{\gamma (\mu^2+A^2)}},\nonumber \\
&&{}\nonumber \\
A^2&=&{}^3g_{\check r\check s}(T^{-1})^{\check ri}\Pi_i
(T^{-1})^{\check sj}\Pi_j,\nonumber \\
B^2&=& \mu^2 (J^{\tau})^2,\nonumber \\
&&{}\nonumber \\
\Rightarrow && X=\sqrt{\gamma} n=\sqrt{\gamma} F({{A^2}\over {\mu^2}},
{{(J^{\tau})^2}\over {\gamma}})=\sqrt{\gamma}
\tilde F({{A^2}\over {\mu^2+A^2}}, {{B^2}\over {\gamma (\mu^2+A^2)}}),
\nonumber \\
&&{}\nonumber \\
\Rightarrow &&\rho_{\mu}=\mu \sqrt{\gamma} \tilde G({{A^2}
\over {\mu^2+A^2}}, {{B^2}\over {\gamma (\mu^2+A^2)}}) l_{\mu}+
J^{\tau} (T^{-1})^{\check ri}\Pi_i z_{\check r\mu}=\nonumber \\
&=&{\cal M} l_{\mu}+ {\cal M}^{\check r} z_{\check r\mu}.
\label{V12}
\eea

Therefore, all the dependence on the metric and on the Lagrangian coordinates
and their momenta is concentrated in the 3 functions $\sqrt{\gamma}$,
$A^2/\mu^2={}^3g_{\check r\check s}(T^{-1})^{\check ri}\Pi_i
(T^{-1})^{\check sj}\Pi_j/\mu^2$, $B^2/\mu^2\gamma =(J^{\tau})^2/\gamma$.

Let us consider various cases.

1) $p=0$, dust. As in Section II the equation for $X$ and the
constraints are

\bea
X^2 && [\mu^2+A^2] = B^2,\nonumber \\
&&{}\nonumber \\
X&=& {B\over {\sqrt{\mu^2+A^2}}}={ {\mu  |J^{\tau}|}\over
{\sqrt{\mu^2+{}^3g_{\check r\check s}(T^{-1})^{\check ri}\Pi_i (T^{-1})^{\check
sj}\Pi_j}}}\nonumber \\
&&{\rightarrow}_{A^2\rightarrow 0}\, |J^{\tau}| [1-{{A^2}\over {2\mu^2}}
+O(A^4)],\nonumber \\
&&{}\nonumber \\
Y^{\check r}&=&-{X\over {\mu}} (T^{-1})^{\check ri} \Pi_i=-{{|J^{\tau}|
(T^{-1})^{\check ri}\Pi_i}\over {
\sqrt{\mu^2+{}^3g_{\check r\check s}(T^{-1})^{\check ri}\Pi_i (T^{-1})^{\check
sj}\Pi_j} }},\nonumber \\
 &&\Downarrow\nonumber \\
\rho_{\mu}&=& |J^{\tau}| \sqrt{\mu^2+{}^3g_{\check r\check s}(T^{-1})^{\check
ri}\Pi_i (T^{-1})^{\check sj}\Pi_j} l_{\mu}+ J^{\tau} (T^{-1})^{\check
ri}\Pi_i z_{\check r\mu}.
\label{V13}
\eea

2) $p=k\rho$, $k\not= -1$. The equation for $X$ is

\bea
X^2&& [\mu^2+{{A^2}\over {(k+1)^2({X\over {\sqrt{\gamma}}})^{2k}}}]
=B^2,\quad or\nonumber \\
({X\over {\sqrt{\gamma}}})^2&& [1+{{A^2}\over {\mu^2+A^2}} ({1\over
{(k+1)^2({X\over {\sqrt{\gamma}}})^{2k}}}-1)]=
{{B^2}\over {\gamma (\mu^2+A^2)}},\nonumber \\
&&{}\nonumber \\
Y^{\check r}&=&-{{\sqrt{\gamma} ({X\over {\sqrt{\gamma}}})^{1-k} (T^{-1})
^{\check ri} \Pi_i}\over {\mu (k+1)}},\nonumber \\
&&{}\nonumber \\
\rho_{\mu}&=&\mu \sqrt{\gamma} ({X\over {\sqrt{\gamma}}})^{k-1}[(k+1)
{{(J^{\tau})^2}\over {\gamma}}-k(){X\over {\sqrt{\gamma}}}^2]l_{\mu}+
J^{\tau}(T^{-1})^{\check ri}\Pi_i z_{\check r\mu}.
\label{V14}
\eea

Let us define $Z$ as the deviation of $X$ from dust (for
$A^2\rightarrow 0$ [$\partial_{\tau} \alpha^i=0$]: we have
$Z\rightarrow 1$.\hfill\break
\hfill\break
$X={{|B|}\over {\sqrt{\mu^2+A^2}}} Z$. Then we get the following
equation for $Z$

\bea
Z^2&& \Big[ {{\mu^2}\over {\mu^2+A^2}} + {{A^2 (\mu^2+A^2)^{k-1}\gamma^k}\over
{(k+1)^2 B^{2k}}} Z^{-2k}\Big] =1,\quad or\nonumber \\
Z^2&& [\alpha^2+\beta_k^2 Z^{-2k}]=1,\nonumber \\
&&{}\nonumber \\
&&\alpha^2={{\mu^2}\over {\mu^2+A^2}} {\rightarrow}_{A^2
\rightarrow 0}\, 1,\nonumber \\
&&\beta^2_k={{A^2 (\mu^2+A^2)^{k-1}\gamma^k}\over {(k+1)^2 B^{2k}}}
{\rightarrow}_{A^2\rightarrow 0}\, 0.
\label{V15}
\eea

We may consider the following subcases:

2a) $k=m\not= -1$, with the equation

\bea
Z_1&=& Z^2,\quad\quad X={{|B|}\over {\sqrt{\mu^2+A^2}}} \sqrt{Z_1},
\nonumber \\
&&{}\nonumber \\
Z_1&& [\alpha^2+\beta^2_m Z_1^{-m}]=1,\quad or\quad \beta^2_mZ_1^{1-m}+
\alpha^2Z_1-1=0.
\label{V16}
\eea

i) $p=\rho$ ($k=m=1$), with the equation

\bea
Z_1&=&{{1-\beta_1^2}\over {\alpha^2}}= {{\gamma
(\mu^2+A^2)(4{{B^2}\over {\gamma}}-A^2)}\over {4\mu^2 B^2}},
\nonumber \\
  X&=&{{\sqrt{4{{B^2}\over {\gamma}}-A^2}}\over {2\mu}}={1\over
{2\mu}}
\sqrt{4\mu^2 {{(J^{\tau})^2}\over {\gamma}}+{}^3g_{\check r\check s}(T^{-1})
^{\check ri}\Pi_i (T^{-1})^{\check sj}\Pi_j},\quad
for\quad A^2 < 4{{B^2}\over {\gamma}}.
\label{V17}
\eea

ii) $p=2\rho$ ($k=m=2$), with the equation

\beq
\alpha^2 Z_1^2-Z_1+\beta^2_2=0,\quad\quad Z_1={1\over {2\alpha^2}}
[1\pm \sqrt{1-4\alpha^2\beta^2_2}],
\label{V18}
\eeq

iii) $p=-2\rho$ ($k=m=-2$), with the equation

\beq
\beta^2_{-2} Z_1^3+\alpha^2 Z_1 -1=0,
\label{V19}
\eeq

2b) $k={1\over m}$, with the equation

\bea
Z_2&=& Z^{{2\over m}},\quad\quad X={{|B|}\over {\sqrt{\mu^2+A^2}}}
Z_2^{{m\over 2}},\nonumber \\
 &&\alpha^2 Z_2^m +\beta^2_{1/m} Z_2^{m-1}-1=0,
\label{V20}
\eea

i) $p={1\over 2} \rho$ ($k={1\over 2}$, $m=2$), with the equation

\bea
&&\alpha^2 Z_2^2+\beta^2_{1/2} Z_2-1=0,\nonumber \\
 Z_2&=&{1\over
{2\alpha^2}} [-\beta^2_{1/2} \pm \sqrt{\beta^4_{1/2}+4\alpha^2}].
\label{V21}
\eea

ii) $p=-{1\over 2}\rho$ ($k=-{1\over 2}$, $m=-2$), with the equation

\beq
\beta^2_{-1/2} (Z_2^{-1})^3 +\alpha^2 (Z_2^{-1})^2 -1=0.
\label{V22}
\eeq

iii) $p={1\over 3}\rho$, photon gas ($k={1\over 3}$, $m=3$),
$\beta^2_{1/3}= {{9A^2\gamma^{1/3}}\over {16
B^{2/3}(\mu^2+A^2)^{2/3}}}$, with the equation

\bea
&&\alpha^2 Z_2^3 +\beta^2_{1/3} Z_2^2 -1=0,\quad\quad or\quad
Z^3_2+pZ^2_2+r=0,\nonumber \\
 && with\quad p={{\beta^2_{1/3}}\over
{\alpha^2}} > 0,\quad r=-{1\over {\alpha^2}} < 0,\nonumber \\
 &&{}\nonumber \\
Z_2&=&Y_2-{1\over 3} p=Y_2- {{\beta^2_{1/3}}\over {3\alpha^2}}\,
{\rightarrow}_{A^2\rightarrow 0}\, 1,\nonumber \\
 X&=&{{|B|}\over
{\sqrt{\mu^2+A^2}}} (Y_2- {{\beta^2_{1/3}}\over
{3\alpha^2}})={{|B|}\over {\sqrt{\mu^2+A^2}}}
(Y_2-{{3A^2(\mu^2+A^2)^{1/3}
\gamma^{1/3}}\over {16 B^{2/3}}} )^{3/2},\nonumber \\
 &&{}\nonumber \\
 && Y_2^3+ a Y_2 +b =0,\nonumber \\
 a&=&-{1\over 3} p^2=-{{\beta^4_{1/3}}\over {3\alpha^4}}=-{{27 A^4
(\mu^2 +A^2)^{2/3}\gamma^{2/3}}\over {2^8 B^{4/3}}} < 0,\nonumber \\
 b&=&{1\over {27}}(2p^3+27 r)={1\over {27
\alpha^2}}(2{{\beta^6_{1/3}}\over {\alpha^4}}-27)={{\mu^2+A^2}\over
{\mu^2}}\Big( {{27A^6\gamma}\over {2^{11}
\mu^4B^2}}-1\Big),\nonumber \\
 &&{{b^2}\over 4}+{{a^3}\over
{27}}={1\over {4\cdot 27 \alpha^4}}(27-4{{\beta^6
_{1/3}}\over {\alpha^4}})={{(\mu^2+A^2)^2}\over {4\mu^4}}
\Big( 1-{{27A^6\gamma}\over {2^{10} \mu^4 B^2}}\Big)\, {\rightarrow}
_{A^2\rightarrow 0}\, {1\over 4} > 0,\nonumber \\
 &&{}\nonumber \\
 Y_2&=&\Big( -{b\over 2}+\sqrt{ {{b^2}\over 4}+{{a^3}\over
{27}}}\,\, \Big)^{1/3}-
\Big( {b\over 2}+\sqrt{ {{b^2}\over 4}+{{a^3}\over {27}}}\,\, \Big)^{1/3}\,
{\rightarrow}_{A^2\rightarrow 0}\, 1,\nonumber \\
  &&{}\nonumber \\
  &&\Downarrow   \nonumber \\
 X&=&|J^{\tau}| \Big[ -{{3A^2\gamma^{1/3}}\over {16
(J^{\tau})^{2/3}}}+{1\over {2^{1/3}}}\Big( 1-{{27A^6\gamma}\over
{2^{11}\mu^6(J^{\tau})^2}}+\sqrt{1- {{27A^6\gamma}\over
{2^{10}\mu^6(J^{\tau})^2}} }\,\, \Big)^{1/3}-\nonumber \\
 &-& {1\over {2^{1/3}}}
\Big( -1+{{27A^6\gamma}\over {2^{11}\mu^6(J^{\tau})^2}}+\sqrt{1-
{{27A^6\gamma}\over {2^{10}\mu^6(J^{\tau})^2}} }\,\, \Big)^{1/3}
\Big]^{3/2}\, {\rightarrow}_{A^2\rightarrow 0} \nonumber \\
 &{\rightarrow}_{A^2\rightarrow 0}\,& |J^{\tau}|
[1-{{9A^2\gamma^{1/3}}\over {32(J^{\tau})^{2/3}}}+O(A^4)],\nonumber \\
 &&\Downarrow \nonumber \\
 \rho_{\mu}&=&{1\over 3} \mu \sqrt{\gamma} ({X\over
{\sqrt{\gamma}}})^{-2/3} [4{{(J^{\tau})^2}\over {\gamma}}-({X\over
{\sqrt{\gamma}}})^2] l_{\mu}+ J^{\tau} (T^{-1})^{\check ri}\Pi_i
z_{\check r\mu}=\nonumber \\
 &=& {\cal M} l_{\mu} + {\cal M}^r z_{r \mu}.
\label{V23}
\eea

In the case of the photon gas we get a closed analytical form for the
constraints.

3) $p=k\rho^{\gamma}$, $\gamma =1+{1\over n}$, $\gamma \not= 1$
($n\not= 0$), with the equation

\bea
X^2&& \Big[ \mu^2+A^2 \Big( 1-k(\mu {X\over {\sqrt{\gamma}}})^{\gamma -1}
\Big)^{ {{2\gamma}\over {\gamma -1}} } \Big] =B^2,\quad or\nonumber \\
X^2&& \Big[ \mu^2+A^2 \Big( 1-k(\mu {X\over {\sqrt{\gamma}}})^{{1\over n}}
\Big)^{2(n+1)}\Big] =B^2,\quad or\nonumber \\
({X\over {\sqrt{\gamma}}})^2&& \Big[ 1+{{A^2}\over {\mu^2+A^2}}\Big( [1-
k(\mu {X\over {\sqrt{\gamma}}})^{{1\over n}}]
^{2(n+1)} -1\Big) \Big] ={{B^2}\over {\gamma (\mu^2+A^2)}},\nonumber \\
&&{}\nonumber \\
Y^{\check r}&=&- {X\over {\mu}} [1-k({X\over {\sqrt{\gamma}}})^{{1\over n}}
]^{n+1} (T^{-1})^{\check ri} \Pi_i,\nonumber \\
&&{}\nonumber \\
\rho_{\mu}&=&\mu \sqrt{\gamma} {{{{(J^{\tau})^2}\over {\gamma}}-k
\mu^{{1\over n}} ({X\over {\sqrt{\gamma}}})^{2+{1\over n}}}\over
{{{X\over {\sqrt{\gamma}}}} [1-k(\mu {X\over
{\sqrt{\gamma}}})^{{1\over n}} ]^{n+1}}} l_{\mu}+ J^{\tau}
(T^{-1})^{\check ri}\Pi_i z_{\check r\mu}.
\label{V24}
\eea

Let us define $Z$ as the deviation of $X$ from dust (for
$A^2\rightarrow 0$ [$\partial_{\tau} \alpha^i=0$]: we have
$Z\rightarrow 1$, $X={{|B|}\over {\sqrt{\mu^2+A^2}}} Z$). Then we get
the following equation for $Z$

\begin{equation}
Z^2 \Big[ {{\mu^2}\over {\mu^2+A^2}}+{{A^2 (\mu^2+A^2)^{k-1}\gamma^k}\over
{(k+1)^2 B^{2k}}}\Big( 1-k({{\mu |B| Z}\over {\sqrt{\gamma}\sqrt{\mu^2+A^2}}})
^{{1\over n}} \Big)^{2(n+1)} \Big] =1.
\label{V25}
\end{equation}

 In conclusion, only in the cases of the dust and of the photon gas we
 get the closed analytic form of the constraints ( i.e. of the density
 of invariant mass ${\cal M}(\tau ,\vec \sigma )$, because for the
 momentum density we have ${\cal M}^{\check r}=J^{\tau}(T^{-1})^{\check ri}\Pi_i$
 independently from the type of perfect fluid).

In all the other cases we have only an implicit form for them
depending on the solution $X$ of Eq.(\ref{V12}) and numerical methods
should be used.

\vfill\eject

\section{Coupling to ADM metric and tetrad gravity.}

Let us now assume to have a globally hyperbolic, asymptotically flat at spatial
infinity spacetime $M^4$ with the spacelike leaves $\Sigma_{\tau}$ of the
foliations associated with its 3+1 splittings, diffeomorphic to $R^3$
\cite{russo1,russo2,russo3,india}.

In $\sigma_{\tau}$-adapted coordinates $\sigma^A=(\sigma^{\tau}=\tau ;\vec
\sigma )$ corresponding to a holonomic basis [$d\sigma^A$, $\partial_A=\partial
 /\partial \sigma^A$] for tensor fields we have
\cite{russo1} [$N$ and $N^r$ are the lapse and shift functions; ${}^3g_{rs}$ is
the 3-metric of $\Sigma_{\tau}$; $l^A(\tau ,\vec \sigma )$ is the unit normal
vector fields to $\Sigma_{\tau}$]

\bea
{}^4g_{AB}&=& g_{AB}=\{ {}^4g_{\tau\tau}=\epsilon
(N^2-{}^3g_{rs}N^rN^s); {}^4g_{\tau r}=-\epsilon          \,
{}^3g_{rs}N^s; {}^4g_{rs}=-\epsilon \, {}^3g_{rs} \}
=\nonumber \\
&=&\epsilon  l_Al_B + \triangle_{AB},\nonumber \\
&&{}\nonumber \\
\triangle_{AB}&=& {}^4g_{AB}-\epsilon l_Al_B.
\label{VI1}
\eea

A set of $\Sigma_{\tau}$-adapted tetrad and cotetrad fields is $(a)=(1),(2),
(3)$; ${}^3e^r_{(a)}$ and ${}^3e^{(a)}_r={}^3e_{(a)r}$ are triad and cotriad
fields on $\Sigma_{\tau}$]

\bea
{}^4_{(\Sigma )}E^A_{(o)}&=&\epsilon l^A = ({1\over N}; -{{N^r}\over N}),
\nonumber \\
{}^4_{(\Sigma )}E^A_{(a)}&=& (0; {}^3e^r_{(a)} ),\nonumber \\
&&{}\nonumber \\
{}^4_{(\Sigma )}E_A^{(o)}&=& l_A= (N; \vec 0),\nonumber \\
{}^4_{(\Sigma )}E_A^{(a)}&=& (N^{(a)}=N^r\, {}^3e^{(a)}_r; {}^3e^{(a)}_r).
\label{VI2}
\eea

In these coordinates the energy-momentum tensor of the perfect fluid is

\beq
T^{AB}=-\epsilon (\rho +p) U^AU^B+ p\, {}^4g^{AB}.
\label{VI3}
\eeq

If we use the notation $\Gamma =\epsilon l_AU^A=\epsilon N U^{\tau}$, we get

\bea
T_{AB}&=&{}^4g_{AC}\, {}^4g_{BC} T^{CD}=[\epsilon
l_Al_C+\triangle_{AC}] [\epsilon l_Bl_D+\triangle_{BD}]
T^{CD}=\nonumber \\
 &=&E l_Al_B +j_Al_B+l_Aj_B +{\cal
S}_{AB},\nonumber \\
 &&{}\nonumber \\
 E&=&T^{AB}l_Al_B= -\epsilon [(\rho +p) \Gamma^2 -p],\nonumber \\
 j_A&=&\epsilon \triangle_{AC}
T^{CB}l_B=-\epsilon (\rho+p) \Gamma \triangle_{AB}U^B,\nonumber \\
 {\cal S}_{AB}&=&\triangle_{AC}\triangle_{BD} T^{CD}=-\epsilon [(\rho +p)
\triangle_{AC}U^C
\triangle_{BD}U^D- p \triangle_{AB}].
\label{VI4}
\eea

The same decomposition can be referred to a non-holonomic basis [$\theta^{\bar
A}=\{ \theta^l=Nd\tau ; \theta^r=d\sigma^r+N^rd\tau \}$, $X_{\bar A}= \{ X_l=
N^{-1} (\partial_{\tau}-N^r\partial_r); \partial_r \}$; $\bar A=(l; r)$] in
which we have [$\bar \Gamma =\epsilon {\bar l}_{\bar A}{\bar U}^{\bar A}=
\epsilon {\bar U}^l=\sqrt{1-{}^3g_{rs}{\bar U}^r{\bar U}^s}$, so that ${\bar
U}^r$ may be interpreted as the generalized boost velocity of ${\bar U}^{\bar
A}$ with respect to ${\bar l}^{\bar A}$\cite{york,smarr}]

\bea
{}^4{\bar g}_{\bar A\bar B}&=& \{ {}^4{\bar g}_{ll}=\epsilon ; {}^4{\bar g}
_{lr}=0; {}^4{\bar g}_{rs}={}^4g_{rs}=-\epsilon \, {}^3g_{rs} \} =\nonumber \\
&=& \epsilon {\bar l}_{\bar A}{\bar l}_{\bar B} +{\bar \triangle}_{\bar A\bar B}
,\nonumber \\
{\bar l}^{\bar A}&=& (\epsilon ; \vec 0),\quad\quad {\bar l}_{\bar A}=(1;\vec 0)
,\nonumber \\
{\bar \triangle}_{ll}&=&
{\bar \triangle}_{lr}=0,\quad\quad {\bar \triangle}_{rs}=-
\epsilon \, {}^3g_{rs},\quad\quad {\bar \triangle}_{\bar A\bar B}{\bar U}
^{\bar B}=(0;-\epsilon \, {}^3g_{rs}),\nonumber \\
 &&{}\nonumber \\
 {\bar T}_{\bar A\bar B}&=& -\epsilon (\rho +p) {\bar U}_{\bar A}{\bar U}_{\bar B}+
p\, {}^4{\bar g}_{\bar A\bar B}=\nonumber \\
  &=&\bar E {\bar l}_{\bar
A}{\bar l}_{\bar B}+{\bar j}_{\bar A} {\bar l}_{\bar B}+{\bar l}_{\bar
A}{\bar j}_{\bar B}+{\bar {\cal S}}_{\bar A\bar B},\nonumber \\
 &&{}\nonumber \\
 {\bar E}&=&{\bar T}^{\bar A\bar B}{\bar l}_{\bar
A}{\bar l}_{\bar B}=-\epsilon [(\rho +p) {\bar \Gamma}^2 -p],
\nonumber \\
 {\bar j}_{\bar A}&=&\epsilon {\bar \triangle}_{\bar A\bar C}{\bar
T}^{\bar C\bar B}l_{\bar B}=({\bar j}_l=0; {\bar j}_r=-\epsilon (\rho
+p)\bar \Gamma \, {}^3g_{rs}{\bar U}^s),\nonumber \\
 {\bar {\cal S}}_{\bar
A\bar B}&=& {\bar \triangle}_{\bar A\bar C}{\bar
\triangle}_{\bar B\bar D}{\bar T}^{\bar C\bar D}=\nonumber \\
&=&({\bar {\cal S}}_{ll}={\bar {\cal S}}_{lr}=0; {\bar {\cal
S}}_{rs}=-\epsilon [-p \, {}^3g_{rs}+(\rho +p){}^3g
_{ru}{\bar U}^u\, {}^3g_{sv}{\bar U}^v)].
\label{VI5}
\eea

$\bar E$ and ${\bar j}_r$ are the energy and momentum densities determined by
the Eulerian observers on $\Sigma_{\tau}$, while ${\bar {\cal S}}_{rs}$ is
called the ``spatial stress tensor".

The non-holonomic basis is used to get the 3+1 decomposition
(projection normal and parallel to $\Sigma_{\tau}$) of Einstein's
equations with matter ${}^4G^{\mu\nu}\, {\buildrel \circ \over =}\,
{{\epsilon c^3}\over {8\pi G}} T^{\mu\nu}$ [when one does not has an
action principle for matter, one cannot use the Hamiltonian ADM
formalism]: in this way one gets four restrictions on the Cauchy data
[ ${}^4{\bar G}^{ll}\, {\buildrel \circ \over =}\, {{\epsilon
c^3}\over {8\pi G}}{\bar T}^{ll}$ and ${}^4{\bar G}^{lr}\, {\buildrel
\circ
\over =}\, {{\epsilon c^3}\over {8\pi G}} {\bar T}^{lr}$; they become
the secondary first class superhamiltonian and supermomentum
constraints in the ADM theory; $k=c^3/8\pi G$]

\bea
&&\Big[ {}^3R+({}^3K)^2-{}^3K_{rs}\, {}^3K^{rs}\Big]
(\tau ,\vec \sigma )\, {\buildrel \circ \over =}\,
{2\over k} \bar E (\tau ,\vec \sigma ),\nonumber \\
&&({}^3K^{rs}-{}^3g^{rs}\, {}^3K)_{|s} (\tau ,\vec \sigma )\, {\buildrel \circ
\over =}\, {1\over k} {\bar J}^r(\tau ,\vec \sigma ),
\label{VI6}
\eea

\noindent and the spatial Einstein's equations
${}^4{\bar G}^{rs}\, {\buildrel \circ \over =}\, {{\epsilon c^3}\over
{8\pi G}} {\bar T}^{rs}$. By introducing the extrinsic curvature
${}^3K_{rs}$, this last equations are written in a first order form
[it corresponds to the Hamilton equations of the ADM theory for
${}^3g_{rs}$ and ${}^3{\tilde \Pi}^{rs}= {{\epsilon c^3}\over {8\pi
G}} \sqrt{\gamma}({}^3K^{rs}-{}^3g^{rs}\, {}^3K)$; ``$|$" denotes the
covariant 3-derivative]

\bea
\partial_{\tau}\, {}^3g_{rs}(\tau ,\vec \sigma )&=&\Big[ N_{r|s}+N_{s|r}-2N\,
{}^3K_{rs}\Big] (\tau ,\vec \sigma ),\nonumber \\
\partial_{\tau}\, {}^3K_{rs}(\tau ,\vec \sigma )\, &{\buildrel \circ \over =}\,&
\Big( N [{}^3R_{rs}+{}^3K\, {}^3K_{rs}-2\, {}^3K_{ru}\, {}^3K^u{}_s]-
\nonumber \\
&-&N_{|s|r}+N^u{}_{|s}\, {}^3K_{ur}+N^u{}_{|r}\, {}^3K_{us}+N^u\, {}^3K_{rs|u}
\Big) (\tau ,\vec \sigma )-\nonumber \\
&-& {1\over {k}} \Big[ ({\bar {\cal S}}_{rs}+{1\over 2} {}^3g_{rs}(\bar E -
{\bar {\cal S}}^u{}_u)\Big] (\tau ,\vec \sigma ).
\label{VI7}
\eea

The matter equations $T^{\mu\nu}{}_{;\nu}\, {\buildrel \circ \over
=}\, 0$ become a generalized continuity equation [entropy conservation
when the particle number conservation law is added to the system]
${\bar l}_{\bar A} T^{\bar A\bar B}{}_{;\bar B}\, {\buildrel \circ
\over =}\, 0$ and generalized Euler equations ${\bar \triangle}_{\bar
A\bar B} T^{\bar B\bar C}{}_{;\bar C}\, {\buildrel \circ \over =}\, 0$
[{${\cal L}_X$ is the Lie derivative with respect to the vector field
X]

\bea
\Big[ \partial_{\tau} \bar E +N {\bar j}^r{}_{|r}\, {\buildrel \circ \over
=}\, \Big[ N ({\bar {\cal S}}^{rs}\, {}^3K_{rs}+\bar E \, {}^3K)-2 {\bar j}^r
N_{|r}+{\cal L}_{\vec N} \bar E\Big] (\tau ,\vec \sigma ),\nonumber \\
\Big[ \partial_{\tau} {\bar j}_r+N {\bar {\cal S}}^{rs}{}_{|s}\Big] (\tau
,\vec \sigma )\, {\buildrel \circ \over =}\, \Big[ N (2\, {}^3K^{rs}{\bar j}_s+
{\bar j}^r\, {}^3K)-{\bar {\cal S}}^{rs}N_{|s}-\bar E N^{|r}+{\cal L}_{\vec N}
{\bar j}^r\Big] (\tau ,\vec \sigma ).
\label{VI8}
\eea

The equation for ${\bar {\cal S}}_{rs}$ would follow from an equation of state
or dynamical equation of the sources [for perfect fluids it is the particle
number conservation].

This formulation is the starting point of many approaches to the
post-Newtonian approximation (see for instance Refs.\cite{damour,asada}) and
to numerical gravity (see for instance Refs.\cite{num1,num2}).

Instead with the action principle for the perfect fluid described with
Lagrangian coordinates (containing the
information on the equation of state and on the particle number and entropy
conservations) coupled to the action for tetrad gravity of Ref.\cite{russo1}
(it is the ADM action of metric gravity re-expressed in terms of a new
parametrization of tetrad fields) we get

\bea
S&=&-\epsilon k\int d\tau d^3\sigma \,
\lbrace N\, {}^3e\, \epsilon_{(a)(b)(c)}\,
{}^3e^r_{(a)}\, {}^3e^s_{(b)}\, {}^3\Omega_{rs(c)}+\nonumber \\
&+&{{{}^3e}\over {2N}} ({}^3G_o^{-1})_{(a)(b)(c)(d)} {}^3e^r_{(b)}(N_{(a) | r}-
\partial_{\tau}\, {}^3e_{(a)r})\, {}^3e^s_{(d)}(N_{(c) | s}-\partial_{\tau}
\, {}^3e_{(c) \ s})\rbrace (\tau ,\vec \sigma ) +\nonumber \\
 &-&\int d\tau d^3\sigma \{ N\sqrt{\gamma} \rho
({{|J|}\over {N\sqrt{\gamma}}}, s) \}(\tau ,\vec \sigma ).
\label{VI9}
\eea

The superhamiltonian and supermomentum constraints of Ref.\cite{russo1} are
modified in the following way by the presence of the perfect fluid

\bea
{\cal H}&=&{\cal H}_o +{\cal M} \approx 0,\nonumber \\ {\cal
H}_r&\approx& \Theta_r =\Theta_{o\, r}+{\cal M}_r\approx 0,
\label{VI10}
\eea

\noindent with ${\cal M}_r=J^{\tau}(T^{-1})^{ri}\Pi_i=-\partial_r\alpha^i \Pi_i$ and
with ${\cal M}$ given by Eq.(\ref{II12}) for the dust and by
Eq.(\ref{V23}) for the photon gas.

In the case of dust the explicit Hamiltonian form of the energy and momentum
densities is

\bea
{\cal M}(\tau ,\vec \sigma )&=&J^{\tau}(\tau ,\vec \sigma )\sqrt{\mu^2+[{}^3g
_{rs}(T^{-1}({\bf \alpha}))^{ri}(T^{-1}({\bf \alpha}))^{sj}\Pi_i\Pi_j]
(\tau ,\vec \sigma )}=\nonumber \\
 &=&-det\, (\partial_r\alpha^i(\tau ,\vec \sigma ))
 \sqrt{\mu^2+{}^3g^{uv}{{\partial_u\alpha^m\partial_v\alpha^n}
 \over {[det\, (\partial_r\alpha^k)]^2}}\Pi_m\Pi_n\Big] (\tau ,\vec \sigma )} ,\nonumber \\
 &&{}\nonumber \\
  {\cal M}_r(\tau ,\vec \sigma )&=& {}^3g_{rs} J^{\tau}(\tau ,\vec \sigma ) [(T^{-1}({\bf
\alpha}))^{si}\Pi_i](\tau ,\vec \sigma )=-\partial_r\alpha^i(\tau ,\vec \sigma )
\Pi_i(\tau ,\vec \sigma ).
\label{VI11}
\eea

In any case, all the dependence of ${\cal M}$ and ${\cal M}^r$ on the
metric and on the Lagrangian coordinates and their momenta is
concentrated in the 3 functions $\sqrt{\gamma}$,
$A^2/\mu^2={}^3g_{\check r\check s}(T^{-1})^{\check ri}\Pi_i
(T^{-1})^{\check
sj}\Pi_j/\mu^2={{{}^3g^{uv}\partial_u\alpha^i\partial_v\alpha^j\Pi_i\Pi-J}\over
{\mu ^2 [det\, (\partial_u\alpha^k)]^2}}$, $B^2/\mu^2\gamma
=(J^{\tau})^2/\gamma = {1\over {\gamma}} [det\, (\partial_r\alpha^i)]^2$.

The study of the canonical reduction to the 3-orthogonal gauges
\cite{russo2,russo3} will be done in a future paper.

\vfill\eject

\section{Non-Dissipative Elastic Materials.}

With the same formalism we may describe relativistic continuum
mechanics [any relativistic material (non-homogeneous,
pre-stressed,...) in the non-dissipative regime] and in particular a
relativistic elastic continuum \cite{magli} [see also
Refs.\cite{magli1,magli2} and their bibliography] in the rest-frame
instant form of dynamics.

Now the scalar fields ${\tilde \alpha}^i(z(\tau ,\vec \sigma ))=\alpha^i(\tau
,\vec \sigma )$ describe the idealized ``molecules" of the material in an
abstract 3-dimensional manifold called the ``material space", while $J^{\check
A}(\tau ,\vec \sigma )=[N \sqrt{\gamma} n U^{\check A}](\tau ,\vec \sigma )$
is the matter number current with future-oriented timelike 4-velocity vector
field $U^{\check A}(\tau ,\vec \sigma )$; $n$ is a scalar field describing the
local rest-frame matter number density. The quantity $\partial_{\check A}\alpha
^i(\tau ,\vec \sigma )=z^{\mu}_{\check A}(\tau ,\vec \sigma ) \partial
_{\mu}{\tilde \alpha}^i(z)$ is called the ``relativistic deformation gradient"
in $\Sigma_{\tau}$-adapted coordinates.

The material space inherits a Riemannian (symmetric and positive
definite) 3-metric from the spacetime $M^4$

\beq
G^{ij}= {}^4g^{\mu\nu} \partial_{\mu}{\tilde \alpha}^i \partial_{\nu}
{\tilde
\alpha}^j={}^4g^{\check A\check B} \partial_{\check A}\alpha^i\partial_{\check
B}\alpha^j.
\label{VII1}
\eeq

Its inverse $G_{ij}$ carries the information about the actual distances of
adjacent molecules in the local rest frame.

For an ideal fluid the 3-form $\eta$ of Eq(\ref{I3}) gives the volume
element in the material space, which is sufficient to describe the mechanical
properties of an ideal fluid. Since we have that $n= {{|J|}\over
{N\sqrt{\gamma}}}$ is a scalar, we can evaluate
$n$ in the local rest frame at $z$ where $U^{\mu}(z)=({1\over {\sqrt{\epsilon \,
{}^4{\bar g}_{\bar o\bar o}}}}; \vec 0)$ and
$\partial_o{\tilde \alpha}^i{|}_z=0$ [in the local
adapted non-holonomic basis we have ${}^4{\bar g}
_{\bar A\bar B}=\epsilon \left( \begin{array}{cc} 1&0\\ 0& -{}^3g_{rs}
\end{array} \right)$, ${}^4{\bar g}^{\bar A\bar B}=\epsilon \left(
\begin{array}{cc}1&0\\ 0&-{}^3g^{rs}\end{array} \right)$, $\sqrt{{}^4{\bar g}}=
\sqrt{\gamma}/\sqrt{\epsilon \, {}^4{\bar g}^{\bar o\bar o}}$, $\sqrt{{}^4{\bar
g}^{-1}}=\sqrt{{}^3{\bar g}^{-1}}/\sqrt{\epsilon \, {}^4{\bar g}_{\bar o\bar
o}}=\sqrt{\epsilon \, {}^4{\bar g}_{\bar o\bar o}}/\sqrt{{}^4{\bar g}}=
\sqrt{{}^4{\bar g}^{-1}}$]. We get at $z$

\beq
n={{J^o}\over {U^o\sqrt{{}^4{\bar g}}}}=\eta_{123}\, det\,
\partial_k{\tilde
\alpha}^i\, {{\sqrt{\epsilon \, {}^4{\bar g}_{\bar o\bar o}}}\over
{\sqrt{{}^4{\bar g}}}}=\eta_{123}\, det\, \partial_k {\tilde
\alpha}^i\,
\sqrt{det\, |{}^4{\bar g}^{rs}|}= \eta_{123} \sqrt{det\, G^{ij}}.
\label{VII2}
\eeq

Therefore, we have $n=\eta_{123} \sqrt{det\, G^{ij}}$. \hfill\break
\hfill\break
Moreover, the material space of elastic materials, which have not only
volume rigidity but also shape rigidity, is equipped with a Riemannian
(symmetric and definite positive) 3-metric
$\gamma^{(M)}_{ij}(\alpha^i)$, the ``material metric", which is frozen
in the material and it is not a dynamical object of the theory. It
describes the ``would be" local rest-frame space distance between
neighbouring ``molecules", measured in the locally relaxed state of
the material. To measure the components $\gamma^{(M)}_{ij}(\alpha^i)$
we have to relax the material at different points $\alpha^i(\tau ,\vec
\sigma )$ separately, since global relaxation of the material may not
be possible [the material space may not be isometric with any
3-dimensional subspace of $M^4$, as in classical non-linear
elastomechanics, when the material exhibits internal stresses frozen
in it]. The components $\gamma^{(M)}_{ij}=\gamma^{(M)}
_{ij}(\alpha^i(\tau ,\vec \sigma ))$ are given functions, which describe
axiomatically the properties of the material (the theory is fully invariant
with respect to reparametrizations of the material space).

For ideal fluids $\gamma^{(M)}_{ij}=\delta_{ij}$; this also holds for
non-pre-stressed materials without ``internal" or ``frozen" stresses.

Now the material space volume element $\eta$ has

\beq
\eta_{123}(\alpha^i)=\sqrt{det\, \gamma^{(M)}_{ij}(\alpha^i)},\quad\quad so\,\, that\quad
n=\eta_{123} \sqrt{det\, G^{ij}}=\sqrt{det\, \gamma^{(M)}_{ij}}
\sqrt{det\, G^{ij}},
\label{VII3}
\eeq

\noindent and it cannot be put $=1$ like for perfect fluids.

The pull-back of the material metric $\gamma^{(M)}_{ij}$ to $M^4$ is

\beq
\gamma^{(M)}_{\check A\check B}=\gamma^{(M)}_{ij} \partial_{\check A}\alpha^i
\partial_{\check B} \alpha^j\quad\quad satisfying\quad \gamma^{(M)}_{\check A\check
B} U^{\check B}=0.
\label{VII4}
\eeq

The next step is to define a measure of the difference between the
induced 3-metric $G_{ij}(\partial \alpha^i)$ and the constitutive
metric $\gamma^{(M)}
_{ij}(\alpha^i)$, to be taken as a measure of the deformation of the material
and as a definition of a ``relativistic strain tensor", locally vanishing when
there is a local relaxation of the material. Some existing proposal for such a
tensor in $M^4$ are\cite{magli1,magli2}:

\beq
i)\quad\quad S^{(1)}_{\check A\check B}={1\over 2}({}^4g_{\check
A\check B}-\epsilon U_{\check A}U_{\check B}-\gamma^{(M)}_{\check
A\check B})
\label{VII5}
\eeq

\noindent
which vanishes at relax and satisfies $S^{(1)}_{\check A\check B} U^{\check B}
=0$ [but it must satisfy the involved matrix inequality $2\, det\, S^{(1)} \geq
det\, ({}^4g-\epsilon UU)$].

\beq
ii)\quad\quad S^{(2) \check A}{}_{\check B}={1\over 2}(K^{\check
A}{}_{\check B}-\delta
^{\check A}_{\check B}),\quad S^{(2)\check A}{}_{\check B}U^{\check B}=0,
\label{VII6}
\eeq

\noindent
with $K^{\check A}{}_{\check B}={}^4g^{\check A\check C}(\gamma^{(M)}_{\check
C\check B}-\epsilon U_{\check C}U_{\check B})$, $K^{\check A}{}_{\check B}
U^{\check B}=U^{\check A}$ [the 4-velocity field is an eigenvector of the
$K$-matrix].

\beq
iii)\quad\quad S^{(3)}=-{1\over 2} ln\, K,
\label{VII7}
\eeq

\noindent with the same $K$-matrix as in ii).

However a simpler proposal \cite{magli} is to define a ``relativistic
strain tensor" in the material space

\beq
S_i{}^j = \gamma^{(M)}_{ik} G^{kj},
\label{VII8}
\eeq

\noindent
with locally $S_i{}^j=\delta_i^j$ when there is local relaxation of the
material (in this case physical spacelike distances between material points
near a point $z^{\mu}(\tau ,\vec \sigma )$ agree with their material distances).
Since $n=\eta_{123} \sqrt{det\, G^{ij}}=\sqrt{det\, \gamma^{(M)}_{ij}}
\sqrt{det\, G^{ij}}$, we have $n=\sqrt{det\, S_i{}^j}$ for the local
rest-frame matter number density.

We can now define the local rest-frame energy per unit volume of the
material $n(\tau ,\vec \sigma ) e(\tau ,\vec \sigma )$, where $e$
denotes the molar local rest-frame energy (moles = number of
particles)

\beq
e(\tau ,\vec \sigma ) = m + u_I(\tau ,\vec \sigma ).
\label{VII9}
\eeq

Here $m$ is the molar local rest mass, $u_I$ is the amount of internal energy
(per mole of the material) of the elastic deformations, accumulated in an
infinitesimal portion during the deformation from the locally relaxed state to
the actual state of strain.

For isotropic media $u_I$ may depend on the deformation only via the invariants
of the strain tensor.

Let us notice that for an anisotropic material (like a crystal) the energy $u_I$
may depend upon the orientation of the deformation with respect to a specific
axis, reflecting the microscopic composition of the material: this information
may be encoded in a vector field $E^i(\tau ,\vec \sigma )$ in the material
space and one may assume $u_I=u_I(G^{-1}_{ij}E^iE^j)$.

The function $e=e[\alpha^i,G^{ij},\gamma^{(M)}_{ij},...]$ describes the
dependence of the energy of the material upon its state of strain and plays
the role of an ``equation of state" or ``constitutive equation" of the material.

In the weak strain approximation of an isotropic elastic continuum
(Hooke approximation) the function $u_I$ depends only on the linear
($h=S_i{}^i$) and quadratic ($q=S_i{}^j S^j{}_i$) invariants of the
strain tensor and coincides with the standard formula of linear
elasticity [$V={1\over n}= {1\over {\sqrt{det\, S_i{}^j}}}$ is the
specific volume],

\beq
u_I=\lambda (V) h^2 + 2 \mu (V) q +O(cubic\, invariants),
\label{VII10}
\eeq

\noindent
where $\lambda$ and $\mu$ are the Lam\'e coefficients.

The action principle for this description of relativistic materials is

\beq
S[{}^4g,\alpha^i ,\partial \alpha^i]=\int d\tau d^3\sigma L(\tau ,\vec
\sigma )
=-\int d\tau d^3\sigma (N\sqrt{\gamma})(\tau ,\vec \sigma)\,  n(\tau ,\vec \sigma )\,
e(\tau ,\vec \sigma ).
\label{VII11}
\eeq

It is shown in Refs.\cite{magli,magli1} that the canonical stress-energy-
momentum tensor $T^{\check A}_{\check B}=p^{\check A}_i\partial_{\check B}
\alpha^i-\delta^{\check A}_{\check B} L$, where $p^{\check A}_i=-{{\partial L}
\over {\partial \partial_{\check A}\alpha^i}}$ is the relativistic
Piola-Kirchhoff momentum density, and coincides with the symmetric
energy-momentum tensor $T_{\check A\check B}=-2{{\partial L}\over
{\partial \, {}^4g^{\check A\check B}}}$, which satisfies $T^{\check
A\check B}{}_{;\check B}=0$ due to the Euler-Lagrange equations. This
energy-momentum tensor may be written in the following form

\beq
T_{\check A\check B}=N \sqrt{\gamma}\, n\, [eU_{\check A\check B} +
Z_{\check A\check B}],
\label{VII12}
\eeq

\noindent
where $Z_{\check A\check B}=Z_{ij}\, \partial_{\check A}\alpha^i
\partial_{\check B}\alpha^j$ is the pull-back from the material space
to $M^4$ of the ``response tensor" of the material

\beq
Z_{ij} = 2\, {{\partial e}\over {\partial G^{ij}}},\quad\quad so\,
that\quad d e(G)={1\over 2} Z_{ij}dG^{ij}.
\label{VII13}
\eeq

The part $\tau_{\check A\check B}=n Z_{\check A\check B}$,
$\tau_{\check A\check B}U^{\check B}=0$, may be called the
relativistic ``stress or Cauchy" tensor and contains the
``stress-strain relation" through the dependence of $Z_{ij}$ on
$S_i{}^j$ implied by the consitutive equation $e=e[\alpha ,\gamma
^{(M)},G,..]$ of the material.

For an isotropic elastic material we get

\beq
Z_{ij}= V\, \Big[ p G^{-1}_{ij} +B \gamma^{(M)}_{ij} +C G_{ij}\Big],
\label{VII14}
\eeq

\noindent
where $p=-{{\partial e}\over {\partial V}}$, $B={2\over V}{{\partial e}\over
{\partial h}}$, $C={2\over V}{{\partial e}\over {\partial q}}$ and we get

\beq
d e(V,h,q)=-p dV +{1\over 2} VB dh +{1\over 2} VC dq.
\label{VII15}
\eeq

The response parameters describe the reaction of the material to the strain:
$p$ is the ``isotropic stress", while $B$ and $C$ give the anisotropic response
as in non-relativistic elesticity [perfect fluids have $e=e(V)$, $B=C=0$,
$Z_{ij}=VpG^{-1}_{ij}$ and $d e(V)=-pdV$ is the Pascal law].

See Ref.\cite{antoci} for a different description of relativistic Hooke law in
linear elasticity: there is a 4-dimensional deformation tensor $S_{\mu\nu}=
{1\over 2}({}^4\nabla_{\mu} \xi_{\nu}+{}^4\nabla_{\nu} \xi_{\mu})$ and the
constitutive equations of the material are given in the form $T^{\mu\nu}=
C^{(\mu\nu )(\alpha\beta )} S_{\alpha\beta}$.

In Ref.\cite{magli} the theory is also extended to the thermodynamics
of isentropic flows (no heat conductivity). The function $e$ is
considered also as a function of entropy $S=S(\alpha )$ and
$de={1\over 2}Z_{ij} dG^{ij}$ is generalized to

\beq
de ={1\over 2} Z_{ij}dG^{ij}-SdT.
\label{VII16}
\eeq

Then $e$ is replaced with the Helmholtz free energy $f=e-TS$ so to obtain

\beq
df={1\over 2}Z_{ij}dG^{ij}-SdT,
\label{VII17}
\eeq

\noindent
[for perfect fluids we get $de(V,S)=-pdV+TdS$, $df(V,T)=-pdV-SdT$]. This
suggests to consider the temperature $T$ as a strain and the entropy $S$ as the
corresponding stress and to introduce an extra scalar field $\alpha^{\tau}(\tau
,\vec \sigma )$ so that $T=const. U^{\check A}\partial_{\check A}\alpha^{\tau}$.
The potential $\alpha^{\tau}(\tau ,\vec \sigma )$ has the microscopic
interpretation as the retardation of the proper time of the molecules with
respect to the physical time calculated over averaged spacetime trajectories
of the idealized continuum material.

In this case the action principle becomes $S=-\int d\tau d^3\sigma \Big[ N
\sqrt{\gamma} n f(G,T)\Big] (\tau ,\vec \sigma )$ and one gets the conserved
energy-momentum tensor $T_{\check A\check B}=N \sqrt{\gamma} n [(f+TS)U_{\check
A}U_{\check B}+Z_{\check A\check B}]$.

In all these cases one can develop the rest-frame instant form just in
the same way as it was done in Section II and III for perfect fluids,
even if it is not possible to obtain a closed form of the invariant
mass.

\vfill\eject

\section{Conclusions.}

In this paper we have studied the Hamiltonian description in Minkowski
spacetime associated with an action principle for perfect fluids with
an equation of state of the form $\rho =\rho (n,s)$ given in
Ref.\cite{brown}, in which the fluid is descrbed only in terms of
Lagrangian coordinates.

This action principle can be reformulated on arbitrary spacelike
hypersurfaces embedded in Minkowski spacetime (covariant 3+1 splitting
of Minkowski spacetime) along the lines of Refs.\cite{lus,india}. At
the Hamiltonian leve the canonical Hamiltonian vanishes and the theory
is governed by four first class constraints ${\cal H}^{\mu}(\tau ,\vec
\sigma )\approx 0$ implying the independence of the description from
the choice of the 3+1 splitting of Minkowski spacetime.

These constraints can be obtained in closed form only for the `dust'
and for the `photon gas'. For other types of perfect fluids one needs
numerical calculations. After the inclusion of the coupling to the
gravitational field one could begin to think to formulate Hamiltonian
numerical gravity with only physical degrees of freedom and hyperbolic
Hamilton equations for them [like the form (\ref{II50}) of the
relativistic Euler equations for the dust].

After the canonical reduction to 3+1 splittings whose leaves are
spacelike hyperplanes, we consider all the configurations of the
perfect fluid whose conserved 4-momentum is timelike. For each of
these configurations we can select the special foliation of Minkowski
spacetime with spacelike hyperplanes orthogonal to the 4-momentum of
the configuration,

This gives rise to the ``Wigner-covariant rest-frame instant form of
dynamics" \cite{lus,india} for the perfect fluids. After a discussion
of the ``external" and ``internal" centers of mass and realizations of
the Poincar\'e algebra, rest-frame Dixon's Cartesian
multipoles\cite{dixon} of the perfect fluid are studied.

It is also shown that the formulation of non-dissipative elastic
materials of Ref.\cite{magli}, based on the use of Lagrangian
coordinates, allows to get the rest-frame instant form for these
materials too.

Finally it is shown how to make the coupling to the gravitational
field by giving the ADM action for the perfect fluid in tetrad
gravity. Now it becomes possible to study the canonical reduction of
tetrad gravity with the perfect fluids as matter along the lines of
Refs.\cite{russo1,russo2,russo3}.

\vfill\eject

\appendix

\section{Relativistic Perfect Fluids.}

As in Ref.\cite{brown} let us consider a perfect fluid in a curved
spacetime $M^4$ with unit 4-velocity vector field $U^{\mu}(z)$,
Lagrangian coordinates ${\tilde \alpha}^i(z)$, particle number density
$n(z)$, energy density $\rho (z)$, entropy per particle $s(z)$,
pressure $p(z)$, temperature $T(z)$. Let $J^{\mu}(z)=\sqrt{{}^4g(z)}
n(z) U^{\mu}(z)$ the densitized particle number flux vector field, so
that we have $n=\sqrt{\epsilon \, {}^4g_{\mu\nu} J^{\mu}J^{\nu}}
/\sqrt{{}^4g}$. Other local thermodynamical variables are the chemical
potential or specific enthalpy (the energy per particle required to
inject a small amount of fluid into a fluid sample, keeping the sample
volume and the entropy per particle $s$ constant)

\beq
\mu = {1\over n} (\rho + p),
\label{a1}
\eeq

\noindent the physical free energy (the injection energy at a constant number
density n and constant total entropy)

\beq
a = {{\rho}\over n} - T s,
\label{a2}
\eeq

\noindent and the chemical free energy (the injection energy at constant volume
and constant total entropy)

\beq
f = {1\over n}(\rho + p)- T s= \mu -T s.
\label{a3}
\eeq

Since the local expression of the first law of thermodynamics is

\beq
d\rho =\mu d n +nT ds,\quad or\quad dp=n d\mu -nT ds,\quad or\quad
d(na)=f dn -ns dT,
\label{a4}
\eeq

\noindent an equation of state for a perfect fluid may be given in one of the
following forms

\beq
\rho = \rho (n,s),\quad or\quad p=p(\mu ,s),\quad or\quad a=a(n,T).
\label{a5}
\eeq

By definition, the stress-energy-momentum tensor for a perfect fluid is

\beq
T^{\mu\nu}=-\epsilon \, \rho U^{\mu}U^{\nu} + p  ({}^4g^{\mu\nu}
-\epsilon U^{\mu}U^{\nu})=-\epsilon (\rho +p) U^{\mu}U^{\nu}+
p\,\,{}^4g^{\mu\nu},
\label{a6}
\eeq

\noindent and its equations of motion are

\beq
T^{\mu\nu}{}_{;\nu}=0,\quad\quad (nU^{\mu})_{;\mu}= {1\over {\sqrt{{}^4g}}}
\partial_{\mu} J^{\mu}=0.
\label{a7}
\eeq

As shown in Ref.\cite{brown} an action functional for a perfect fluid
depending upon $J^{\mu}(z)$, ${}^4g_{\mu\nu}(z)$, $s(z)$ and ${\tilde
\alpha}^i(z)$ requires the introduction of the following Lagrange
multipliers to implement all the required properties:\hfill\break
\hfill\break
i) $\theta (z)$: it is a scalar field named `thermasy'; it is
interpreted as a potential for the fluid temperature $T={1\over n}
{{\partial \rho}\over {\partial s}}{|}_n$.  In the Lagrangian it is
interpreted as a Lagrange multiplier for implementing the ``entropy
exchange constraint" $(sJ^{\mu})
_{,\mu}=0$.\hfill\break
\hfill\break
ii) $\varphi (z)$: it is a scalar field; it is interpreted as a
potential for the chemical free energy $f$. In the Lagrangian it is
interpreted as a Lagrange multipliers for the ``particle number
conservation constraint" $J^{\mu} {}_{,\mu}=0$.\hfill\break
\hfill\break
iii) $\beta_i(z)$: they are three scalar fields; in the Lagrangian
they are interpreted as Lagrange multipliers for the ``constraint"
${\tilde
\alpha}^i{}_{,\mu} J^{\mu}=0$ that restricts the fluid 4-velocity
vector to be directed along the flow lines ${\tilde \alpha}^i=const.$

Given an arbitrary equation of state of the type $\rho =\rho (n,s)$, the action
functional is

\begin{eqnarray}
S[{}^4g_{\mu\nu}, J^{\mu}, s, {\bf {\tilde \alpha}}, \varphi ,\theta , \beta_i]
&=& \int d^4z \{ -\sqrt{{}^4g} \rho ({{|J|}\over {\sqrt{{}^4g}}},
s)+\nonumber \\
&+&J^{\mu}[\partial_{\mu}\varphi +s\partial_{\mu}\theta +\beta_i\partial
_{\mu}{\tilde \alpha}^i] \} .
\label{a8}
\end{eqnarray}

By varying the 4-metric we get the standard stress-energy-momentum tensor

\beq
T^{\mu\nu}={2\over {\sqrt{{}^4g}}} {{\delta S}\over {\delta \, {}^4g_{\mu\nu}}}
=-\epsilon \rho U^{\mu}U^{\nu}+ p  ({}^4g^{\mu\nu}-\epsilon U^{\mu}U^{\nu})=
-\epsilon (\rho +p)U^{\mu}U^{\nu}+ p\, {}^4g^{\mu\nu},
\label{a9}
\eeq

\noindent where the pressure is given by

\beq
p = n {{\partial \rho}\over {\partial n}}{|}_s - \rho .
\label{a10}
\eeq

The Euler-Lagrange equations for the fluid motion are

\bea
{{\delta S}\over {\delta J^{\mu}}}&=& \mu U_{\mu} +\partial_{\mu}\varphi +s
\partial_{\mu} \theta +\beta_i \partial_{\mu} {\tilde \alpha}^i=0,\nonumber \\
{{\delta S}\over {\delta \varphi}}&=&-\partial_{\mu} J^{\mu}=0,\nonumber \\
{{\delta S}\over {\delta \theta}}&=&-\partial_{\mu}(s J^{\mu})=0,\nonumber \\
{{\delta S}\over {\delta s}}&=&-\sqrt{{}^4g} {{\partial \rho}\over {\partial s}}
+J^{\mu} \partial_{\mu} \theta =0,\nonumber \\
{{\delta S}\over {\delta {\tilde \alpha}^i}}&=&-\partial_{\mu} (\beta_i
J^{\mu})=0,\nonumber \\
{{\delta S}\over {\delta \beta_i}}&=& J^{\mu} \partial_{\mu}{\tilde \alpha}^i=0.
\label{a11}
\eea

The second equation is the particle number conservation, the third one
the entropy exchange constraint and the last one restricts the fluid
4-velocity vector to be directed along the flow lines ${\tilde
\alpha}^i=const.$. The first equation gives the Clebsch or
velocity-potential representation of the 4-velocity $U_{\mu}$ (the
scalar fields in this representation are called Clebsch or velocity
potentials). The fifth equations imply the constancy of the
$\beta_i$'s along the fluid flow lines, so that these Lagrange
multipliers can be expressed as a function of the Lagrangian
coordinates. The fourth equation, after a comparison with the first
law of thermodynamics, leads to the identification
$T=U^{\mu}\partial_{\mu} \theta={1\over n} {{\partial \rho}\over
{\partial s}}{|}_n$ for the fluid temperature.

Moreover, one can show that the Euler-Lagrange equations imply the
conservation of the stress-energy-momentum tensor $T^{\mu\nu}{}_{;\nu}=0$.
This equations can be split in the projection along the fluid flow lines and in
the one orthogonal to them:\hfill\break
\hfill\break
i) The projection along the fluid flow lines plus the particle number
conservation give $U_{\mu} T^{\mu\nu}{}_{;\nu}=- {{\partial \rho}\over
{\partial s}} U^{\mu}\partial_{\mu}s=0$, which is verified due to the
entropy exchange constraint. Therefore, the fluid flow is locally
adiabatic, that is the entropy per particle along the fluid flow lines
is conserved.\hfill\break
\hfill\break
ii) The projection orthogonal to the fluid flow lines gives the Euler equations,
relating the fluid acceleration to the gradient of pressure

\beq
({}^4g_{\mu\nu}-\epsilon U_{\mu}U_{\nu})
T^{\nu\alpha}{}_{;\alpha}=-\epsilon (\rho +p) U_{\mu
;\nu}U^{\nu}- (\delta^{\nu}_{\mu}-\epsilon
U_{\mu}U^{\nu})\partial_{\nu} p.
\label{a12}
\eeq

\noindent By using $p=n{{\partial \rho}\over {\partial n}}{|}_s -\rho$, it is
shown in Ref.\cite{brown} that these equations can be rewritten as

\beq
2(\mu U_{[\mu})_{;\nu ]} U^{\nu}= -\epsilon (\delta^{\nu}_{\mu}-
U_{\mu}U^{\nu}) {1\over n} {{\partial
\rho}\over {\partial s}}{|}_n \partial_{\nu}s.
\label{a13}
\eeq

The use of the entropy exchange constraint allows the rewrite the
equations in the form

\beq
2 V_{[\mu ;\nu ]} U^{\nu} = T \partial_{\mu} s,
\label{a14}
\eeq

\noindent
where $V_{\mu}= \mu U_{\mu}$ is the Taub current (important for the
description of circulation and vorticity), which can be identified
with the 4-momentum per particle of a small amount of fluid to be
injected in a larger sample of fluid without changing the total fluid
volume or the entropy per particle. Now from the Euler-Lagrange we get

\beq
2 V_{[\mu ;\nu ]} U^{\nu} =-2 (\partial_{[\mu}\varphi
+s\partial_{[\mu}\theta +\beta_i \partial_{[\mu}{\tilde
\alpha}^i)_{;\nu ]} U^{\nu}= (s\partial
_{[\mu}\theta)_{;\nu ]} U^{\nu}= T \partial_{\mu}s,
\label{a15}
\eeq

\noindent
and this result implies the validity of the Euler equations.

In the non-relativistic limit $(nU^{\mu})_{;\mu}=0$,
$T^{\mu\nu}{}_{;\nu}=0$ become the particle number (or mass)
conservation law, the entropy conservation law and the Euler-Newton
equations. See Refs.\cite{damour,asada} for the post-Newtonian
approximation.

We refer to Ref.\cite{brown} for the complete discussion.
The previous action has the advantage on other actions that the canonical
momenta conjugate to $\varphi$ and $\theta$ are the particle number density
and entropy density seen by Eulerian observers at rest in space. The action
evaluated on the solutions of the equations of motion is $\int d^4z
\sqrt{{}^4g(z)} p(z)$.

In Ref.\cite{brown} there is a study of a special class of global Noether
symmetries of this action associated with arbitrary functions $F({\bf {\tilde
\alpha}}, \beta_i, s)$. It is shown that for each $F$ there is a conservation
equation $\partial_{\mu}(FJ^{\mu})=0$ and a Noether charge $Q[F]=\int_{\Sigma}
d^3\sigma \sqrt{\gamma} n\, (\epsilon l_{\mu}U^{\mu}) F({\bf {\tilde
\alpha}}, \beta_i, s)$ [$\Sigma$ is a spacelike hypersurface with future
pointing unit normal $l_{\mu}$ and with a 3-metric with determinant
$\sqrt{\gamma}$]. For $F=1$ inside a volume V in $\Sigma$ we get the
conservation of particle number within a flow tube defined by the
bundle of flow lines contained in the volume V. The factor $\epsilon
l_{\mu}U^{\mu}$ is the relativistic `gamma factor' characterizing a
boost from the Lagrangian observers with 4-velocity $U^{\mu}$ to the
Eulerian observers with 4-velocity $l^{\mu}$; thus $ n (\epsilon
l_{\mu}U^{\mu})$ is the particle number density as seen from the
Eulerian observers. These symmetries describe the changes of
Lagrangian coordinates ${\tilde \alpha}^i$ and the fact that both the
Lagrange multipliers $\varphi$ and $\theta$ are constant along each
flow line (so that it is possible to transform any solution to the
fluid equations of motion into a solution with $\varphi =\theta
=0$ on any given spacelike hypersurface).

However, the Hamiltonian formulation associated with this action is
not trivial, because the many redundant variables present in it give
rise to many first and second class constraints. In particular we
get:\hfill\break
 1) second class constraints:\hfill\break
 $\quad\quad$ A) $\pi_{J^{\tau}}\approx 0$, $J^{\tau}-\pi_{\varphi}\approx
0$;\hfill\break
 $\quad\quad$ B) $\pi_s\approx 0$, $sJ^{\tau}-\pi_{\theta}\approx
0$; \hfill\break
 $\quad\quad$ C) $\pi_{\beta_s}\approx 0$,
$\beta_sJ^{\tau}-\pi_{\alpha^s}\approx 0$.
\hfill\break
 2) first class constraints: $\pi_{J^r}\approx 0$, so that the $J^r$'s are
gauge variables.\hfill\break
 Therefore the physical variables are the
five pairs: $\varphi$, $\pi_{\varphi}$; $\theta$, $\pi_{\theta}$;
${\tilde \alpha}^i$, $\pi_{\alpha^r}$ and one could study the
associated canonical reduction.

In Ref.\cite{brown}  [see its rich bibliography for the references]
there is a systematic study of the action principles associated to the
three types of equations of state present in the literature, first by
using the Clebsch potentials and the associated Lagrange multipliers,
then only in terms of the Lagrangian coordinates by inserting the
solution of some of the Euler-Lagrange equations in the original
action and eventually by adding surface terms.

1) Equation of state $\rho =\rho (n,s)$. One has the action

\begin{equation}
S[n,U^{\mu},\varphi ,\theta ,s,{\tilde
\alpha}^r,\beta_r;{}^4g_{\mu\nu}]
=-\int d^4x \sqrt{{}^4g} \Big[ \rho (n,s)-nU^{\mu}(\partial_{\mu}
\varphi -\theta \partial_{\mu}s+\beta_r\partial_{\mu}{\tilde \alpha}^r)\Big]
\label{a16}
\end{equation}

If one knows $s=s({\tilde \alpha}^r)$ and $J^{\mu}=J^{\mu}({\tilde
\alpha}^r)=-\sqrt{{}^4g}\epsilon^{\mu\nu\rho\sigma}\partial_{\nu}
{\tilde \alpha}^1\partial_{\rho}{\tilde \alpha}^2
\partial_{\sigma}{\tilde \alpha}^3 \eta_{123}({\tilde \alpha}^r)$, one can define
$\tilde S=S-\int d^4x \partial_{\mu}[(\varphi +s\theta )J^{\mu}]$, and
one can show that it has the form

\begin{equation}
\tilde S=\tilde S[{\tilde \alpha}^r]=-\int d^4x\sqrt{{}^4g} \rho ({{|J|}
\over {\sqrt{{}^4g}}},s).
\label{a17}
\end{equation}

2) Equation of state: $p=p(\mu ,s)$ [$V^{\mu}=\mu U^{\mu}$ Taub vector]

\begin{eqnarray}
S_{(p)}&=&s_{(p)}[V^{\mu},\varphi ,\theta , s,{\tilde
\alpha}^r,\beta_r;{}^4g_{\mu\nu}]=
\nonumber \\
&=&\int d^4x\sqrt{{}^4g} \Big[ p(\mu ,s)-{{\partial p}\over {\partial
\mu}} \Big(|V|-{{V^{\mu}}\over {|V|}}(\partial_{\mu}\varphi +
s\partial_{\mu}\theta +\beta_r \partial_{\mu}{\tilde \alpha}^r)\Big)
\Big]
\label{a18}
\end{eqnarray}

\noindent or by using one of its EL equations $V_{\mu}{\buildrel \circ \over =}
\,-(\partial_{\mu}\varphi +s\partial_{\mu}\theta +\beta_r\partial_{\mu}{\tilde \alpha}^r
)$ to eliminate $V^{\mu}$ one gets Schutz's action [$\mu$ determined by $\mu^2
=-V^{\mu}V_{\mu}$]

\begin{equation}
{\tilde S}_{(p)}[\varphi ,\theta ,s,{\tilde
\alpha}^r,\beta_r;{}^4g_{\mu\nu}]
=\int d^4x\sqrt{{}^4g} p(\mu ,s)
\label{a19}
\end{equation}

3) Equation of state $a=a(n,T)$. The action is

\begin{equation}
S_{(a)}[J^{\mu},\varphi ,\theta ,{\tilde
\alpha}^r,\beta_r;{}^4g_{\mu\nu}]
=\int d^4x \Big[ |J|
a({{|J|}\over {\sqrt{{}^4g}}},\partial_{\mu}\theta J^{\mu}) -J^{\mu}(\partial
_{\mu}\varphi +\beta_r\partial_{\mu}{\tilde \alpha}^r )\Big]
\label{a20}
\end{equation}

\noindent or ${\tilde S}_{(a)}[\varphi ,\theta ,s,{\tilde \alpha}^r,
\beta_r]=S_{(a)}-\int d^4x \Big[ |J| a({{|J|}\over {\sqrt{{}^4g}}},
{{J^{\mu}}\over {|J|}}\partial_{\mu}\theta )\Big]$.

At the end of Ref.\cite{brown} there is the action for ``isentropic" fluids
and for their particular case of a ``dust" (used in Ref.\cite{brk}
as a reference fluid in canonical gravity).

The isentropic fluids have equation of state $a(n,T)={{\rho (n)}\over
n}-sT$ with $s=const.$ (constant value of the entropy per particle).
By introducing $\varphi^{'}=\varphi +s\theta$, the action can be
written in the form

\begin{equation}
S_{(isentrpic)}[J^{\mu},\varphi^{'}, {\tilde
\alpha}^r,\beta_r,{}^4g^{\mu\nu}]=\int d^4x \Big[ -\sqrt{{}^4g}
\rho ({{|J|}\over {\sqrt{{}^4g}}})+J^{\mu}
(\partial_{\mu}\varphi^{'}+\beta_r\partial_{\mu}{\tilde
\alpha}^r)\Big]
\label{a21}
\end{equation}

\noindent or

\begin{equation}
{\tilde S}_{(isentropic)}[{\tilde \alpha}^r;{}^4g_{\mu\nu}]
=-\int d^4x\sqrt{{}^4g} \rho ({{|J|}\over {\sqrt{{}^4g}}})
\label{a22}
\end{equation}

The dust has equation of state $\rho (n)=\mu n$, namely $a(n,T)=\mu
-sT$ so that we get zero pressure
$p=n{{\partial \rho}\over {\partial n}}-\rho =0$. Again with
$\varphi^{'}=\varphi +s\theta$ the action becomes

\begin{equation}
S_{(dust)}J^{\mu},\varphi^{'}, {\tilde
\alpha}^r,\beta_r,{}^4g^{\mu\nu}]
=\int d^4x \Big[ -\mu |J|+J^{\mu}(\partial_{\mu}\varphi^{'}+
\beta_r\partial_{\mu}{\tilde \alpha}^r)\Big]
\label{a23}
\end{equation}

\noindent or with $U_{\mu}=-{1\over {\mu}}(\partial_{\mu}\varphi^{'}+\beta_r
\partial_{\mu}{\tilde \alpha}^r)$  [In Ref.\cite{brk}: $M=\mu n$ rest mass (energy)
density and $T=\varphi^{'}/\mu$, $W_r=-\beta_r$, $Z^r={\tilde
\alpha}^r$; $U_{\mu}=-\partial_{\mu}T+W_r\partial_{\mu}Z^r$]

\begin{equation}
S^{'}_{(dust)}[T,Z^r,M,W_r;{}^4g_{\mu\nu}]=-{1\over 2} \int d^4x
\sqrt{{}^4g} (\mu n) \Big(U_{\mu}\, {}^4g^{\mu\nu} U_{\nu}-\epsilon \Big) ,
\label{a24}
\end{equation}

\noindent or

\begin{equation}
{\tilde S}_{(dust)}[{\tilde \alpha}^r;{}^4g_{\mu\nu}]=-\int d^4x \mu
|J|
\label{a25}
\end{equation}

In Ref.\cite{brk} there is a study of the action (\ref{a24}) since the
dust is used as a reference fluid in general relativity. At the
Hamiltonian level one gets:\hfill\break
 i) 3 pairs of second class constraints [$\pi^r_{\vec W}(\tau ,\vec
 \sigma )\approx 0$, $\pi_{\vec Z\, r}(\tau ,\vec
\sigma )-W_r(\tau ,\vec \sigma )\pi_T(\tau ,\vec \sigma )\approx 0$],
which allow the elimination  of $W_r(\tau ,\vec \sigma )$ and
$\pi^r_{\vec W}(\tau ,\vec \sigma )$;\hfill\break
 ii) a pair of second class constraints [
$\pi_M(\tau ,\vec \sigma )\approx 0$ plus the secondary $M(\tau ,\vec
\sigma )- {{\pi^2_T}\over {\sqrt{\gamma}
\sqrt{\pi_T^2+{}^3g^{rs}(\pi_T\partial_rT+
\pi_{\vec Z\, u}\partial_rZ^u)(\pi_T\partial_sT+\pi_{\vec Z\, v}\partial_sZ^v)}
}}(\tau ,\vec \sigma ) \approx 0$], which allow the  elimination of
$M, \pi_M$.

\vfill\eject

\section{Covariant Relativistic Thermodynamics of Equilibrium and
Non-Equilibrium.}

In this Appendix we shall collect some results on relativistic fluids which
are well known but scattered in the specialized literature. We shall use
essentially Ref.\cite{israel}, which has to be consulted for the relevant
bibliography. See also Ref.\cite{roy}.

Firstly we remind some notions of covariant thermodynamics of equilibrium.

Let us remember that given the stress-energy-momentum tensor of a continuous
medium $T^{\mu\nu}$, the densities of energy and momentum are $T^{oo}$ and
$c^{-1} T^{ro}$ respectively [so that $dP^{\mu}=c^{-1} \eta T^{\mu\nu} d\Sigma
_{\nu}$ is the 4-momentum that crosses the 3-area element $d\Sigma_{\nu}$ in
the sense of its normal ($\eta =-1$ if the normal is spacelike, $\eta =+1$ if
it is timelike)]; instead, $cT^{or}$ is the energy flux in the positive $r$
direction, while $T^{rs}$ is the $r$ component of the stress in the plane
perpendicular to the $s$ direction (a pressure, if it is positive). A local
observer with timelike 4-velocity $u^{\mu}$ ($u^2=\epsilon c^2$) will measure
energy density $c^{-2} T^{\mu\nu}u_{\mu}u_{\nu}$ and energy flux $\epsilon
T^{\mu\nu}u_{\mu}n_{\nu}$ along the direction of a unit vector $n^{\mu}$ in
his rest frame.

For a fluid at thermal equilibrium with $T^{\mu\nu}=\rho
U^{\mu}U^{\nu}-\epsilon {p\over {c^2}} ({}^4g^{\mu\nu}-\epsilon
U^{\mu}U^{\nu})$ [$U^{\mu}$ is the hydrodynamical 4-velocity of the
fluid]  with particle number density $n$, specific volume $V={1\over
n}$ and entropy per particle $s={{k_BS} \over n}$ ($k_B$ is
Boltzmann's constant) in its rest frame, the energy density is

\beq
\rho c^2 = n (mc^2+e),
\label{b1}
\eeq

\noindent
where $e$ is the mean internal (thermal plus chemical) energy per particle and
$m$ is particle's rest mass.

From a non-relativistic point of view, by writing the equation of
state in the form $s=s(e,V)$ the temperature and the pressure emerge
as partial derivatives from the first law of thermodynamics in the
form (Gibbs equation)

\beq
d s(e,V)={1\over T} (d e +p dV).
\label{b2}
\eeq

If $\mu_{clas}=e+pV-Ts$ is the non-relativistic chemical potential per
particle, its relativistic version is

\beq
\mu^{'}=mc^2+\mu_{class}=\mu -Ts,
\label{b3}
\eeq

\noindent
[$\mu = {{\rho c^2+p}\over n}$ is the specific enthalpy, also called
chemical potential as in Appendix A] and we get

\bea
\mu^{'} n &=& \rho c^2 + p -n T s = \rho c^2 + p -k_B T S,\nonumber \\
 k_B T dS &=& d(\rho c^2) -\mu^{'} d n =d(\rho c^2) -(\mu -Ts) d n,\quad\quad
or\nonumber \\
 d(\rho c^2) &=&\mu^{'} d n + T d (n s) =\mu d n + n T d s.
\label{b4}
\eea

By introducing the ``thermal potential" $\alpha ={{\mu^{'}}\over
{k_BT}}= {{\mu -Ts}\over {k_BT}}$ and the inverse temperature $\beta
={{c^2}\over {k_BT}}$, these two equations take the form

\bea
S&=&{{n s}\over {k_B}}= \beta (\rho + {p\over {c^2}})-\alpha
n,\nonumber \\
 dS&=&\beta d \rho -\alpha d n.
 \label{b5}
 \eea

Let us remark that in Refs.\cite{damour,smarr,carter} one uses
different notations, some of which are given in the following equation
(in Ref. \cite{damour} $\rho$ is denoted $e$ and $\rho^{'}$ is denoted
$r$)

\beq
\rho +{p\over {c^2}}=n(mc^2+e)+{p\over {c^2}}=\rho^{'} h =\rho^{'} (c^2+e^{'}
+{p\over {c^2\rho^{'}}})=\rho^{'} (c^2+h^{'}),
\label{b6}
\eeq

\noindent
where $\rho^{'}=nm$ is the rest-mass density
[$r_{*}=\sqrt{{}^4g}\rho^{'}$ is called the coordinate rest-mass
density] and $e^{'}=e/m$ is the specific internal energy [so that
$\rho^{'} h$ is the ``effective inertial mass of the fluid; in the
post-Newtonian approximation of Ref.\cite{damour} it is shown that
$\sigma =c^{-2}(T^{oo}+\sum_sT^{ss})+O(c^{-4})=c^{-2} \sqrt{{}^4g}
(-T^o_o+T^s_s)+O(c^{-4})$ has the interpretation of equality of the
``passive" and the ``active" gravitational mass]. For the specific
enthalpy or chemical potential we get [$\mu /m=h=c^2+h^{'}$ is called
enthalpy]

\beq
\mu ={1\over n} (\rho +{p\over {c^2}})={m\over {\rho^{'}}}(\rho +{p\over
{c^2}}) = m h =m (c^2+h^{'}).
\label{b7}
\eeq

See Ref.\cite{smarr} for a richer table of conversion of notations.

Relativistically, we must consider, besides the stress-energy-momentum
tensor $T^{\mu\nu}$ and the associated 4-momentum $P^{\mu}=\int_V
d^3\Sigma_{\nu} T^{\mu\nu}$, a particle flux density $n^{\mu}$ (one
$n^{\mu}_a$ for each constituent $a$ of the system) and the entropy
flux density $s^{\mu}$. At thermal equilibrium all these a priori
unrelated 4-vectors must all be parallel to the hydrodynamical
4-velocity

\beq
n^{\mu}=nU^{\mu}\quad\quad,\quad\quad s^{\mu}=sU^{\mu},\quad\quad
P^{\mu}=P U^{\mu},
\label{b8}
\eeq

Analogously, we have $V^{\mu}=VU^{\mu}$ ($V=1/n$ is the specific
volume), $\beta^{\mu}=\beta U^{\mu}={{c^2}\over {k_BT}}U^{\mu}$ [a
related 4-vector is the equilibrium parameter 4-vector
$i^{\mu}=\mu^{'} \beta^{\mu}$].

Since $\epsilon U_{\mu} T^{\mu\nu}=\rho U^{\nu}$, we get the final
manifestly covariant form of the previous two equations (now the
hydrodynamical 4-velocity is considered as an extra thermodynamical
variable)

\bea
S^{\mu} &=& S U^{\mu} = {{ns}\over {k_B}} U^{\mu} = {p\over {c^2}}
\beta^{\mu}
- \alpha n^{\mu} -\epsilon \beta_{\nu} T^{\nu\mu},\nonumber \\
 dS^{\mu} &=& -\alpha dn^{\mu} -\epsilon \beta_{\nu}
dT^{\nu\mu}.
\label{b9}
\eea

Global thermal equilibrium imposes

\beq
\partial_{\mu} \alpha =\partial_{\mu} \beta_{\nu}+\partial_{\nu} \beta_{\mu}
=0.
\label{b10}
\eeq

As a consequence we get

\beq
d({p\over {c^2}} \beta^{\mu}) = n^{\mu} d\alpha +\epsilon T^{\nu\mu}
d\beta_{\nu},
\label{b11}
\eeq

\noindent
namely the basic variables $n^{\mu}$, $T^{\nu\mu}$ and $S^{\mu}$ can
all be generated from partial derivatives of the ``fugacity" 4-vector
(or ``thermodynamical potential")

\bea
\phi^{\mu}(\alpha ,\beta_{\lambda})&=& {p\over {c^2}} \beta^{\mu},\nonumber \\
 n^{\mu}&=&{{\partial \phi^{\mu}}\over {\partial \alpha}},\quad\quad
T^{\nu\mu}_{(mat)}={{\partial \phi^{\mu}}\over {\partial
\beta_{\nu}}},\quad\quad  S^{\mu}=\phi^{\mu}-\alpha
n^{\mu}-\beta_{\nu}T^{\nu\mu}_{(mat)},
\label{b12}
\eea

\noindent
once the equation of state is known. Here $T^{\nu\mu}_{(mat)}$ is the canonical
or material (in general non symmetric) stress tensor, ensuring that reversible
flows of field energy are not accompanied by an entropy flux.

This final form remains valid (at least to first order in deviations) for
``states that deviate from equilibrium", when the 4-vectors $S^{\mu}$,
$n^{\mu}$,... are no more parallel; the extra information in this equation is
precisely the standard linear relation between entropy flux and heat flux. The
second law of thermodynamics for relativistic systems is $\partial_{\mu}
S^{\mu} \geq 0$, which becomes a strict equality in equilibrium.

The fugacity 4-vector $\phi^{\mu}$ is evaluated by using the covariant
relativistic statistical theory for thermal equilibrium \cite{israel}
starting from a grand canonical ensemble with density matrix $\hat
\rho$ by maximizing the entropy $S=- Tr (\hat \rho ln\, \hat \rho )$
subject to the constraints $Tr \hat \rho =1$, $Tr (\hat \rho \hat
n)=n$, $Tr (\hat \rho {\hat P}^{\lambda})=P^{\lambda}$: this gives (in
the large volume limit)

\bea
\hat \rho &=&Z^{-1} e^{\alpha \hat n +\beta_{\mu} {\hat P}^{\mu}},\nonumber \\
 &&with \nonumber \\
 ln\, Z &=&\int_{\triangle \Sigma} \epsilon \phi^{\mu}
d\Sigma_{\mu},\quad\quad n=\int_{\triangle \Sigma} \epsilon n^{\mu}
d\Sigma_{\mu},\quad\quad P^{\mu}=\int_{\triangle \Sigma} \epsilon
T^{\mu\nu}_{(mat)} d\Sigma_{\nu},
\label{b13}
\eea

\noindent
[it is assumed that the members of the ensemble are small (macroscopic)
subregions of one extended body in thermal equilibrium, whose worldtubes
intersect an arbitrary spacelike hypersurface in small 3-areas $\triangle
\Sigma$]. Therefore, one has to find the grand canonical partition function

\beq
Z(V_{\mu},\beta_{\mu},i_{\mu})= \sum_n e^{i_{\mu}n^{\mu}}
Q_n(V_{\mu},\beta
_{\mu}),\quad\quad where\quad  Q_n(V_{\mu},\beta_{\mu})=\int_{V_{\mu}} d\sigma_n(q,p)
e^{-\beta_{\mu}P^{\mu}},
\label{b14}
\eeq

\noindent
is the canonical partition function for fixed volume $V_{\mu}$ and
$d\sigma_n$ is the invariant microcanonical density of states. For an
ideal Boltzmann gas of N free particles of mass $m$ [see Section III
for its equation of state] it is

\beq
d\sigma_n(p,m)={1\over {N!}} \int \delta^4(P-\sum_{i=1}^Np_i)
\prod_{i=1}
^N 2V_{\mu}p_i^{\mu} \theta (p_i^o) \delta (p_i^2-\epsilon m^2) d^4p_i.
\label{b15}
\eeq

Following Ref.\cite{karsch} [using a certain type of gauge fixings to
the first class constraints $p^2_i-\epsilon m^2\approx 0$] in
Ref.\cite{lus} $Q_n$ was evaluated in the rest-frame instant form on
the Wigner hyperplane (this method can be extended to a gas of
molecules, which are N-body bound states):

\beq
Q_n={1\over {N!}} \Big[ {{Vm^2}\over {2\pi^2\beta}} K_2(m\beta)
\Big]^N.
\label{b16}
\eeq

The same results may be obtained by starting from the covariant
relativistic kinetic theory of gas [see Ref.\cite{stewart,hakim}; in
Ref.\cite{israel} there is a short review] whose particles interact
only by collisions by using Synge's invariant distribution function
$N(q,p)$\cite{synge} [the number of particle worldlines with momenta
in the range $(p_{\mu},d\omega )$ that cross a target 3-area
$d\Sigma_{\mu}$ in $M^4$ in the direction of its normal is given by
$dN=N(q,p) d\omega \eta v^{\mu}d\Sigma_{\mu}$ ($=Nd^3qd^3p$ for the
3-space $q^o=const.$); $d\omega =d^3p/v^o \sqrt{{}^4g}$ is the
invariant element of 3-area on the mass-shell]. One arrives at a
transport equation for $N$,${{dN}\over {d\tau}}=\nabla_{\mu}
(Nv^{\mu})$ [$v^{\mu}$ is the particle velocity obtained from the
Hamilton equation implied by the one-particle Hamiltonian
$H=\sqrt{\epsilon \, {}^4g
^{\mu\nu}(q)p_{\mu}p_{\nu}}=m$ (it is the energy after the gauge fixing $q^o
\approx \tau$ to the first class contraint ${}^4g^{\mu\nu}(q)p_{\mu}p_{\nu}-
\epsilon m^2\approx 0$); $\nabla_{\mu}$ is the covariant gradient holding the
4-vector $p_{\mu}$ (not its components) fixed]
 with a ``collision term" $C[N]$ describing the collisions; for a dilute
simple gas dominated by binary collisions one arrives at the Boltzmann
equation [for $C[N]=0$ one solution is the relativistic version of the
Maxwell-Boltzmann distribution function, i.e. the classical
J\"uttner-Synge one $N= const. e^{-\beta_{\mu}P^{\mu}}/4\pi m^2
K_2(m\beta)$ for the Boltzmann gas\cite{synge}]. The H-theorem
[$\nabla_{\mu} S^{\mu} \geq 0$, where $S^{\mu}(q)=-\int [N ln
(Nh^3)-N]v^{\mu}d\omega$ is the entropy flux] and the results at
thermal equilibrium emerge [from the balance law $\nabla_{\mu}(\int N
f v^{\mu} d\omega )=\int f C[N] d\omega$ ($f$ is an arbitrary
tensorial function) one can deduce the conservation laws
$\nabla_{\mu}n^{\mu}=\nabla_{\mu}T^{\nu\mu}=0$, where $n^{\mu}=\int N
v^{\mu} d\omega$, $T_{\nu}{}^{\mu}=\int N p_{\nu}v^{\mu} d\omega$; the
vanishing of entropy production at local thermal equilibrium gives
$N_{eq}(q,p)=h^{-3}e^{\alpha (q)+\beta_{\nu}(q)P^{\nu}}$ in the case
of Boltzmann statistic and one gets ($U^{\mu}=\beta^{\mu}/\beta$)
$n^{\mu}_{eq}=\int N_{eq}v^{\mu}d\omega = n U^{\mu}$,
$T^{\nu\mu}_{eq}=\rho U^{\nu}U^{\mu}-\epsilon p({}^4g^{\nu\mu}-
\epsilon U^{\nu}U^{\mu})$, $S^{\mu}_{eq}=p\beta^{\mu}-\alpha n^{\mu}_{eq}-\beta
_{\nu}T^{\mu\nu}_{eq}$; one obtains the equations for the Boltzmann ideal gas
given in Section III].

One can study the small deviations from thermal equilibrium
[$N=N_{eq}(1+f)$, where $N_{eq}$ is an arbitrary local equilibrium
distribution] with the linearized Boltzmann equation and then by using
either the Chapman-Enskog ansatz of quasi-stationarity of small
deviations (this ignores the gradients of $f$ and gives the standard
Landau-Lifshitz and Eckart phenomenological laws; one gets Fourier
equation for heat conduction and the Navier-Stokes equation for the
bulk and shear stresses; however one has parabolic and not hyperbolic
equations implying non-causal propagation) or with the Grad method in
the 14-moment approximation. This method retains the gradients of $f$
[there are 5 extra thermodynamical variables, which can be explicitly
determined from 14 moments among the infinite set of moments $\int
Np^{\mu}p^{\nu}p^{\rho}...d^4p$ of kinetic theory; no extra auxiliary
state variables are introduced to specify a non-equilibrium state
besides $T^{\mu\nu}$, $n^{\mu}$, $S^{\mu}$] and gives phenomenological
laws which are the kinetic equivalent of M\"uller extended
thermodynamics and its various developments; now the equations are
hyperbolic, there is no causality problem but there are problems with
shock waves. See Ref.\cite{israel} for the bibliography and for a
review of the non-equilibrium phenomenological laws (see also Ref.
\cite{wee}) of Eckart, Landau-Lifshitz, of the various formulations of
extended thermodynamics, of non-local thermodynamics.

While in Ref.\cite{lind} it is said that the difference between causal
hyperbolic theories and acausual parabolic one is unobservable, in Ref.
\cite{anil}[see also Ref.\cite{an}] there is a discussion of the cases in
which hyperbolic theories are relevant. See also the numerical codes of
Refs.\cite{num,num1,num2}.

In phenomenological theories the starting point are the equations $\partial
_{\mu}T^{\mu\nu}=\partial_{\mu}n^{\mu}=0$, $\partial_{\mu}S^{\mu} \geq 0$.
There is the problem of how to define a 4-velocity and a rest-frame for a
given non-equilibrium state. Another problem is how to specify a
non-equilibrium
state completely at the macroscopic level: a priori one could need an infinite
number of auxiliary quantities (vanishing at equilibrium) and an equation of
state depending on them. The basic postulate of extended thermodynamics is the
absence of such variables.

Regarding the rest frame problem there are two main solutions in the
literature connected with the relativistic description of ``heat
flow":

i) Eckart theory. One considers a local observer in a simple fluid who
is at rest with respect to the average motion of the particles: its
4-velocity $U^{\mu}_{(eck)}$ is parallel by definition to the particle
flux $n^{\mu}$, namely

\beq
n^{\mu}= n_{(eck)} U^{\mu}_{(eck)}.
\label{b17}
\eeq

This local observer sees ``heat flow" as a flux of energy in his rest
frame:

\bea
\epsilon U_{(eck)\mu} T^{\mu\nu} &=& \rho_{(eck)} U^{\mu}_{(eck)}+q^{\mu}
_{(eck)},\nonumber \\
 &&\quad\quad so\, that\, we\, get\nonumber \\
 T^{\mu\nu} &=& \rho_{(eck)} U^{\mu}_{(eck)}U^{\nu}_{(eck)} +
q^{\mu}_{(eck)} U^{\nu}_{(eck)}+U^{\mu}_{(eck)} q_{(eck)}^{\nu}
+P^{\mu\nu}_{(eck)},\nonumber \\
 &&{}\nonumber \\
 P^{\mu\nu}_{(eck)}&=&P^{\nu\mu}_{(eck)}= \epsilon
(p+\pi_{(eck)})({}^4g^{\mu\nu}-
\epsilon U^{\mu}_{(eck)}U^{\nu}_{(eck)}+\pi^{\mu\nu}_{(eck)},\nonumber \\
 &&{}\nonumber \\
 &&P^{\mu\nu}_{(eck)} U_{(eck)\nu}= q_{(eck)\mu}U^{\mu}_{(eck)}=0,
\quad\quad \pi_{(eck)\mu\nu}({}^4g^{\mu\nu}-\epsilon U^{\mu}_{(eck)}U^{\nu}
_{(eck)})=0,
\label{b18}
\eea

\noindent
where $p$ is the thermodynamic pressure, $\pi_{(eck)}$ the bulk viscosity and
$\pi^{\mu\nu}_{(eck)}$ the shear stress.

This description has the particle conservation law $\partial_{\mu}n^{\mu}=0$.

ii) Landau-Lifshitz theory. One considers a different observer
(drifting slowly in the direction of heat flow with a 3-velocity
${\vec v}_D=\vec q /nmc^2$) whose 4-velocity $U^{\mu}_{(ll)}$ is by
definition such to give a vanishing ``heat flow", i.e. there is no net
energy flux in his rest frame: $U^{\mu}_{(ll)} T^{\nu}_{\mu}
n_{\nu}=0$ for all vectors $n_{\mu}$ orthogonal to $U^{\mu}_{(ll)}$.
This implies that $U^{\mu}_{(ll)}$ is the timelike eigenvector of
$T^{\mu\nu}$, $T^{\mu\nu}U_{(ll)\nu}=\epsilon \rho_{(ll)}
U^{\mu}_{(ll)}$, which is unique if $T^{\mu\nu}$ satisfies a positive
energy condition. Now we get

\bea
T^{\mu\nu}&=& \rho_{(ll)}
U^{\mu}_{(ll)}U^{\nu}_{(ll)}+P^{\mu\nu}_{(ll)},\nonumber \\
 &&{}\nonumber \\
 P^{\mu\nu}_{(ll)}&=&P^{\nu\mu}_{(ll)}= \epsilon
(p+\pi_{(ll)})({}^4g^{\mu\nu}-
\epsilon U^{\mu}_{(ll)}U^{\nu}_{(ll)}+\pi^{\mu\nu}_{(ll)},\nonumber \\
 &&{}\nonumber \\
 &&P^{\mu\nu}_{(ll)} U_{(ll)\nu}= 0,
\quad\quad \pi_{(ll)\mu\nu}({}^4g^{\mu\nu}-\epsilon U^{\mu}_{(ll)}U^{\nu}
_{(ll)})=0,\nonumber \\
 &&{}\nonumber \\
 n^{\mu}&=&n_{(ll)} U^{\mu}_{(ll)}+j_{(ll)}^{\mu},\quad\quad
j_{(ll)\mu} U^{\mu}_{(ll)}=0\quad\quad (\vec j=-n{\vec v}_D=-\vec
q/mc^2).
\label{b19}
\eea

This observer in his rest frame does not see a heat flow but a particle drift.
This description has the simplest form of the energy-momentum tensor.

One has $n_{(eck)}=n_{(ll)} ch\, \varphi$, $\rho_{(eck)}=\rho_{(ll)} ch^2\,
\varphi +p_{(ll)} sh^2\, \varphi =\pi^{\mu\nu} j_{\mu}j_{\nu}/n^2_{(eck)}$,
with $ch\, \varphi = U^{\mu}_{(ll)} U_{(eck)\mu}$ [the difference is a
Lorentz factor $\sqrt{1-{\vec v}_D^2/c^2}$, so that there are
insignificant differences for many practical purposes if deviations
from equilibrium are small]. The angle $\varphi \approx j/n \approx
v_D/c \approx q/nmc^2$ is a dimensionless measure of the deviation
from equilibrium [$n_{(ll)}-n_{(eck)}$ and $\rho
_{(ll)}-\rho_{(eck)}$ are of order $\varphi^2$].

One can decompose $T^{\mu\nu}$, $n^{\mu}$, in terms of any 4-velocity
$U^{\mu}$ that falls within a cone of angle $\approx \varphi$
containing $U^{\mu}_{(eck)}$ and $U^{\mu}_{(ll)}$; each choice
$U^{\mu}$ gives a particle density $n(U)=
\epsilon u_{\mu}n^{\mu}$ and energy density $\rho (U)=U_{\mu}U_{\nu}T^{\mu\nu}$
which are independent of $U^{\mu}$ if one neglects terms of order $\varphi^2$.
Therefore, one has:\hfill\break
\hfill\break
i) if $S_{eq}(\rho (U),n(U))$ is the equilibrium entropy density, then
$S(U)=\epsilon U_{\mu}S^{\mu}=S_{eq}+O(\varphi^2)$;\hfill\break
\hfill\break
ii) if $p(U)=-{{\partial \rho /n}\over {\partial 1/n}}{|}_{S/n}$ is the
(reversible) thermodynamical pressure defined as work done in an isentropic
expansion (off equilibrium this definition allows to sepate it from the bulk
stress $\pi (U)$ in the stress-energy-momentum tensor) and $p_{eq}$ is the
pressure at equilibrium, then $p(U)=P_{eq}(\rho (U),n(U))+O(\varphi^2)$.

By postulating that the covariant Gibbs relation remains valid for
arbitrary infinitesimal displacements $(\delta n^{\mu}, \delta
T^{\mu\nu},..)$ from an equilibrium state, one gets a covariant
off-equilibrium thermodynamics based on the equation

\bea
S^{\mu} &=& p(\alpha ,\beta ) \beta^{\mu}- \alpha n^{\mu} -\beta_{\nu}
T^{\mu\nu} -Q^{\mu}(\delta n^{\nu}, \delta T^{\nu\rho},..),
\nonumber \\
 \nabla_{\mu} S^{\mu}&=&-\delta n^{\mu}\partial_{\mu} \alpha -\delta T^{\mu\nu}
\nabla_{\nu}\beta_{\mu}-\nabla_{\mu} Q^{\mu} \geq 0,
\label{b20}
\eea

\noindent
with $Q^{\mu}$ of second order in the displacements and $\alpha$, $\beta_{\mu}$
arbitrary. At equilibrium one recovers $S^{\mu}_{eq}=p\beta^{\mu}-\alpha
n^{\mu}_{eq}-\beta_{\nu}T^{\mu\nu}_{eq}$, $\nabla_{\mu}S^{\mu}_{eq}=0$
[with $U^{\mu}=\beta^{\mu}/\beta$ and (for viscous heat-conducting
fluids, but not for superfluids) $\partial_{\mu}\alpha = \nabla_{\mu}\beta
_{\nu}+\nabla_{\nu}\beta_{\mu}=0$].

If we choose $\beta^{\mu}=U^{\mu}/k_BT$ parallel to $n^{\mu}$ of the given
off-equilibrium state, we are in the ``Eckart frame", $U^{\mu}=U^{\mu}
_{(eck)}$, and we get

\bea
S&=&\epsilon U_{(eck)\mu}S^{\mu}=S_{eq}+\epsilon
U_{(eck)\mu}Q^{\mu},\nonumber \\
 &&{}\nonumber \\
 \sigma^{\mu}_{(eck)}&=&({}^4g^{\mu\nu}-\epsilon U^{\mu}_{(eck)}U^{\nu}_{(eck)})
S_{\nu}=\beta q_{(eck)}^{\mu} -({}^4g^{\mu\nu}-\epsilon U^{\mu}_{(eck)}U^{\nu}
_{(eck)}) Q_{\nu},\nonumber \\
 &&{}\nonumber \\
 q_{(eck)}^{\mu}&=&-({}^4g^{\mu\nu}-\epsilon
U^{\mu}_{(eck)}U^{\nu}_{(eck)}) T_{\nu}^{\rho} U_{(eck)\rho},
\label{b21}
\eea

\noindent
so that to linear order we get the standard relation between entropy
flux ${\vec \sigma}_{(eck)}$ and heat flux ${\vec q}_{(eck)}$

\beq
{\vec \sigma}_{(eck)}={{ {\vec q}_{(eck)}}\over {k_BT}} +(possible\,\,
2nd\,\, order \,\, term).
\label{b22}
\eeq

If we choose $U^{\mu}=U^{\mu}_{(ll)}$, the timelike eigenvector of
$T^{\mu\nu}$, so that
$U_{(ll)\mu}T^{\mu}_{\nu}({}^4g_{\rho}^{\nu}-\epsilon U^{\nu}_{(ll)}
U_{(ll)\rho})=0$, we are in the ``Landau-Lifshitz frame" and we get

\bea
\sigma^{\mu}_{(ll)}&=&({}^4g^{\mu\nu}-\epsilon U^{\mu}_{(ll)}U^{\nu}_{(ll)})
S_{\nu}=-\alpha j^{\mu}_{(ll)}-({}^4g^{\mu\nu}-\epsilon U^{\mu}_{(ll)}U^{\nu}
_{(ll)}) Q_{\nu},\nonumber \\
 &&{}\nonumber \\
j^{\mu}_{(ll)}&=&({}^4g^{\mu\nu}-\epsilon
U^{\mu}_{(ll)}U^{\nu}_{(ll)}) n_{\nu},
\label{b23}
\eea

\noindent
so that at linear order we get the standard relation between entropy flux ${\vec
\sigma}_{(ll)}$ and diffusive flux ${\vec j}_{(ll)}$

\beq
{\vec \sigma}_{(ll)}=-{{\mu}\over {k_BT}} {\vec
j}_{(ll)}+(possible\,\, 2nd\,\, order\,\, term).
\label{b24}
\eeq

In the Landau-Lifschitz frame heat flow and diffusion are englobed in
the diffusive flux ${\vec j}_{(ll)}$ relative to the mean mass-energy
flow.
\hfill\break
\hfill\break
\hfill\break
The entropy inequality becomes (each term is of second order in the
deviations from local equilibrium)

\beq
0 \leq \nabla_{\mu}S^{\mu}=- \delta n^{\mu}\partial_{\mu}\alpha
-\delta T^{\mu\nu}
\nabla_{\nu}\beta_{\mu}-\nabla_{\mu}Q^{\mu},
\label{b25}
\eeq

\noindent
with the fitting conditions $\delta n^{\mu} U_{\mu}=\delta T^{\mu\nu}U_{\mu}U
_{\nu}=0$, which contain all information about the viscous stresses, heat
flow and diffusion in the off-equilibrium state (they are dependent on the
arbitrary choice of the 4-velocity $U^{\mu}$).\hfill\break
\hfill\break
Once a detailed form of $Q^{\mu}$ is specified, linear relations between
irreversible fluxes $\delta T^{\mu\nu}$, $\delta n^{\mu}$ and gradients $\nabla
_{(\mu}\beta_{\nu )}$, $\partial_{\mu}\alpha$ follow.\hfill\break
\hfill\break
A) $Q^{\mu}=0$ (like in the non-relativistic case).\hfill\break
\hfill\break
The spatial entropy flux $\vec \sigma$ is only a strictly linear function of
heat flux $\vec q$ and diffusion flux $\vec j$. In this case the off-equilibrium
entropy density $S=\epsilon U_{\mu}S^{\mu}$ is given by the equilibrium
equation of state $S=S_{eq}(\rho ,n)$. We have $0 \leq \nabla_{\mu}S^{\mu}=-
\delta n^{\mu}\partial_{\mu}\alpha -\delta T^{\mu\nu}\nabla_{\nu}\beta_{\mu}$
with fitting conditions $\delta n^{\mu}U_{\mu}=\delta T^{\mu\nu}U_{\mu}U_{\nu}
=0$ and with $U^{\mu}$ still arbitrary at first order.\hfill\break
\hfill\break
A1) Landau-Lifschitz frame and theory. $U^{\mu}=U^{\mu}_{(ll)}$ is the
timelike eigenvector of $T^{\mu\nu}$. This and the fitting conditions imply
$\delta T^{\mu\nu}U_{(ll)\nu}=0$. The shear and bulk stresses $\pi_{(ll)}
^{\mu\nu}$, $\pi_{(ll)}$ are identified by the decomposition

\beq
\delta T^{\mu\nu}=\pi^{\mu\nu}_{(ll)}+\pi_{(ll)} ({}^4g^{\mu\nu}-\epsilon
U^{\mu}_{(ll)}U^{\nu}_{(ll)}), \quad\quad
\pi^{\mu\nu}_{(ll)}U_{(ll)\nu}=
\pi_{(ll)}^{\mu}{}_{\mu}=0.
\label{b26}
\eeq

The inequality $\nabla_{\mu}S^{\mu} \geq 0$ becomes

\bea
&&-j^{\mu}_{(ll)}\partial_{\mu}\alpha -\beta \pi^{\mu\nu}_{(ll)} <
\nabla_{\nu}
\beta_{\mu} > -\beta \pi_{(ll)} \nabla_{\mu}U^{\mu}_{(ll)} \geq 0, \quad
\quad j^{\mu}_{(ll)}=\delta n^{\mu},\nonumber \\
 &&{}\nonumber \\
< X_{\mu\nu} > &=& [({}^4g^{\alpha}_{\mu}-\epsilon
U^{\alpha}_{(ll)}U_{(ll)\mu}) ({}^4g^{\beta}_{\nu}-\epsilon
U^{\beta}_{(ll)}U_{(ll)\nu})-\nonumber \\
 &-&{1\over 3}({}^4g
_{\mu\nu}-\epsilon U_{(ll)\mu}U_{(ll)\nu})({}^4g^{\alpha\beta}-\epsilon
U^{\alpha}_{(ll)}U^{\beta}_{(ll)})] X_{\alpha\beta},
\label{b27}
\eea

\noindent
[the $< .. >$ operation extracts the purely spatial, trace-free part of any
tensor].\hfill\break
\hfill\break
If the equilibrium state is isotropic (Curie's principle) and if we assume that
$(j^{\mu}_{(ll)}, \pi^{\mu\nu}_{(ll)}, \pi_{(ll)})$ are ``linear and purely
local" functions of the gradiants, $\nabla_{\mu}S^{\mu} \geq 0$ implies

\beq
j^{\mu}_{(ll)}=- \kappa ({}^4g^{\mu\nu}-\epsilon
U^{\mu}_{(ll)}U^{\nu}_{(ll)})
\partial_{\mu}\alpha, \quad \kappa > 0,
\label{b28}
\eeq

\noindent [it is a mixture of Fourier's law
of heat conduction and of Fick's law of diffusion, stemming from the
relativistic mass-energy equivalence],\hfill\break
\hfill\break
and the standard Navier-Stokes equations ($\zeta_S$, $\zeta_V$ are
shear and bulk viscosities)

\beq
\pi_{(ll)\mu\nu}=-2 \zeta_S < \nabla_{\nu}\beta_{\mu} > ,\quad\quad
\pi_{(ll)}={1\over 3} \zeta_V \nabla_{\mu}U^{\mu}_{(ll)}.
\label{b29}
\eeq

A2) Eckart frame and theory. $U^{\mu}_{(eck)}$ parellel to $n^{\mu}$.
Now we have the fitting condition $\delta n^{\mu}=0$. The heat flux
appears in the decomposition of $\delta T^{\mu\nu}$
[$a_{(eck)\mu}=U^{\nu}_{(eck)}\nabla_{\nu} U_{(eck)\mu}$ is the
4-acceleration]

\beq
\delta T^{\mu\nu}=q^{\mu}_{(eck)}U^{\nu}_{(eck)}+U^{\mu}_{(eck)}q^{\nu}_{(eck)}
+\pi^{\mu\nu}_{(eck)}+\pi_{(eck)} ({}^4g^{\mu\nu}-\epsilon
U^{\mu}_{(eck)} U^{\nu}_{(eck)}).
\label{b30}
\eeq

The inequality $\nabla_{\mu}S^{\mu} \geq 0$ becomes

\beq
q^{\mu}_{(eck)} (\partial_{\mu}\alpha -\beta a_{(eck)\mu})-\beta
(\pi^{\mu\nu}
_{(eck)}\nabla_{\nu}U_{(eck)\mu}+\pi_{(eck)}\nabla_{\mu}U^{\mu}_{(eck)})
\geq 0.
\label{b31}
\eeq

With the simplest assumption of linearity and locality, we obtain
Fourier's law of heat conduction [it is not strictly equivalent to the
Landau-Lifshitz one, because they differ by spatial gradients of the
viscous stresses and the time-derivative of the heat flux]

\beq
q^{\mu}_{(eck)}=-\kappa ({}^4g^{\mu\nu}-\epsilon
U^{\mu}_{(eck)}U^{\nu}
_{(eck)}) (\partial_{\nu} T +T \partial_{\tau}U_{(eck)\nu}),
\label{b32}
\eeq

\noindent
[the term depending on the acceleration is sometimes referred to as an
effect of the ``inertia of heat"], and the same form of the
Navier-Stokes equations for $\pi^{\mu\nu}_{(eck)}$, $\pi_{(eck)}$
(they are not strictly equivalent to the Landau-Lifshitz ones, because
they differ by gradients of the drift ${\vec v}_D
=\vec q/nmc^2$).\hfill\break
\hfill\break
For a simple fluid Fourier's law and Navier-Stokes equations (9 equations) and
the conservation laws $\nabla_{\mu}T^{\mu\nu}=\nabla_{\mu}n^{\mu}=0$ (5
equations) determine the 14 variables $T^{\mu\nu}$, $n^{\mu}$ from suitable
initial data. However, these equations are of mixed parabolic-hyperbolic-
elliptic type and, as said, one gets acausality and instability.\hfill\break
\hfill\break
Kinetic theory gives

\beq
Q^{\mu}=-{1\over 2} \int N_{eq} f^2 p^{\mu} d\omega \not= 0,
\label{b33}
\eeq

\noindent
for a gas up to second order in the deviation $(N-N_{eq})=N_{eq}f$ [$Q^{\mu}=0$
requires small gradients and quasi-stationary processes]. Two alternative
classes of phenomenological theories are\hfill\break
\hfill\break
B) Linear non-local thermodynamics (NLT).\hfill\break
\hfill\break
This theory gives a rheomorphic rather than causal description of the
phenomenological laws: transport coefficients at an event $x$ are taken to
depend, not on the entire causal past of $x$, but only on the past history of a
``comoving local fluid element". It is a linear theory restricted to small
deviations from equilibrium, which can be derived from the linearized Boltzmann
equation by projector-operator techniques (and probably inherits its causality
properties). Instead of writing $(\delta n_{\mu}(x),\delta T_{\mu\nu}(x))=
\sigma (U,T) (-\partial_{\mu}\alpha (x), -\nabla_{(\mu}\beta_{\nu )}(x))$,
this local phenomenological law is generalized to $(\delta n_{\mu}(\vec x,x^o),
\delta T_{\mu\nu}(\vec x,x^o))=\int^{\infty}_{-\infty} dx^{o {'}}
\sigma (x^o-x^{o {'}}) (-\partial_{\mu}\alpha (\vec xx^{o {'}}), -\nabla
_{(\mu}\beta_{\nu )}(\vec xx^{o {'}}))$.\hfill\break
\hfill\break
\hfill\break
C) Local non-linear extended thermodynamics (ET).\hfill\break
\hfill\break
It is more relevant for relativistic astrophysics, where correlation
and memory effects are not of primary interest and, instead, one needs
a tractable and consistent transport theory coextensive at the
macroscopic level with Boltzmann's equation. It is assumed that the
second order term $Q^{\mu}(\delta n^{\nu},\delta T^{\nu\rho},..)$ does
not depend on auxiliary variables vanishing at equilibrium: this
ansatz is the phenomenological equivalent of Grad's 14-moment
approximation in kinetic theory. These theories are called
``second-order theories" and many of them are analyzed in
Ref.\cite{ger}; when the dissipative fluxes are subject to a
conservation equation, these theories are called of causal
``divergence type" like the ones of Refs.\cite{liu,gl,calzetta}.
Another type of theory (extended irreversible thermodynamics; in
general these theories are not of divergence type) was developed in
Refs.\cite{mu,isr,is,jou}: in it there are transport equations for the
dissipative fluxes rather than conservation laws.

For small deviations one retains only the quadratic terms in the
Taylor expansion of $Q^{\mu}$ (leading to ``linear" phenomenological
laws): this implies 5 new undetermined coefficients

\beq
Q^{\mu}={1\over 2} U^{\mu} [\beta_o \pi^2+\beta_1
q^{\mu}q_{\mu}+\beta_2 \pi
^{\mu\nu}\pi_{\mu\nu}] -\alpha_o \pi q^{\mu} -\alpha_1 \pi^{\mu\nu} q_{\nu},
\label{b34}
\eeq

\noindent
with $\beta_i > 0$ from $\epsilon U_{\mu}Q^{\mu}>0$ (the $\beta_i$'s are
`relaxation times'). A first-order change of rest frame produces a second-order
change in $Q^{\mu}$ [going from the Landau-Lifshitz frame to the Eckart one,
one gets $\alpha_{(eck)i}-\alpha_{(ll)i}=\beta_{(ll)1}-\beta_{(eck)1}=[(\rho
+p)T]^{-1}$, $\beta_{(ll)0}=\beta_{(eck)o}$, $\beta_{(ll)2}=\beta_{(eck)2}$,
and the phenomenological laws are now invariant to first order].\hfill\break
\hfill\break
In the ``Eckart frame" the phenomenological laws take the form

\bea
q^{\mu}_{(eck)}&=&-\kappa T ({}^4g^{\mu\nu}-\epsilon
U^{\mu}_{(eck)}U^{\nu}
_{(eck)}) [ T^{-1}\partial_{\nu}T +a_{(eck)\nu} +\beta_{(eck)1} \partial_{\tau}
q_{(eck)\nu}-\nonumber \\
 &-&\alpha_{(eck)o}\partial_{\nu}\pi_{(eck)}-\alpha_{(eck)1}
\nabla_{\rho} \pi^{\rho}_{(eck)\nu}],\nonumber \\
 &&{}\nonumber \\
 \pi_{(eck)\mu\nu}&=&-2 \zeta_S [< \nabla_{\nu}U_{(eck)\mu}
>+\beta_{(eck)2}
\partial_{\tau}\pi_{(eck)\mu\nu}-\alpha_{(eck)1} < \nabla_{\nu} q_{(eck)\mu} >
],\nonumber \\
 &&{}\nonumber \\
 \pi_{(eck)}&=& -{1\over 3} \zeta_V [ \nabla_{\mu}
U^{\mu}_{(eck)}+\beta_{(eck)o}
\partial_{\tau}\pi_{(eck)}-\alpha_{(eck)o}\nabla_{\mu}q^{\mu}_{(eck)}],
\label{b35}
\eea

\noindent
which reduce to the equation of the standard Eckart theory if the 5
relaxation ($\beta_i$) and coupling ($\alpha_i$) coefficients are put
equal to zero. See for instance Ref.\cite{ingo} for a complete
treatment and also Ref. \cite{appl}. For appropriate values of these
coefficients these equations are hyperbolic and, therefore, causal and
stable.  The transport equations can be understood \cite{anil} as
evolution equations for the dissipative variables as they describe how
these fluxes evolve from an initial arbitrary state to a final steady
one [the time parameter $\tau$ is usually interpreted as the
relaxation time of the dissipative processes]. In the case of a gas
the new coefficients can be found explicitly \cite{is} [see also
Ref.\cite{uzan} for a recent approach to relativistic interacting
gases starting from the Boltzmann equation], and they are purely
thermodynamical functions. Wave front speeds are finite and comparable
with the speed of sound. A problem with these theories is that they do
not admit a regular shock structure (like the Navier-Stokes equations)
once the speed of the shock front exceeds the highest characteristic
velocity (a ``subshock" will form within a shock layer for speeds
exceeding the wave-front velocities of thermo-viscous effects). The
situation slowly ameliorates if more moments are taken into account
\cite{ingo}.
\hfill\break
\hfill\break
In the approach reviewed in Ref.\cite{ingo} the extra indeterminacy
associated to the new 5 coefficients is eliminated (at the price of
high non-linearity) by annexing to the usual conservation and entropy
laws a new phenomenological assumption (in this way one obtains a
causal divergence type theory):

\beq
\nabla_{\rho}A^{\mu\nu\rho}=I^{\mu\nu},
\label{b36}
\eeq

\noindent
in which $A^{\rho\mu\nu}$ and $I^{\mu\nu}$ are symmetric tensors with
the following traces

\beq
A^{\mu\nu}{}_{\nu}=-n^{\mu}, \quad\quad I^{\mu}{}_{\mu}=0.
\label{b37}
\eeq

These conditions are modelled on kinetic theory, in which $A^{\rho\mu\nu}$
represents the third moment of the distribution function in momentum space,
and $I^{\mu\nu}$ the second moment of the collision term in Boltzmann's
equation. The previous equations are central in the determination of the
distribution function in Grad's 14-moment approximation. The phenomenological
theory is completed by the postulate that the state variables $S^{\mu}$,
$A^{\rho\mu\nu}$, $I^{\mu\nu}$ are invariant functions of $T^{\mu\nu}$,
$n^{\mu}$ only.
The theory is an almost exact phenomenological counterpart of
the Grad approximation. See Ref.\cite{carter} for the beginning (only non
viscous heta conducting materials are treated) of a derivation
of extended thermodynamics from a variational principle.

Everything may be rephrased in terms of the Lagrangian
coordinates of the fluid used in this paper. What is lacking
in the non-dissipative case of heat conduction is the functional
form of the off-equilibrium equation of state reducing to $\rho =\rho (n,s)$
at thermal equilibrium. In the dissipative case the system is open and
$T^{\mu\nu}$, $P^{\mu}$, $n^{\mu}$ are not conserved.

See Ref.\cite{morri} for attempts to define a classical theory of
dissipation in the Hamiltonian framework and Ref.\cite{nose} about
Hamiltonian molecular dynamics for the addition of an extra degree of
freedom to an N-body system to transform it into an open system (with
the choice of a suitable potential for the extra variable the
equilibrium distribution function of the N-body subsystem is exactly
the canonical ensemble).

However, the most constructive procedure is to get (starting from an
action principle) the Hamiltonian form of the energy-momentum of a
closed system, like it has been done in Ref.\cite{crater} for a system
of N charged scalar particles, in which the mutual
action-at-a-distance interaction is the complete Darwin potential
extracted from the Lienard-Wiechert solution in the radiation gauge
(the interactions are momentum- and, therefore, velocity-dependent).
In this case one can define an open (in general dissipative) subsystem
by considering a cluster of $n < N$ particles and assigning to it a
non-conserved energy-momentum tensor built with all the terms of the
original energy-momentum tensor which depend on the canonical
variables of the $n$ particles (the other $N-n$ particles are
considered as external fields).

\vfill\eject

\section{Notations on spacelike hypersurfaces.}

Let us first review some preliminary results from Refs.\cite{lus}
needed in the description of physical systems on spacelike hypersurfaces.

Let $\lbrace \Sigma_{\tau}\rbrace$ be a one-parameter family of spacelike
hypersurfaces foliating Minkowski spacetime $M^4$ with 4-metric $\eta_{\mu\nu}
=\epsilon (+---)$, $\epsilon =\pm$ [$\epsilon =+1$ is the particle physics
convention; $\epsilon =-1$ the general relativity one]
and giving a 3+1 decomposition
of it. At fixed $\tau$, let
$z^{\mu}(\tau ,\vec \sigma )$ be the coordinates of the points on $\Sigma
_{\tau }$ in $M^4$, $\lbrace \vec \sigma \rbrace$ a system of coordinates on
$\Sigma_{\tau}$. If $\sigma^{\check A}=(\sigma^{\tau}=\tau ;\vec \sigma
=\lbrace \sigma^{\check r}\rbrace)$ [the notation ${\check A}=(\tau ,
{\check r})$ with ${\check r}=1,2,3$ will be used; note that ${\check A}=
\tau$ and ${\check A}={\check r}=1,2,3$ are Lorentz-scalar indices] and
$\partial_{\check A}=\partial /\partial \sigma^{\check A}$,
one can define the vierbeins

\begin{equation}
z^{\mu}_{\check A}(\tau ,\vec \sigma )=\partial_{\check A}z^{\mu}(\tau ,\vec
\sigma ),\quad\quad
\partial_{\check B}z^{\mu}_{\check A}-\partial_{\check A}z^{\mu}_{\check B}=0,
\label {c1}
\end{equation}

\noindent so that the metric on $\Sigma_{\tau}$ is

\begin{eqnarray}
&&g_{{\check A}{\check B}}(\tau ,\vec \sigma )=z^{\mu}_{\check A}(\tau ,\vec
\sigma )\eta_{\mu\nu}z^{\nu}_{\check B}(\tau ,\vec \sigma ),\quad\quad
\epsilon g_{\tau\tau}(\tau ,\vec \sigma ) > 0,\nonumber \\
&&g(\tau ,\vec \sigma )=-det\, ||\, g_{{\check A}{\check B}}(\tau ,\vec
\sigma )\, || ={(det\, ||\, z^{\mu}_{\check A}(\tau ,\vec \sigma )\, ||)}^2,
\nonumber \\
&&\gamma (\tau ,\vec \sigma )=-det\, ||\, g_{{\check r}{\check s}}(\tau ,\vec
\sigma )\, ||=det\, ||{}^3g_{\check r\check s}(\tau ,\vec \sigma )||,
\label{c2}
\end{eqnarray}

\noindent where $g_{\check r\check s}=-\epsilon \, {}^3g_{\check r\check s}$
with ${}^3g_{\check r\check s}$ having positive signature $(+++)$.

If $\gamma^{{\check r}{\check s}}(\tau ,\vec \sigma )=-\epsilon \,
{}^3g^{\check r\check s}$ is the inverse of the
3-metric $g_{{\check r}{\check s}}(\tau ,\vec \sigma )$ [$\gamma^{{\check r}
{\check u}}(\tau ,\vec \sigma )g_{{\check u}{\check s}}(\tau ,\vec
\sigma )=\delta^{\check r}_{\check s}$], the inverse $g^{{\check A}{\check B}}
(\tau ,\vec \sigma )$ of $g_{{\check A}{\check B}}(\tau ,\vec \sigma )$
[$g^{{\check A}{\check C}}(\tau ,\vec \sigma )g_{{\check c}{\check b}}(\tau ,
\vec \sigma )=\delta^{\check A}_{\check B}$] is given by

\begin{eqnarray}
&&g^{\tau\tau}(\tau ,\vec \sigma )={{\gamma (\tau ,\vec \sigma )}\over
{g(\tau ,\vec \sigma )}},\nonumber \\
&&g^{\tau {\check r}}(\tau ,\vec \sigma )=-[{{\gamma}\over g} g_{\tau {\check
u}}\gamma^{{\check u}{\check r}}](\tau ,\vec \sigma )=\epsilon
[{{\gamma}\over g} g_{\tau \check u}\, {}^3g^{\check u\check r}](\tau ,\vec
\sigma ),\nonumber \\
&&g^{{\check r}{\check s}}(\tau ,\vec \sigma )=\gamma^{{\check r}{\check s}}
(\tau ,\vec \sigma )+[{{\gamma}\over g}g_{\tau {\check u}}g_{\tau {\check v}}
\gamma^{{\check u}{\check r}}\gamma^{{\check v}{\check s}}](\tau ,\vec \sigma )
=\nonumber \\
&=&-\epsilon \, {}^3g^{\check r\check s}(\tau ,\vec \sigma )+[{{\gamma}\over g}
g_{\tau \check u}g_{\tau \check v}\, {}^3g{\check u\check r}\, {}^3g^{\check v
\check s}](\tau ,\vec \sigma ),
\label{c3}
\end{eqnarray}

\noindent so that $1=g^{\tau {\check C}}(\tau ,\vec \sigma )g_{{\check C}\tau}
(\tau ,\vec \sigma )$ is equivalent to

\begin{equation}
{{g(\tau ,\vec \sigma )}\over {\gamma (\tau ,\vec \sigma )}}=g_{\tau\tau}
(\tau ,\vec \sigma )-\gamma^{{\check r}{\check s}}(\tau ,\vec \sigma )
g_{\tau {\check r}}(\tau ,\vec \sigma )g_{\tau {\check s}}(\tau ,\vec \sigma ).
\label{c4}
\end{equation}

We have

\begin{equation}
z^{\mu}_{\tau}(\tau ,\vec \sigma )=(\sqrt{ {g\over {\gamma}} }l^{\mu}+
g_{\tau {\check r}}\gamma^{{\check r}{\check s}}z^{\mu}_{\check s})(\tau ,
\vec \sigma ),
\label{c5}
\end{equation}

\noindent and

\begin{eqnarray}
\eta^{\mu\nu}&=&z^{\mu}_{\check A}(\tau ,\vec \sigma )g^{{\check A}{\check B}}
(\tau ,\vec \sigma )z^{\nu}_{\check B}(\tau ,\vec \sigma )=\nonumber \\
&=&(l^{\mu}l^{\nu}+z^{\mu}_{\check r}\gamma^{{\check r}{\check s}}
z^{\nu}_{\check s})(\tau ,\vec \sigma ),
\label{c6}
\end{eqnarray}

\noindent where

\begin{eqnarray}
l^{\mu}(\tau ,\vec \sigma )&=&({1\over {\sqrt{\gamma}} }\epsilon^{\mu}{}_{\alpha
\beta\gamma}z^{\alpha}_{\check 1}z^{\beta}_{\check 2}z^{\gamma}_{\check 3})
(\tau ,\vec \sigma ),\nonumber \\
&&l^2(\tau ,\vec \sigma )=1,\quad\quad l_{\mu}(\tau ,\vec \sigma )z^{\mu}
_{\check r}(\tau ,\vec \sigma )=0,
\label{c7}
\end{eqnarray}

\noindent is the unit (future pointing) normal to $\Sigma_{\tau}$ at
$z^{\mu}(\tau ,\vec \sigma )$.

For the volume element in Minkowski spacetime we have

\begin{eqnarray}
d^4z&=&z^{\mu}_{\tau}(\tau ,\vec \sigma )d\tau d^3\Sigma_{\mu}=d\tau [z^{\mu}
_{\tau}(\tau ,\vec \sigma )l_{\mu}(\tau ,\vec \sigma )]\sqrt{\gamma
(\tau ,\vec \sigma )}d^3\sigma=\nonumber \\
&=&\sqrt{g(\tau ,\vec \sigma )} d\tau d^3\sigma.
\label{c8}
\end{eqnarray}

Let us remark that according to the geometrical approach of
Ref.\cite{ku},one can use Eq.(\ref{c5}) in the form \hfill\break
\hfill\break
$z^{\mu}_{\tau}(\tau ,\vec \sigma )=N(\tau ,
\vec \sigma )l^{\mu}(\tau ,\vec \sigma )+N^{\check r}(\tau ,\vec \sigma )
z^{\mu}_{\check r}(\tau ,\vec \sigma )$, \hfill\break
\hfill\break
where $N=\sqrt{g/\gamma}=\sqrt{g_{\tau\tau}
-\gamma^{{\check r}{\check s}}g_{\tau{\check r}}g_{\tau{\check s}}}
=\sqrt{g_{\tau\tau}+\epsilon
{}^3g^{{\check r}{\check s}}g_{\tau{\check r}}g_{\tau{\check s}}}$
and $N^{\check r}=g_{\tau \check s}\gamma^{\check s\check r}=-\epsilon
g_{\tau \check s}\, {}^3g^{\check s\check r}$ \hfill\break
\hfill\break
are the standard lapse and shift functions $N_{[z](flat)}$, $N^{\check r}
_{[z](flat)}$ of the Introduction, so that \hfill\break
\hfill\break
$g_{\tau \tau}=\epsilon N^2+g_{\check r\check s}N^{\check r}N^{\check s}=
\epsilon [N^2-{}^3g_{\check r\check s}N^{\check r}N^{\check s}]$, \hfill\break
$g_{\tau \check r}=g_{\check r\check s}N^{\check s}=-\epsilon
\, {}^3g_{\check r\check s}N^{\check s}$,\hfill\break
$g^{\tau \tau}=\epsilon N^{-2}$, \hfill\break
$g^{\tau \check r}=-\epsilon N^{\check r}/N^2$, \hfill\break
$g^{\check r\check s}=\gamma^{\check r\check s}+\epsilon
{{N^{\check r}N^{\check s}}\over {N^2}}=-\epsilon
[{}^3g^{\check r\check s}- {{N^{\check r}N^{\check s}}\over {N^2}}]$,
\hfill\break
\hfill\break
${{\partial}\over {\partial z^{\mu}_{\tau}}}=l_{\mu}\, {{\partial}\over
{\partial N}}+z_{{\check s}\mu}\gamma^{{\check s}{\check r}} {{\partial}\over
{\partial N^{\check r}}}=l_{\mu}\, {{\partial}\over {\partial N}}-\epsilon
z_{{\check s}\mu}\, {}^3g^{{\check s}{\check r}} {{\partial}\over
{\partial N^{\check r}}}$, \hfill\break
$d^4z=N\sqrt{\gamma}d\tau d^3\sigma$.

The rest frame form of a timelike fourvector $p^{\mu}$ is $\stackrel
{\circ}{p}{}^{\mu}=\eta \sqrt{\epsilon p^2} (1;\vec 0)= \eta^{\mu o}\eta
\sqrt{\epsilon p^2}$,
$\stackrel{\circ}{p}{}^2=p^2$, where $\eta =sign\, p^o$.
The standard Wigner boost transforming $\stackrel{\circ}{p}{}^{\mu}$ into
$p^{\mu}$ is

\begin{eqnarray}
L^{\mu}{}_{\nu}(p,\stackrel{\circ}{p})&=&\epsilon^{\mu}_{\nu}(u(p))=
\nonumber \\
&=&\eta^{\mu}_{\nu}+2{ {p^{\mu}{\stackrel{\circ}{p}}_{\nu}}\over {\epsilon
p^2}}-
{ {(p^{\mu}+{\stackrel{\circ}{p}}^{\mu})(p_{\nu}+{\stackrel{\circ}{p}}_{\nu})}
\over {p\cdot \stackrel{\circ}{p} +\epsilon p^2} }=\nonumber \\
&=&\eta^{\mu}_{\nu}+2u^{\mu}(p)u_{\nu}(\stackrel{\circ}{p})-{ {(u^{\mu}(p)+
u^{\mu}(\stackrel{\circ}{p}))(u_{\nu}(p)+u_{\nu}(\stackrel{\circ}{p}))}
\over {1+u^o(p)} },\nonumber \\
&&{} \nonumber \\
\nu =0 &&\epsilon^{\mu}_o(u(p))=u^{\mu}(p)=p^{\mu}/\eta
\sqrt{\epsilon p^2}, \nonumber \\
\nu =r &&\epsilon^{\mu}_r(u(p))=(-u_r(p); \delta^i_r-{ {u^i(p)u_r(p)}\over
{1+u^o(p)} }).
\label{c9}
\end{eqnarray}

The inverse of $L^{\mu}{}_{\nu}(p,\stackrel{\circ}{p})$ is $L^{\mu}{}_{\nu}
(\stackrel{\circ}{p},p)$, the standard boost to the rest frame, defined by

\begin{equation}
L^{\mu}{}_{\nu}(\stackrel{\circ}{p},p)=L_{\nu}{}^{\mu}(p,\stackrel{\circ}{p})=
L^{\mu}{}_{\nu}(p,\stackrel{\circ}{p}){|}_{\vec p\rightarrow -\vec p}.
\label{c10}
\end{equation}

Therefore, we can define the following vierbeins [the $\epsilon^{\mu}_r(u(p))$'s
are also called polarization vectors; the indices r, s will be used for A=1,2,3
and $\bar o$ for $A=o$]

\begin{eqnarray}
&&\epsilon^{\mu}_A(u(p))=L^{\mu}{}_A(p,\stackrel{\circ}{p}),\nonumber \\
&&\epsilon^A_{\mu}(u(p))=L^A{}_{\mu}(\stackrel{\circ}{p},p)=\eta^{AB}\eta
_{\mu\nu}\epsilon^{\nu}_B(u(p)),\nonumber \\
&&{} \nonumber \\
&&\epsilon^{\bar o}_{\mu}(u(p))=\eta_{\mu\nu}\epsilon^{\nu}_o(u(p))=u_{\mu}(p),
\nonumber \\
&&\epsilon^r_{\mu}(u(p))=-\delta^{rs}\eta_{\mu\nu}\epsilon^{\nu}_r(u(p))=
(\delta^{rs}u_s(p);\delta^r_j-\delta^{rs}\delta_{jh}{{u^h(p)u_s(p)}\over
{1+u^o(p)} }),\nonumber \\
&&\epsilon^A_o(u(p))=u_A(p),
\label{c11}
\end{eqnarray}

\noindent which satisfy

\begin{eqnarray}
&&\epsilon^A_{\mu}(u(p))\epsilon^{\nu}_A(u(p))=\eta^{\mu}_{\nu},\nonumber \\
&&\epsilon^A_{\mu}(u(p))\epsilon^{\mu}_B(u(p))=\eta^A_B,\nonumber \\
&&\eta^{\mu\nu}=\epsilon^{\mu}_A(u(p))\eta^{AB}\epsilon^{\nu}_B(u(p))=u^{\mu}
(p)u^{\nu}(p)-\sum_{r=1}^3\epsilon^{\mu}_r(u(p))\epsilon^{\nu}_r(u(p)),
\nonumber \\
&&\eta_{AB}=\epsilon^{\mu}_A(u(p))\eta_{\mu\nu}\epsilon^{\nu}_B(u(p)),\nonumber
\\
&&p_{\alpha}{{\partial}\over {\partial p_{\alpha}} }\epsilon^{\mu}_A(u(p))=
p_{\alpha}{{\partial}\over {\partial p_{\alpha}} }\epsilon^A_{\mu}(u(p))
=0.
\label{c12}
\end{eqnarray}

The Wigner rotation corresponding to the Lorentz transformation $\Lambda$ is

\begin{eqnarray}
R^{\mu}{}_{\nu}(\Lambda ,p)&=&{[L(\stackrel{\circ}{p},p)\Lambda^{-1}L(\Lambda
p,\stackrel{\circ}{p})]}^{\mu}{}_{\nu}=\left(
\begin{array}{cc}
1 & 0 \\
0 & R^i{}_j(\Lambda ,p)
\end{array}
\right) ,\nonumber \\
{} && {}\nonumber \\
R^i{}_j(\Lambda ,p)&=&{(\Lambda^{-1})}^i{}_j-{ {(\Lambda^{-1})^i{}_op_{\beta}
(\Lambda^{-1})^{\beta}{}_j}\over {p_{\rho}(\Lambda^{-1})^{\rho}{}_o+\eta
\sqrt{\epsilon p^2}} }-\nonumber \\
&-&{{p^i}\over {p^o+\eta \sqrt{\epsilon p^2}} }[(\Lambda^{-1})^o{}_j-
{ {((\Lambda^{-1})^o
{}_o-1)p_{\beta}(\Lambda^{-1})^{\beta}{}_j}\over {p_{\rho}(\Lambda^{-1})^{\rho}
{}_o+\eta \sqrt{\epsilon p^2}} }].
\label{c13}
\end{eqnarray}

The polarization vectors transform under the
Poincar\'e transformations $(a,\Lambda )$ in the following way

\begin{equation}
\epsilon^{\mu}_r(u(\Lambda p))=(R^{-1})_r{}^s\, \Lambda^{\mu}{}_{\nu}\,
\epsilon^{\nu}_s(u(p)).
\label{c14}
\end{equation}

\vfill\eject

\section{More on Dixon's Multipoles.}

Let us add other forms of the Dixon multipoles.

In  the case of the fluid configurations treated in Section II and IV,
the Hamilton equations generated by the Dirac Hamiltonian (\ref{II49})
in the gauge ${\vec q}_{sys}\approx 0$ [$\vec \lambda (\tau )=0$]
imply [in Ref.\cite{dixon} this is a consequence of
$\partial_{\mu}T^{\mu\nu}\, {\buildrel \circ
\over
=}\, 0$]

\begin{eqnarray}
{{dp_T^{\mu}(T_s)}\over {dT_s}}\, &{\buildrel \circ \over =}\,& 0,\quad
for\, n=0,\nonumber \\
{{d p_T^{\mu_1...\mu_n\mu}(T_s)}\over {dT_s}}\, &{\buildrel \circ \over =}\,&
-nu^{(\mu_1}(p_s) p_T^{\mu_2...\mu_n)\mu}(T_s)+n t_T^{(\mu_1...\mu_n)\mu}(T_s),
\quad n\geq 1.
\label{d1}
\end{eqnarray}

Let us define for $n \geq 1$

\begin{eqnarray}
b_T^{\mu_1...\mu_n\mu}(T_s)&=&p_T^{(\mu_1...\mu_n\mu )}(T_s)=\nonumber \\
&=&
\epsilon^{(\mu_1}_{r_1}(u(p_s)).... \epsilon^{\mu_n}_{r_n}(u(p_s))\epsilon^{\mu )}_A(u(p_s))
I_T^{r_1..r_nA\tau}(T_s) ,\nonumber \\
 &&{}\nonumber \\
 \epsilon^{r_1}_{\mu_1}(u(p_s))....\epsilon^{r_n}_{\mu_n}(u(p_s)) b_T^{\mu_1...\mu_n\mu}(T_s)
 &=&{1\over {n+1}} u^{\mu}(p_s) I_T^{r_1...r_n\tau\tau}(T_s)+\epsilon^{\mu}_r(u(p_s))
 I_T^{(r_1...r_nr)\tau}(T_s),\nonumber \\
 &&{}\nonumber \\
c_T^{\mu_1...\mu_n\mu}(T_s)&=&c_T^{(\mu_1...\mu_n
)\mu}(T_s)=p_T^{\mu_1...
\mu_n\mu}(T_s)-p_T^{(\mu_1...\mu_n\mu )}(T_s)=\nonumber \\
&=&[\epsilon^{\mu_1}_{r_1}(u(p_s))...\epsilon^{\mu_n}_{r_n}\epsilon^{\mu}_A(u(p_s))-
    \nonumber \\
 &-&\epsilon^{(\mu_1}_{r_1}(u(p_s))...\epsilon^{\mu_n}_{r_n}(u(p_s))
\epsilon_A^{\mu )}(u(p_s))] I_T^{r_1..r_nA\tau}(T_s),\nonumber \\
 &&{}\nonumber \\
&&c_T^{(\mu_1...\mu_n\mu )}(T_s)=0,\nonumber \\
 \epsilon^{r_1}_{\mu_1}(u(p_s))....\epsilon^{r_n}_{\mu_n}(u(p_s)) c_T^{\mu_1...\mu_n\mu}(T_s)
 &=& {n\over {n+1}}u^{\mu}(p_s) I_T^{r_1...r_n\tau\tau}(T_s) +\nonumber \\
 &+&\epsilon^{\mu}_r(u(p_s)) [ I_T^{r_1...r_nr\tau}(T_s) - I_T^{(r_1...r_nr)\tau}(T_s)],
\label{d2}
\end{eqnarray}

\noindent and then for $n\geq 2$

\begin{eqnarray}
d_T^{\mu_1...\mu_n\mu\nu}(T_s)&=&d_T^{(\mu_1...\mu_n)(\mu\nu )}(T_s)=
t_T^{\mu_1...\mu_n\mu\nu}(T_s)-\nonumber \\
 &-&{{n+1}\over
n}[t_T^{(\mu_1...\mu_n\mu )\nu}(T_s)+t_T^{(\mu_1...\mu_n\nu )\mu}
(T_s)]+\nonumber \\
 &+&{{n+2}\over n}t_T^{(\mu_1...\mu_n\mu\nu
)}(T_s)=\nonumber
\\ &=&\Big[ \epsilon^{\mu_1}_{r_1} ... \epsilon^{\mu_n}_{r_n}
\epsilon^{\mu}_A \epsilon^{\nu}_B-
 {{n+1}\over n}\Big( \epsilon^{(\mu_1}_{r_1} ... \epsilon^{\mu_n}_{r_n} \epsilon^{\mu )}_A
 \epsilon^{\nu}_B+\nonumber \\
 &+&\epsilon^{(\mu_1}_{r_1} ... \epsilon^{\mu_n}_{r_n} \epsilon^{\nu )}_B
 \epsilon^{\mu}_A\Big) +{{n+2}\over n} \epsilon^{(\mu_1}_{r_1} .. \epsilon^{\mu_n}_{r_n}
 \epsilon^{\mu}_A \epsilon^{\nu )}_B\Big] (u(p_s))\nonumber \\
 && I_T^{r_1..r_nAB}(T_s),\nonumber \\
 &&{}\nonumber \\
 &&d_T^{(\mu_1...\mu_n\mu )\nu}(T_s)=0,\nonumber \\
 &&{}\nonumber \\
 \epsilon^{r_1}_{\mu_1}(u(p_s))....\epsilon^{r_n}_{\mu_n}(u(p_s)) d_T^{\mu_1...\mu_n\mu\nu}(T_s)
 &=& {{n-1}\over {n+1}} u^{\mu}(p_s)u^{\nu}(p_s) I_T^{r_1...r_n\tau\tau}(T_s)+\nonumber \\
 &+&{1\over n} [u^{\mu}(p_s)\epsilon^{\nu}_r(u(p_s))+u^{\nu}(p_s)\epsilon^{\mu}_r(u(p_s))]
   \nonumber \\
 &&[(n-1) I_T^{r_1...r_nr\tau}(T_s)+ I_T^{(r_1...r_nr)\tau}(T_s)]+\nonumber \\
 &+&\epsilon^{\mu}_{s_1}(u(p_s))\epsilon^{\nu}_{s_2}(u(p_s)) [I_T^{r_1...r_ns_1s_2}(T_s)
 -\nonumber \\
  &-&{{n+1}\over n}( I_T^{(r_1...r_ns_1)s_2}(T_s)+ I_T^{(r_1...r_ns_2)s_1}(T_s)) +\nonumber \\
 &+& I_T^{(r_1...r_ns_1s_2)}(T_s)].
\label{d3}
\end{eqnarray}

Then Eqs.(\ref{d1}) may be rewritten in the form

\begin{eqnarray}
&&1)\quad n=1\nonumber \\
 &&{}\nonumber \\
t^{\mu\nu}_T(T_s)&=&t^{(\mu\nu )}_T(T_s)\, {\buildrel \circ \over =}\,
p_T^{\mu}(T_s)u^{\nu }(p_s)+{1\over 2}{d\over
{dT_s}}(S^{\mu\nu}_T(T_s)[\alpha ]+2b_T
^{\mu\nu}(T_s)),\nonumber \\
&&\Downarrow \nonumber \\
 t^{\mu\nu}_T(T_s)\, &{\buildrel \circ \over
=}\,&p_T^{(\mu}(T_s)u^{\nu )}(p_s) +{d\over
{dT_s}}b_T^{\mu\nu}(T_s)=P^{\tau}u^{\mu}(p_s)u^{\nu}(p_s)+
P^ru^{(\mu}(p_s)\epsilon^{\nu )}_r(u(p_s))+\nonumber \\
&+&\epsilon^{(\mu}_r(u(p_s))u^{\nu )}(p_s) I_T^{r\tau\tau}(T_s)
+\nonumber \\
 &+&\epsilon^{(\mu}_r(u(p_s))\epsilon^{\nu )}_s(u(p_s))
 I_T^{rs\tau}(T_s),\nonumber \\
 {d\over {dT_s}}S^{\mu\nu}_T(T_s)[\alpha ]\, &{\buildrel
\circ \over =}\,&2p_T^{[\mu}(T_s) u^{\nu
]}(p_s)=2P^r_{\phi}\epsilon^{[\mu}_r(u(p_s))u^{\nu ]}(p_s) \approx
0,\nonumber \\
 &&{}\nonumber \\
 &&2)\quad n=2\quad [identity\,\,
t_T^{\rho\mu\nu}=t_T^{(\rho\mu )\nu}+t_T^{(\rho
\nu )\mu}+t_T^{(\mu\nu )\rho}]\nonumber \\
 &&{}\nonumber \\
2t_T^{(\rho\mu )\nu}(T_s)\, &{\buildrel \circ \over =}\,& 2u^{(\rho}(p_s)b_T
^{\mu )\nu}(T_s)+u^{(\rho}(p_s)S_T^{\mu )\nu}(T_s)[\alpha ]
+{d\over {dT_s}}(b_T^{\rho\mu\nu}
(T_s)+c_T^{\rho\mu\nu}(T_s)),\nonumber \\
 &&\Downarrow \nonumber \\
t_T^{\rho\mu\nu}(T_s)\, &{\buildrel \circ \over =}\,&u^{\rho}(p_s)b_T
^{\mu\nu}(T_s)+S_T^{\rho (\mu}(T_s)[\alpha ]
u^{\nu )}(p_s)+{d\over {dT_s}}({1\over 2}b_T
^{\rho\mu\nu}(T_s)-c_T^{\rho\mu\nu}(T_s)),\nonumber \\
&&{}\nonumber \\
&&3) \quad n \geq 3 \nonumber \\
&&{}\nonumber \\
t_T^{\mu_1...\mu_n\mu\nu}(T_s)\, &{\buildrel \circ \over =}\,& d_T^{\mu_1...
\mu_n\mu\nu}(T_s)+u^{(\mu_1}(p_s)b_T^{\mu_2...\mu_n)\mu\nu}(T_s)+2u^{(\mu
_1}(p_s)c_T^{\mu_2...\mu_n)(\mu\nu )}(T_s)+\nonumber \\
&=&{2\over n}c_T^{\mu_1...\mu_n(\mu}(T_s)u^{\nu )}(p_s)+{d\over {dT_s}}
[{1\over {n+1}}b_T^{\mu_1...\mu_n\mu\nu}(T_s)+{2\over n}c_T^{\mu_1...\mu
_n(\mu\nu )}(T_s)],
\label{d4}
\end{eqnarray}

This allows \cite{dixon} to rewrite $< T^{\mu\nu},f >$ in the following form

\begin{eqnarray}
< T^{\mu\nu},f > &=&\int dT_s \int {{d^4k}\over {(2\pi )^4}} \tilde
f(k) e^{-ik\cdot x_s(T_s)} \Big[ u^{(\mu}(p_s)p_T^{\nu
)}(T_s)-ik_{\rho}S^{\rho (\mu}_T(T_s)[\alpha  ]u^{\nu
)}(p_s)+\nonumber
\\ &+&\sum_{n=2}^{\infty}{{(-i)^n}\over {n!}} k_{\rho_1}...k_{\rho_n}
{\cal I}_T^{\rho_1...\rho_n\mu\nu}(T_s)\Big],
\label{d5}
\end{eqnarray}

\noindent with

\begin{eqnarray}
{\cal I}_T^{\mu_1...\mu_n\mu\nu}(T_s)&=&{\cal I}_T^{(\mu_1...\mu_n
)(\mu\nu )}(T_s)= d_T^{\mu_1...\mu_n\mu\nu}(T_s)-\nonumber \\
&-&{2\over {n-1}}u^{(\mu_1}(p_s) c_T^{\mu_2...\mu_n )(\mu\nu
)}(T_s)+{2\over n} c_T^{\mu_1...\mu_n(\mu}(T_s)u^{\nu
)}(p_s)=\nonumber \\
 &=&\Big[ \epsilon^{\mu_1}_{r_1}...\epsilon^{\mu_n}_{r_n} \epsilon^{\mu}_A \epsilon^{\nu}_B-
 {{n+1}\over n}\Big( \epsilon^{(\mu_1}_{r_1}...\epsilon^{\mu_n}_{r_n} \epsilon^{\mu )}_A
 \epsilon^{\nu}_B+\nonumber \\
 &+&\epsilon^{(\mu_1}_{r_1}...\epsilon^{\mu_n}_{r_n} \epsilon^{\nu )}_B \epsilon^{\mu}_A\Big) +
 {{n+2}\over n} \epsilon^{(\mu_1}_{r_1}...\epsilon^{\mu_n}_{r_n}
 \epsilon^{\mu}_A \epsilon^{\nu )}_B\Big] (u(p_s)0 I_T^{r_1..r_nAB}(T_s)-\nonumber \\
 &-&\Big[ {2\over {n-1}} u^{(\mu_1}(p_s) \Big( \epsilon^{\mu_2}_{r_1}...
 \epsilon^{\mu_n)}_{r_{n-1}} \epsilon^{(\mu}_{r_n} \epsilon^{\nu )}_A-
 \epsilon^{(\mu_2}_{r_1}...\epsilon^{\mu_n)}_{r_{n-1}} \epsilon^{(\mu}_{r_n}
 \epsilon^{\nu ))}_A\Big)-\nonumber \\
 &-&{2\over n} \Big( \epsilon^{\mu_1}_{r_1}...\epsilon^{\mu_n}_{r_n} \epsilon^{(\mu}_A-
 \epsilon^{(\mu_1}_{r_1}... \epsilon^{\mu_n}_{r_n} \epsilon^{(\mu )}_A u^{\nu )}(p_s)
 \big] (u(p_s)0 I_T^{r_1..r_nA\tau}(T_s),\nonumber \\
 &&{}\nonumber \\
 &&{\cal I}_T^{(\mu_1...\mu_n\mu )\nu}(T_s)=0,\nonumber \\
 &&{}\nonumber \\
 \epsilon^{r_1}_{\mu_1}(u(p_s))....\epsilon^{r_n}_{\mu_n}(u(p_s))
 {\cal I}_T^{\mu_1...\mu_n\mu\nu}(T_s)
 &=& {{n+3}\over {n+1}} u^{\mu}(p_s)u^{\nu}(p_s) I_T^{r_1...r_n\tau\tau}(T_s)+\nonumber \\
 &+&{1\over n}[u^{\mu}(p_s)\epsilon^{\nu}_r(u(p_s))+u^{\nu}(p_s)\epsilon^{\mu}_r(u(p_s))]
 I_T^{r_1...r_nr\tau}(T_s)+\nonumber \\
 &+&\epsilon^{\mu}_{s_1}(u(p_s))\epsilon^{\nu}_{s_2}(u(p_s)) [I_T^{r_1...r_ns_1s_2}(T_s)
 -\nonumber \\
 &-&{{n+1}\over n}( I_T^{(r_1...r_ns_1)s_2}(T_s)+ I_T^{(r_1...r_ns_2)s_1}(T_s)) +
    \nonumber \\
 &+&I_T^{(r_1...r_ns_1s_2)}(T_s)].
\label{d6}
\end{eqnarray}

Finally, a set of multipoles equivalent to the ${\cal
I}_T^{\mu_1...\mu_n\mu\nu}$ is

\begin{eqnarray}
&&n \geq 0\nonumber \\
 &&{}\nonumber \\
J_T^{\mu_1...\mu_n\mu\nu\rho\sigma}(T_s)&=&J_T^{(\mu_1...\mu_n)[\mu\nu ][\rho
\sigma ]}(T_s)= {\cal I}_T^{\mu_1...\mu_n[\mu [\rho\nu ]\sigma ]}(T_s)=\nonumber \\
&=&t_T^{\mu_1...\mu_n[\mu [\rho\nu ]\sigma ]}(T_s)-{1\over {n+1}}\Big[
u^{[\mu}(p_s)p_T^{\nu ]\mu_1...\mu_n[\rho\sigma ]}(T_s)+\nonumber \\
 &+&u^{[\rho}(p_s)p_T
^{\sigma ]\mu_1...\mu_n[\mu\nu ]}(T_s)\Big] =\nonumber \\
 &=&\Big[ \epsilon^{\mu_1}_{r_1} .. \epsilon^{\mu_n}_{r_n} \epsilon^{[\mu}_r
 \epsilon^{[\rho}_s \epsilon^{\nu ]}_A \epsilon^{\sigma ]}_B\Big] (u(p_s))
 I_T^{r_1..r_nAB}(T_s)-\nonumber \\
  &-&{1\over {n+1}}\Big[ u^{[\mu}(p_s) \epsilon^{\nu ]}_r(u(p_s)) \epsilon^{[\rho}_s(u(p_s))
  \epsilon^{\sigma ]}_A(u(p_s))+\nonumber \\
   &+&u^{[\rho}(p_s) \epsilon^{\sigma ]}_r(u(p_s))
  \epsilon^{[\mu}_s(u(p_s)) \epsilon^{\nu ]}_A(u(p_s))\Big]\nonumber \\
  &&\epsilon^{\mu_1}_{r_1}(u(p_s))...\epsilon^{\mu_n}_{r_n}(u(p_s))
  I_T^{rr_1..r_nsA\tau}(T_s),\nonumber \\
 &&{}\nonumber \\
 &&[ (n+4)(3n+5)\, linearly\, independent\, components],\nonumber \\
&&{}\nonumber \\
 && n \geq 1\nonumber \\
  &&{}\nonumber \\
u_{\mu_1}(p_s)&&J_T^{\mu_1...\mu_n\mu\nu\rho\sigma}(T_s)=
J_T^{\mu_1...\mu_{n-1}(\mu_n\mu\nu )\rho\sigma}(T_s)=0,\nonumber \\
 &&{}\nonumber \\
 && n \geq 2 \nonumber \\
 &&{}\nonumber \\
{\cal I}_T^{\mu_1...\mu_n\mu\nu}(T_s)&=&{{4(n-1)}\over {n+1}}
J_T^{(\mu_1...\mu_{n-1}|
\mu |\mu_n)\nu}(T_s),\nonumber \\
 &&{}\nonumber \\
  \epsilon^{r_1}_{\mu_1}(u(p_s))....\epsilon^{r_n}_{\mu_n}(u(p_s))
  J_T^{\mu_1...\mu_n\mu\nu\rho\sigma}(T_s) &=& \Big[ \epsilon^{[\mu}_r
 \epsilon^{[\rho}_s \epsilon^{\nu ]}_A \epsilon^{\sigma ]}_B\Big] (u(p_s))
 I_T^{r_1..r_nAB}(T_s)-\nonumber \\
  &-&{1\over {n+1}}\Big[ u^{[\mu}(p_s) \epsilon^{\nu ]}_r(u(p_s)) \epsilon^{[\rho}_s(u(p_s))
  \epsilon^{\sigma ]}_A(u(p_s))+\nonumber \\
   &+&u^{[\rho}(p_s) \epsilon^{\sigma ]}_r(u(p_s))
  \epsilon^{[\mu}_s(u(p_s)) \epsilon^{\nu ]}_A(u(p_s))\Big]
  I_T^{rr_1..r_nsA\tau}(T_s).\nonumber \\
  &&{}
\label{d7}
\end{eqnarray}

The $J_T^{\mu_1...\mu_n\mu\nu\rho\sigma}$ are the Dixon
``$2^{n+2}$-pole inertial moment tensors" of the extended system: they
[or equivalently the ${\cal I}_T^{\mu_1...
\mu_n\mu\nu}$'s] determine its
energy-momentum tensor together with the monopole $p^{\mu}_T$ and the
spin dipole $S^{\mu\nu}_T$. The equations $\partial_{\mu} T^{\mu\nu}\,
{\buildrel \circ \over =}\, 0$ are satisfied due to the equations of
motion (\ref{d4}) for $P^{\mu}_T$ and $S^{\mu\nu}_T$ [the so called
Papapetrou-Dixon-Souriau equations given in Eqs.(\ref{IV18})] without
the need of the equations of motion for the
$J_T^{\mu_1...\mu_n\mu\nu\rho\sigma}$. When all the multipoles
$J_T^{\mu_1...\mu_n\mu\nu\rho\sigma}$ are zero [or negligible] one
speaks of a pole-dipole field configuration of the perfect fluid.

\vfill\eject


\begin{references}

\bibitem{brown}J.D.Brown, Class.Quantum Grav. {\bf 10}, 1579 (1993).
\bibitem {israel} W.Israel, ``Covariant Fluid Mechanics and Thermodynamics: An
Introduction", in ``Relativistic Fluid Dynamics", eds. A.Anile and
Y.Choquet-Bruhat, Lecture Notes in Math. n. 1385 (Springer, Berlin,
1989).
\bibitem{lin}C.C.Lin, `Hydrodynamics of Helium II', in ``Liquid Helium", ed.
G.Careri (Academic Press, New York, 1963).\hfill\break
J.Serrin, `Mathematical Principles of Classical Fluid Mechanics', in ``Handbuch
der Physik" vol. 8, eds. S.Fl\"ugge and C.Truesdell (Springer, Berlin, 1959).
\bibitem{tul}J.Kijowski and W.M.Tulczyjew, ``A Symplectic Framework for Field
Theories", Lecture Notes in Physics vol. 107 (Springer, Berlin, 1979);
`Relativistic Hydrodynamics of Isentropic Flows", Mem.Acad.Sci.Torino V {\bf 6},
3 (1982).
\bibitem{tul1}H.P.K\"unzle and J.M.Nester, J.Math.Phys. {\bf 25}, 1009 (1984).
\bibitem{magli}J.Kijowski and G.Magli, Class. Quantum Grav. {\bf 15}, 3891
(1998).
\bibitem{adler}S.Adler and T.Buchert, Astron.Astrophys. {\bf 343}, 317 (1999)
ASTRO-PH/9806320.
\bibitem{bao}D.Bao, J.Marsden and R.Walton, Commun.Math.Phys. {\bf 99}, 319
(1985).\hfill\break
D.D.Holm, ``Hamiltonian Techniques for Relativistic Fluid Dynamics and Stability
Theory", in ``Relativistic Fluid Dynamica", eds. A.Anile and Y.Choquet-Bruhat
(Springer, Berlin, 1989).
\bibitem{brk}J.D.Brown and K.Kuchar, Phys.Rev. {\bf D51}, 5600 (1995).
\bibitem{lus}L.Lusanna, Int.J.Mod.Phys. {\bf 12A}, 645 (1997).
\bibitem{india}L.Lusanna, ``Towards a Unified Description of the Four
Interactions in Terms of Dirac-Bergmann Observables", invited
contribution to the book of the Indian National Science Academy for
the International Mathematics Year 2000 AD (HEP-TH/9907081). ``Tetrad
Gravity and Dirac's Observables", talk given at the Conf. ``Constraint
Dynamics and Quantum Gravity 99", Villasimius 1999
(GR-QC/9912091).``The Rest-Frame Instant Form of Dynamics and Dirac's
Observables", talk given at the Int.Workshop``Physical Variables in
Gauge Theories", Dubna 1999.


\bibitem{ku}K.Kuchar, J.Math.Phys. {\bf 17}, 777, 792, 801 (1976).
\bibitem{pauri}M.Pauri and M.Prosperi, J.Math.Phys. {\bf 16}, 1503 (1975).
\bibitem{mate}L.Lusanna and M.Materassi, ``The Canonical Decomposition in
Collective and Relative Variables of a Klein-Gordon Field in the
Rest-Frame Wigner-Covariant Instant Form", to appear in
Int.J.Mod.Phys. A (HEP-TH/9904202).
\bibitem{lon}G.Longhi and M.Materassi, Int.J.Mod.Phys. {\bf A14}, 3397
(1999)(HEP-TH/9809024); J.Math.Phys. {\bf 40}, 480 (1999) (HEP-TH/9803128).
\bibitem{iten}D.Alba, L.Lusanna and M.Pauri, ``Center of Mass, Rotational
Kinematics and Multipolar Expansions for the Relativistic and
Non-Relativistic N-Body Problems in the Rest-Frame Instant Form", in
preparation.
\bibitem{crater}H.Crater and L.Lusanna, ``The Rest-Frame Darwin Potential
from the Lienard-Wiechert Solution in the Radiation Gauge", Firenze
univ. preprint 2000 (HEP-TH/0001046).
 \bibitem{bini}D.Bini, G.Gemelli and R.Ruffini, ``Spining Test Particles in
 General Relativity: Nongeodesic Motion in the Reissner-Nordstr\"om Spacetime"
 talk at the III  W.Fairbank Meeting and I ICRA Network Workshop
 ``The Lense-Thirring Effect", Roma-Pescara 1998.
 \bibitem{bei}W.Beiglb\"ock, Commun.Math.Phys. {\bf 5}, 106 (1967). J.Ehlers and
 E.Rudolph, Gen.Rel.Grav. {\bf 8}, 197 (1977). R.Schattner, Gen.Rel.Grav.
 {\bf 10}, 377 and 395 (1978).
\bibitem{dixon}W.G.Dixon, J.Math.Phys. {\bf 8}, 1591 (1967). ``Extended Objects
in General Relativity: their Description and Motion", in ``Isolated Gravitating
Systems in General Relativity", ed. J.Ehlers (North-Holland, Amsterdam,
1979).
\bibitem{thorne}K.S.Thorne, ``The Theory of Gravitational Radiation: an
Introductory Review" in `Gravitational Radiation' eds. N.Deruelle and
T.Piran, 1982 NATO-ASI School at Les Houches (North Holland,
Amsterdam, 1983); Rev.Mod.Phys. {\bf 52}, 299 (1980).
\bibitem{ryan}M.P.Ryan Jr. and L.C.Shepley, ``Homogeneous Relativistic
Cosmologies" (Princeton Univ. Press, Princeton, 1975).
\bibitem{damour}L.Blanchet, T.Damour and G.Sch\"afer, Mon.Not.R.astr.Soc.
{\bf 242}, 289 (1990).
\bibitem{an}A.M.Anile, ``Relativistic Fluids and Magnetofluids" (Cambridge
Univ.Press, Cambridge, 1989).

\bibitem{russo1}L.Lusanna and S.Russo, ``Tetrad Gravity I): A New Formulation",
Firenze Univ. preprint 1998 (GR-QC/9807073).
\bibitem{russo2}L.Lusanna and S.Russo, ``Tetrad Gravity II): Dirac's
Observables", Firenze Univ. preprint 1998 (GR-QC/9807074).
\bibitem{russo3}R.DePietri and L.Lusanna, ``Tetrad Gravity III): Asymptotic
Poincar\'e Charges, the Physical Hamiltonian and Void Spacetimes",
(GR-QC/9909025).

\bibitem{york}L.Smarr and J.W.York jr., Phys.Rev. {\bf D17}, 2529 (1978).
\hfill\break
J.W.York jr., ``Kinematics and Dynamics of General Relativity", in ``Sources
of Gravitational Radiation", ed. L.L.Smarr (Cambridge Univ.Press, Cambridge,
1979).
\bibitem{smarr} L.Smarr, C.Taubes and J.R.Wilson, ``General Relativistic
Hydrodynamics: the Comoving, Eulerian and Velocity Potential Formalisms", in
``Essays in General Relativity: a Festschrift for Abraham Taub", ed. F.J.Tipler
(Academic Press, New York, 1980).
\bibitem{asada}H.Asada, M.Shibata and T.Futamase, ``Post-Newtonian Hydrodynamic
Equations Using the (3+1) Formalism in General Relativity", Osaka Univ.
preprint (GR-QC/9606041).
\bibitem{num1}J.A.Font, N.Stergioulas and K.D.Kokkotas, ``Nonlinear
Hydrodynamical Evolution of Rotating Relativistic Stars: Numerical Methods and
Code Tests" (GR-QC/9908010).
\bibitem{num2}T.W.Baumgarte, S.A.Hughes, L.Rezzolla, S.L.Shapiro and
M.Shibata, ``Implementing Fully relativistic Hydrodynamics in Three Dimensions"
(GR-QC/9907098).
\bibitem{magli1}J.Kijowski and G.Magli, Rep.Math.Phys. {\bf 39}, 99 (1997).
\bibitem{magli2}J.Kijowski and G.Magli,J.Geom.Phys. {\bf 9}, 207 (1992).
\bibitem{antoci}S.Antoci and L.Mihic, ``A Four-Dimensional Hooke's Law can
Encompass Linear Elasticity and Inertia" (GR-QC/9906094).
\bibitem{roy}R.Maartens, ``Causal Thermodynamics in Relativity", Lectures at
the H.Rund Workshop on `Relativity and Thermodynamics' 1996 (ASTRO-PH/9609119).
\bibitem{carter}B.Carter, ``Covariant Theory of Conductivity in Ideal Fluid or
Solid Media", in ``Relativistic Fluid Dynamics", eds. A.Anile and
Y.Choquet-Bruhat, Lecture Notes in Math. n. 1385 (Springer, Berlin, 1989).
\bibitem {karsch}F.Karsch and D.E.Miller, Phys.Rev. {\bf D24}, 2564 (1981).
\bibitem{stewart}J.M.Stewart, ``Non-Equilibrium Relativistic Kinetic Theory",
Lecture Notes Phys. n.10 (Springer, Berlin, 1971).
\bibitem{hakim}R.Hakim, J.Math.Phys. {\bf 8}, 1315, 1379 (1967).
\bibitem{synge}J.L.Synge, ``The Relativistic Gas" (North-Holland, Amsterdam,
1957).
\bibitem{wee}Ch.G.van Weert, ``Some Problems in Relativistic Hydrodynamics",
in ``Relativistic Fluid Dynamics", eds. A.Anile and
Y.Choquet-Bruhat, Lecture Notes in Math. n. 1385 (Springer, Berlin, 1989).
\bibitem{lind}L.Lindblom, Ann.Phys.(N.Y.) {\bf 247}, 1 (1996).
\bibitem{anil} A.M.Anile, D.Pav\'on and V.Romano, ``The Case for Hyperbolic
Theories of Dissipation in Relativistic Fluids", (GR-QC/9810014).
\bibitem{num}J.M.Mart\'i and E.M\"uller, ``Numerical Hydrodynamics in Special
Relativity", (ASTRO-PH/9906333).
\bibitem{ger}R.Geroch and L.Lindblom, Ann.Phys.(N.Y.) {\bf 207}, 394 (1991).
\bibitem{liu}I.Liu, I.M\"uller and T.Ruggeri, Ann.Phys.(N.Y.) {\bf 169}, 191
(1986).
\bibitem{gl}R.Geroch and L.Lindblom, Phys.Rev. {\bf D41}, 1855 (1990).
\bibitem{calzetta}E.Calzetta, Class.Quantum Grav. {\bf 15}, 653 (1998),
(GR-QC/9708048).
\bibitem{mu}I.M\"uller, Z.Phys. {\bf 198}, 329 (1967).
\bibitem{isr}W.Israel, Ann.Phys.(N.Y.) {\bf 100}, 310 (1976).
\bibitem{is}W.Israel and J.M.Stewart, Ann.Phys.(N.Y.) {\bf 118}, 341 (1979).
\bibitem{jou}D.Jou, G.Leblon and J.Casas Vasquez, ``Extended Thermodynamics" 2nd
edition (Springer, Heidelberg, 1996).
\bibitem{ingo}I.M\"uller and T.Ruggeri, ``Rational Extended Thermodynamics",
2nd edition (Springer, Berlin, 1998).
\bibitem{appl}J.Peitz and S.Appl, ``3+1 Formulation of Non-Ideal Hydrodynamics",
(GR-QC/9710107). Class.Quantum Grav. {\bf 16}, 979 (1999).
\bibitem{uzan}J.P.Uzan, Class.Quantum Grav. {\bf 15}, 1063 and 3737 (1998).
\hfill\break
J.M.Stewart, Class.Quantum Grav. {\bf 15}, 3731 (1998).
\bibitem{morri}P.J.Morrison, Physica {\bf 18D}, 410 (1986).
\bibitem{nose}S.Nose', J.Chem.Phys. {\bf 81}, 511 (1984).





\end{references}
\end{document}